\documentclass[12pt,a4paper]{conference}

\def\Chapter#1#2#3{\chapter{#1}\lhead{#3}\rhead{#2}}

\usepackage{fancyhdr}
\usepackage{epsfig}
\usepackage{makeidx,showidx}
\usepackage{multind}
\makeindex{author}
\makeindex{subject}

\newcommand{\beq}{\begin{equation}}
\newcommand{\eeq}[1]{\label{#1}\end{equation}}
\newcommand{\eeqn}{\end{equation}}

\newcommand{\beqa}{\begin{eqnarray}}
\newcommand{\eeqa}[1]{\label{#1}\end{eqnarray}}
\newcommand{\eeqan}{\end{eqnarray}}

\let\bar=\overbar

\renewcommand{\L}{{\cal L}}

\newcommand{\Dslash}{\not{\hbox{\kern-4pt $D$}}}
\newcommand{\dslash}{\not{\hbox{\kern-2pt $\del$}}}

\newcommand{\msb}{{\bar{\ssstyle M \kern -1pt S}}}

\begin{document}

%  FRONTMATTER:

\emptyheads

\begin{titlepage}

%\hbox to\hsize{\null}

\vfill
\begin{center}

\begin{Large}
{\bf
%Abstract Book of the \\[1ex]
XV International Conference on  \\[1ex]
Gravitational Microlensing \\[4ex]
{\LARGE Conference Book}}
\end{Large}

\vfill

\vfill\vfill

\begin{Large}
January 20--22, 2011\\[2ex]
University of Salerno, Italy\\

\vfill\vfill \vfill

Editors\\
  V. Bozza, S. Calchi Novati, L. Mancini, G. Scarpetta \\
\end{Large}
\hspace{0.2cm} {\large (Local Organizing Committee)}%
\vfill\vfill

Organized by the Astrophysics Group of the \\
Physics Department of the University of Salerno \\[2ex]

Sponsored by IIASS (International Institute for Advanced Scientific Studies), \\
INFN (Istituto Nazionale di Fisica Nucelare), International Ph.D.
in Astrophysics, and University of Salerno \\[2ex]

http://smc2011.physics.unisa.it \\

\end{center}

\end{titlepage}

\thispagestyle{empty}
\frontmatter\setcounter{page}{3}

\begin{center}
\begin{large}
{\bf Foreword}
\end{large}
\end{center}
\bigskip

This volume collects the abstracts in extended format of the
$15^{\mathrm{th}}$ Microlensing Conference held in the University
of Salerno on January 20-22, 2011. The Conference has gathered 68
scientists from 17 countries confirming microlensing as a mature
and established tool of research over a broad range of
astrophysical issues, from dark matter searches to the detection
of new extrasolar planets of very low mass, down to Earth-size or
below. The 3-days Conference has been preceeded by a 2-days school
on ``Modelling Planetary Microlensing Events'' focused on this
last issue. The abstracts collected here offer an updated snapshot
of the current researches in this field. The topics include: the
status of current surveys, planetary events, dark matter searches,
cosmological microlensing, theoretical investigations and an
outlook towards the future, with in particular a discussion on the
possible role to be played by microlensing searches for exoplanets
in the forecoming space missions, WFIRST and EUCLID. Finally, the
conference has been enriched by a series of topical speeches on
related issues from a non-microlensing point of view: the physics
of giant planet accretion and evolution, ``new physics'' and dark
matter in the LHC era and an update on the GAIA mission and its
potential to characterize planetary systems  with high-precision
astrometry.

\blankpage

\begin{center}
\begin{large}
{\bf Scientific Rationale}
\end{large}
\end{center}
\bigskip

25 years after the seminal intuition by Bohdan Paczynski,
microlensing has rapidly evolved into a new promise for modern
astrophysics. Indeed, the detection of celestial bodies by their
gravitational effects on the light of background sources has
proved to be a very powerful tool for the study of many aspects of
our Galaxy and beyond.

The original proposal of using microlensing to estimate the amount
of baryonic dark matter in the form of compact substellar bodies
is still topical. The observation of microlensing events toward
nearby galaxies still represents a hard challenge for present
observational facilities. Very strong efforts are under way for
upgrading the current strategies in particular toward the galaxy
M31.

Hundreds of microlensing events are discovered toward the Galactic
Bulge every year. This considerable amount of statistics can be
used to characterize the stellar populations of the bulge and the
disc of our Galaxy. After several years of observations, the
significance of the microlensing sample in the characterization of
our Galaxy is growing more and more.

Yet, the most intriguing perspective offered by microlensing is
represented by its power to discover new extrasolar planets of
very low mass, down to Earth-size or below. Several planetary
events have already been detected and studied, while many more are
expected as soon as future dedicated telescopes become
operational. Interestingly, microlensing is already being used to
estimate the abundance of planets around stars in the disc of our
Galaxy and to characterize their distributions in distance and
mass. As the planetary anomalies typically last only a few hours,
the cooperation among all observing groups is mandatory, in order
to characterize the events properly and maximize the scientific
achievements. Even amateur astronomers are now giving their
fundamental contribution. In this respect, microlensing stands as
a perfect example of how science can unite the whole mankind in a
common path toward pure knowledge.

Salerno Microlensing Conference 2011 will gather all people active
in this field, providing the state of the art of microlensing
searches and the perspectives opened by new methodologies and new
observational and computational facilities. Colloquia on dark
matter searches and planet formation theories are also foreseen as
a central part of the conference. The three-days conference will
be preceded by a school dedicated to the delicate issue of
efficient modelling of planetary microlensing events, which
requires major efforts and new ideas from new talents in order to
get access to the precious physical information hidden in
microlensing light curves. For one week in January 2011, Salerno
will thus be the place in which the present and the future of
microlensing will be unveiled.

\begin{tabbing}
this is the time for all sentient life forms to \= \kill
  \>Valerio Bozza  \\
  \>Local Organizing Committee \\
\end{tabbing}

%\blankpage

\begin{center}
\begin{large}
{\bf Welcome address}
\end{large}
\end{center}
\bigskip

On behalf of the Rector of Salerno University, professor Raimondo
Pasquino, the Dean of the Science Faculty, professor Mariella
Transirico, the Decan of the Department of Physics, professor
Ferdinando Mancini and the Organizing Committee, I welcome all the
participants to this exciting Conference.

This University has the peculiar status to be at the same time one
of the youngest Universities in Italy and the oldest one.

The re-establishment of Salerno University in the Modern Era dates
back to about forty years ago, and as you can see, in a very short
time it has grown in the Irno valley as a university campus, with
about forty thousand students, more than one thousand professors
and researchers, distributed in ten faculties.

Five years ago the Campus was completed with the Medical Science
Faculty, realizing the ideal link with one of the oldest Medical
Schools of the West, the so called ``Scuola Medica Salernitana''.

Salerno was not the seat of the first Medieval University, but the
Medical School was the first to organize studies and diffuse
culture on international scale during the medieval period.

According to the legend, the ``Scuola Medica Salernitana'' traces
its origins to four erudite: the Greek Pontus, the Jew Helinus,
the Arab Abdela, and the Latin Salernus. Thanks to the encounter
of these four cultures the Medical School established its
knowledge divulgating the Greek, Jewish, Arab and Latin medical
knowledge. The divulgation in the West of the Islamic and Greek
Medical Science is certainly due to Constantino the African (XI
century), thanks to his translations into Latin of the most
important Arabic and Greek medical treatises.

I like to read some words extracted from the nobel lecture of
Abdus Salam (1979)

``Scientific thought and its creation is the common and shared
heritage of mankind. In this respect, the history of science, like
the history of all civilization, has gone through cycles. Perhaps
I can illustrate this with an actual example.

Seven hundred and sixty years ago, a young Scotsman left his
native glens to travel south to Toledo in Spain. His name was
Michael, [...] Michael reached Toledo in 1217 AD. [...] From
Toledo, Michael travelled to Sicily, to the Court of Emperor
Frederick II. Visiting the medical school at Salerno, chartered by
Frederick in 1231, Michael met the Danish physician, Henrik
Harpestraeng [...] Henrik had come to Salerno to compose his
treatise on blood-letting and surgery. Henrik's sources were the
medical canons of the great clinicians of Islam, Al-Razi and
Avicenna, which only Michael the Scot could translate for him.
Toledo's and Salerno's schools, representing as they did the
finest synthesis of Arabic, Greek, Latin and Hebrew scholarship,
were some of the most memorable of international assays in
scientific collaboration''.

The ``Scuola Medica Salernitana'' was at the height of its fame in
the XIV century and carried on its teaching for nine centuries,
until Giocchino Murat's decree, dated the 25th of January 1812,
prescribed the closure.

After two centuries, we all, coming here in the Salerno University
from east and west countries, renew in some sense the old
tradition, establishing an ideal link with the cultural heritage
of the ``Scuola Medica Salernitana'' .

I hope you will enjoy your stay in Salerno and I wish you good and
useful work.

\begin{tabbing}
this is the time for all sentient life forms to \= \kill
  \>Gaetano Scarpetta  \\
  \>Chair of the Local Organizing Committee \\
\end{tabbing}

%\newpage
%\input acknowledgements.tex
\blankpage

\begin{center}
\begin{large}
{\bf Local Organizing Committee}
\end{large}
\end{center}
\bigskip
\begin{tabbing}
xxxxxxxxxx \= xxxxxxxxxxxxxxxxxx \=   \kill
\>  V. Bozza                \> \\
\>  S. Calchi Novati        \> \\
\>  L. Mancini              \> \\
\>  G. Scarpetta   (Chair)  \> \\

\end{tabbing}
\bigskip\bigskip
\parbox[t]{2.5in}{
\begin{tabular}[t]{ll}
\multicolumn{2}{c}{\bf Scientific Organizing Committee}\\[2ex]
 J.-P. Beaulieu &France\\
 D. Bennett &USA\\
 I. Bond &New Zealand\\
 S. Calchi Novati (Chair) &Italy\\
 M. Dominik &UK \\
 A. Gould &USA \\
 C. Han &South Korea \\
 Ph. Jetzer &Switzerland \\
 E. Kerins &UK \\
 M. Moniez &France \\
 T. Sumi &Japan \\
 Y. Tsapras &UK \\
 A. Udalski &Poland \\
 \L. Wyrzykowski &UK
\end{tabular}}
\hspace{1.0in}
\parbox[t]{2.5in}{\begin{tabular}[t]{ll}
\multicolumn{2}{c}{\bf Scientific Secretary}\\[2ex]
O. De Pasquale  \\
V. Di Marino  \\
T. Nappi    \\
S. Russo \\
\end{tabular}}
\vfill
\newpage

\blankpage

\def\headentry#1#2{\bigskip\noindent{\Large\sffamily#1}\hfill #2\\ }
\def\chapterentry#1#2{\noindent{\bf #1}\hfill{#2}\\ }
\def\authorentry#1{\hbox{\hspace{.2in}}#1 \\ }

\noindent
{\bf \Huge CONTENTS}
\bigskip

\vspace{1.35in}
\noindent
\chapterentry{Foreword}{iii}
\chapterentry{Scientific Rationale}{v}
\chapterentry{welcome}{vi}
%\chapterentry{Acknowledgements}{v}
\chapterentry{Organization}{ix}
\chapterentry{Participants}{xvi}
\chapterentry{School Program}{xix}
\chapterentry{Conference Program}{xx}

%-------------------------------------------------------
\headentry{School on Modelling Planetary Microlensing Events}{1}
%-------------------------------------------------------

\chapterentry{Introduction to microlensing}{2}
\authorentry{{\it Philippe Jetzer}}
\\
\chapterentry{From microlensing observations to science}{3}
\authorentry{{\it Martin Dominik}}
\\
\chapterentry{The theory and phenomenology of planetary microlensing}{4}
\authorentry{{\it Scott Gaudi}}
\\
\chapterentry{From raw images to lightcurves: how to make sense of your data}{5}
\authorentry{{\it Yiannis Tsapras}}
\\
\chapterentry{The efficient modeling of planetary microlensing events}{6}
\authorentry{{\it David Bennett}}
\\
\chapterentry{Contour integration and downhill fitting}{7}
\authorentry{{\it Valerio Bozza}}
\\
\chapterentry{Microlensing modelling and high performance computing}{8}
\authorentry{{\it Ian Bond}}

\vspace{1.0cm}
%-------------------------------------------------------
\headentry{Conference topical speeches}{9}
%-------------------------------------------------------

\chapterentry{Giant planet accretion and dynamical
evolution: considerations \\
on systems around small-mass stars}{10}
\authorentry{{\it Alessandro Morbidelli}}
\\
\chapterentry{The dark matter –- LHC endeavour to unveil TeV new
physics}{14}
\authorentry{{\it Antonio Masiero}}
\\
\chapterentry{Characterization of planetary systems with
high-precision \\
astrometry: the Gaia potential}{15}
\authorentry{{\it Alessandro Sozzetti}}

\bigskip

%-------------------------------------------------------
\headentry{Status of current surveys}{18}
%-------------------------------------------------------

\chapterentry{Status of the OGLE-IV survey}{19}
\authorentry{{\it Andrzej Udalski}}
\\
\chapterentry{MOA-II observation in 2010 season}{20}
\authorentry{{\it Takahiro Sumi}}
\\
%\chapterentry{The 2010 MicroFUN season}{?}
%\authorentry{{\it Andrew Gould}}
%\\
\chapterentry{The RoboNet 2010 season}{22}
\authorentry{{\it Yiannis Tsapras}}

\bigskip
%-------------------------------------------------------
\headentry{Cosmological Microlensing}{25}
%-------------------------------------------------------

\chapterentry{Dark matter determinations from Chandra observations of \\
quadruply lensed quasars}{26}
\authorentry{{\it David Pooley}}
\\
\chapterentry{Cosmic equation of state from strong gravitational lensing systems}{27}
\authorentry{{\it Marek Biesiada \& Beata Malec}}
\\

\bigskip
\vspace{2.0cm}
%-------------------------------------------------------
\headentry{Galactic Microlensing: the dark matter search}{31}
%-------------------------------------------------------

\chapterentry{PAndromeda - the Pan-STARRS M31 survey for dark matter}{32}
\authorentry{{\it Arno Riffeser, Stella
Seitz, Ralf Bender, C.-H. Lee, Johannes Koppenhoefer}}
\\
\chapterentry{Final OGLE-II and OGLE-III results on microlensing towards  \\
the LMC and SMC}{34}
\authorentry{{\it {\L}ukasz Wyrzykowski}}
\\
\chapterentry{Analysis of microlensing events towards the LMC}{37}
\authorentry{{\it Luigi Mancini \& Sebastiano Calchi Novati}}
\\
\chapterentry{Simulation of short time scale pixel lensing towards the Virgo cluster}{39}
\authorentry{{\it Sedighe Sajadian \& Sohrab Rahvar}}
\\
\chapterentry{M31 pixel lensing and the PLAN project}{40}
\authorentry{{\it Sebastiano Calchi Novati}}

\bigskip
%-------------------------------------------------------
\headentry{Planetary events}{44}
%-------------------------------------------------------

\chapterentry{MOA-2009-BLG-266LB: the first cold Neptune with a measured \\
mass}{45}
\authorentry{{\it David Bennett}}
\\
\chapterentry{Increasing the detection rate of low-mass planets in high-magnification  \\
events and MOA-2006-BLG-130}{46}
\authorentry{{\it Julie Baudry \& Philip Yock}}
\\
\chapterentry{The complete orbital solution for OGLE-2008-BLG-513}{49}
\authorentry{{\it Jennifer Yee}}
\\
\chapterentry{Planetary microlensing event MOA-2010-BLG-328}{51}
\authorentry{{\it Kei Furusawa}}
\\
\chapterentry{Binary microlensing event OGLE-2009-BLG-020 gives orbit predictions \\
verifiable by follow-up observations}{53}
\authorentry{{\it Jan Skowron}}

\bigskip
\vspace{1.0cm}
%-------------------------------------------------------
\headentry{Theoretical investigations}{59}
%-------------------------------------------------------

\chapterentry{Microlensing and planet populations - What do we know, \\
and how could we learn more}{60}
\authorentry{{\it Martin Dominik}}
\\
\chapterentry{The Frequency of extrasolar planet detections with microlensing \\
simulations}{65}
\authorentry{{\it Rieul Gendron \& Shude Mao}}
\\
\chapterentry{A semi-analytical model for gravitational microlensing events}{66}
\authorentry{{\it Denis Sullivan, Paul Chote, Michael Miller}}
\\
\chapterentry{GPU-assisted contouring for modeling binary microlensing events}{70}
\authorentry{{\it Markus Hundertmark, Frederic V. Hessman, Stefan Dreizler }}
\\
\chapterentry{Red noise effect in space-based microlensing observations}{74}
\authorentry{{\it Achille Nucita, Daniele Vetrugno, Francesco De
Paolis, Gabriele Ingrosso,\\ Berlinda M. T. Maiolo, Stefania
Carpano}}
\\
\chapterentry{Light curve errors introduced by limb-darkening models}{76}
\authorentry{{\it David Heyrovsky}}
\\
\chapterentry{Isolated, stellar-mass black holes through microlensing}{77}
\authorentry{{\it Kailash Sahu, Howard E. Bond, Jay Anderson, Martin Dominik, \\
Andrzej Udalski, Philip Yock}}
\\
\chapterentry{The observability of isolated compact remnants with microlensing}{80}
\authorentry{{\it Nicola Sartore \& Aldo Treves}}
\\
\chapterentry{Gravitational microlensing by the Ellis wormhole}{82}
\authorentry{{\it Fumio Abe}}
\\
\chapterentry{The deflection of light ray in strong field: a material medium \\
approach}{100}
\authorentry{{\it Asoke Kumar Sen}}
%\\
%\chapterentry{How to stop a runaway (Monte Carlo Markov) chain}{11}
%\authorentry{{\it Keith Horne}}
\\
\chapterentry{Rapidly rotating lenses - repeating orbital motion features in
\\ close binary microlensing}{104}
\authorentry{{\it Matthew Penny, Eamonn Kerins, Shude Mao}}

\bigskip
%-------------------------------------------------------
\headentry{Towards the future: {\large new
facilities/instrumentation/procedures}}{107}
%-------------------------------------------------------

\chapterentry{Microlensing with the SONG global network}{108}
\authorentry{{\it Uffe G. J{\o}rgensen, Kennet B. W. Harps{\o}e,
Per K. Rasmussen, Michael I.\,Andersen, Anton N. S{\o}rensen,
J{\o}rgen Christensen-Dalsgaard, S{\o}ren Frandsen, Frank
Grundahl, Hans Kjeldsen}}
\\
\chapterentry{Next-generation microlensing pilot planet search and the frequency
of planetary systems}{112}
\authorentry{{\it Dan Maoz \& Yossi Shvartzvald}}
\\
\chapterentry{Kohyama Astronomical Observatory: current status}{116}
\authorentry{{\it Atsunori Yonehara, Mizuki Isogai, Akira Arai, Hiroki Tohyama}}
\\
\chapterentry{Optimal imaging for gravitational microlensing}{118}
\authorentry{{\it Kennet B.W. Harps{\o}e, Uffe G. J{\o}rgensen, Per K. Rasmussen,
Michael I. Andersen, Anton N. S{\o}rensen, J{\o}rgen
Christensen-Dalsgaard, S{\o}ren Frandsen, Frank Grundahl, Hans
Kjeldsen}}
\\
\chapterentry{IPAC's role as the science center for NASA's WFIRST mission}{121}
\authorentry{{\it Kaspar von Braun}}
\\
\chapterentry{EUCLID microlensing planet hunt}{122}
\authorentry{{\it Jean-Philippe Beaulieu \& Matthew Penny}}
\\
\chapterentry{Space-based microlensing exoplanet survey: WFIRST and/or Euclid}{124}
\authorentry{{\it David Bennett}}
\\
\chapterentry{Microlensing with Gaia satellite}{125}
\authorentry{{\it {\L}ukasz Wyrzykowski}}

\bigskip
%-------------------------------------------------------
\headentry{Poster session}{127}
%-------------------------------------------------------

\chapterentry{Critical curve topology in special triple lens configurations}{128}
\authorentry{{\it Kamil Danek}}
\\
\chapterentry{PAndromeda - a dedicated deep survey of M31 with \\Pan-STARRS 1}{129}
\authorentry{{\it Chien-Hsiu Lee, Arno Riffeser, Stella Seitz, Ralf Bender, Johannes Koppenhoefer}}
\\

%\chapterentry{Author Index}{27}
%\chapterentry{Subject Index}{29}

%\blankpage

%%%%%%%%%%%%%%%   Conference Program

\begin{center}
\begin{large}
{\bf SMC2011 Participants}
\end{large}
\end{center}
\smallskip

\begin{small}
\begin{tabbing}
Perez-Angon, Miguel-Angelxxxxx \=   Rutherford Appleton Laboratory \kill    %
Abe, Fumio                  \> Nagoya University, Japan                                   \\      %
Baudry, Julie               \> University of Orsay, France                                \\      %
Bachelet, Etienne           \> University of Toulouse, France                             \\      %
Beaulieu, Jean-Philippe     \> Institut d'Astrophysique de Paris, France                  \\      %
Bennett, David              \> University of Notre Dame, USA                              \\      %
Bond, Ian                   \> Massey University, New Zealand                             \\      %
Bonino, Donata              \> INAF - Turin Astronomical Observatory, Italy               \\      %
Bozza, Valerio              \> University of Salerno, Italy                               \\      %
Browne, Paul                \> University of St Andrews, UK                               \\      %
Calchi Novati, Sebastiano   \> University of Salerno, Italy                               \\      %
Danek, Kamil                \> Charles University, Czech Republic                         \\      %
De Paolis, Francesco        \> University of Salento, Italy                               \\      %
Dominik, Martin             \> University of St Andrews, UK                               \\      %
Dominis Prester, Dijana     \> University of Rijeka, Croatia                              \\      %
Fouqu$\acute{\mathrm{e}}$, Pascal      \> University of Toulouse, France                  \\      %
Furusawa, Kei               \> Nagoya University, Japan                                   \\      %
Gardiol, Daniele            \> INAF - Turin Astronomical Observatory, Italy               \\      %
Gaudi, Scott                \> Ohio State University, USA                                 \\      %
Gendron, Riel               \> University of Manchester, UK                               \\      %
Gould, Andrew               \> Ohio State University, USA                                 \\      %
Harps{\o}e, Kennet          \> Niels Bohr Institute, Denmark                              \\      %
Harris, Pauline             \> Victoria University of Wellington, New Zealand             \\      %
Henderson, Calen            \> Ohio State University, USA                                 \\      %
Heyrovsky, David            \> Charles University, Czech Republic                         \\      %
Horne, Keith                \> University of St Andrews, UK                               \\      %
Hundertmark, Markus         \> University of G\"{o}ttingen, Germany                       \\      %
Jetzer, Philippe            \> University of Zurich, Switzerland                          \\      %
J{\o}rgensen, Uffe G.       \> Niels Bohr Institute, Denmark                              \\      %
Lambiase, Gaetano           \> University of Salerno, Italy                               \\      %
Lee, Chien-Hsiu             \> University Observatory Munich                              \\      %
Liebig, Christine           \> University of St Andrews                                   \\      %
Lubini, Mario               \> University of Zurich, Switzerland                          \\      %
Malec, Beata                \> Copernicus Center for Interdisciplinary Studies, Poland    \\      %
Mancini, Luigi              \> University of Salerno, Italy                               \\      %
Maoz, Dan                   \> Tel-Aviv University, Israel                                \\      %
Masiero, Antonio            \> University of Padova, Italy                                \\      %
Mirzoyan, Sergey            \> University of Salerno, Italy                               \\      %
Miyake, Noriyuki            \> Nagoya University, Japan                                   \\      %
Morbidelli, Alessandro      \> Observatoire de la Cote d'Azur, France                     \\      %
Muraki, Yasushi             \> Konan University, Japan                                    \\      %
Nucita, Achille             \> University of Salento, Italy                               \\      %
Orio, Marina                \> INAF - Padova Astronomical Observatory, Italy              \\      %
Paulin-Henriksson, St$\acute{\mathrm{e}}$phane  \> CEA - Paris, France                    \\      %
Payandeh, Farrin            \> Payame Noor University Tabriz, Iran                        \\      %
Penny, Matthew              \> University of Manchester, UK                               \\      %
Pooley, David               \> Eureka Scientific, USA                                     \\      %
Retana Montenegro, Edwin F. \> Universidad de Costa Rica, Costa Rica                      \\      %
Riffeser, Arno              \> Max Planck Institute for Extraterrestrial Physics, Germany \\      %
Sahu, Kailash               \> Space Telescope Science Institute, USA                     \\      %
Sajadian, Sedighe           \> Sharif University of Technology, Iran                      \\      %
Sartore, Nicola             \> INAF - IASF Milano, Italy                                  \\      %
Scarpetta, Gaetano          \> University of Salerno, Italy                               \\      %
Sen, Asoke Kumar            \> Sen Assam University, India                                \\      %
Shvartzvald, Yossi          \> Tel-Aviv University, Israel                                \\      %
Skowron, Jan                \> Ohio State University, USA                                 \\      %
Sozzetti, Alessandro        \> INAF - Turin Astronomical Observatory, Italy               \\      %
Sullivan, Denis             \> Victoria University of Wellington, New Zealand             \\      %
Sumi, Takahiro              \> Nagoya University, Japan                                   \\      %
Tortora, Crescenzo          \> University of Zurich, Switzerland                          \\      %
Tsapras, Yiannis            \> Queen Mary University, UK                                  \\      %
Udalski, Andrzej            \> Warsaw University Observatory, Poland                      \\      %
Vetrugno, Daniele           \> University of Salento, Italy                               \\      %
Vilasi, Gaetano             \> University of Salerno, Italy                               \\      %
von Braun, Kaspar           \> California Institute of Technology, USA                    \\      %
Yee, Jennifer               \> Ohio State University, USA                                 \\      %
Yock, Philip                \> University of Auckland, New Zealand                        \\      %
Yonehara, Atsunori          \> Kyoto Sangyo University, Japan                             \\      %
Wyrzykowski, {\L}ukasz      \> University of Cambridge, UK                                \\      %
\end{tabbing}
\end{small}

\blankpage \clearpage

\begin{center}
\begin{large}
{\bf School on Modelling Planetary Microlensing Events \\
\vspace{1.0cm}{\Large Program}}
\end{large}
\end{center}
\bigskip

\begin{small}
\begin{tabbing}
9:00xx\= Reports from the B factories xxxxxxxxx \=   \kill
{\large Tuesday, January 18, 2011}  \\
\\
09:00\>  Introduction to microlensing              \> \hspace{2cm} Philippe Jetzer \\
\\
10:50     \>{\it Coffee Break}\\
\\
11:15     \>  From microlensing observations to science         \> \hspace{2cm} Martin Dominik\\
\\
13:00   \> {\it Lunch Break} \\
\\
14:45\>  The theory and phenomenology of planetary   \> \hspace{2cm} Scott Gaudi \\
\>microlensing\\
\\
16:45     \>{\it Coffee Break}\\
\\
17:15\>  From raw images to lightcurves: how to make   \> \hspace{2cm} Yiannis Tsapras \\
\>sense of your data\\

\\
\\
9:00xx\= Reports from the B factories xxxxxxxxx \=   \kill
{\large Wednesday, January 19, 2011}  \\
\\
09:00\>  The efficient modeling of planetary microlensing       \> \hspace{2cm} David Bennett \\
\>events\\
\\
10:50     \>{\it Coffee Break}\\
\\
11:15     \>  Contour integration and downhill fitting         \> \hspace{2cm} Valerio Bozza\\
\\
13:00   \> {\it Lunch Break} \\
\\
14:45\>  Microlensing modelling and high performance   \> \hspace{2cm} Ian Bond \\
\>computing\\
\\
16:45     \>{\it Coffee Break}\\
\\
\end{tabbing}
\end{small}

%\blankpage

\begin{center}
\begin{large}
{\bf SMC2011 Program}
\end{large}
\end{center}
\bigskip

\begin{small}
\begin{tabbing}
9:00xx\= Reports from the B factories xxxxxxxxx \=   \kill
{\large Thursday, January 20, 2011}  \\
\\
09:00\> Registration           \\
09:30\> {\it Greetings}         \\
09:40     \>   Communications     \> \\
\\
\\
\>    \> {\bf Chair: Gaetano Scarpetta}\\
\\
{\bf Status of current Surveys I}\>    \>\\
09:50     \>  Status of the OGLE-IV survey              \> \hspace{2cm} Andrzej Udalski \\
10:20     \>  MOA-II observation in 2010 season         \> \hspace{2cm} Takahiro Sumi\\
\\
10:50     \>{\it Coffee Break}\\
\\
{\bf Topical Speech I}\>    \>\\
11:20     \>  Formation and evolution of our        \> \hspace{2cm} Alessandro Morbidelli \\
\>solar system \\
\\
{\bf Theoretical investigations I}\>    \>\\
12:10     \>  Microlensing and planet populations:          \> \hspace{2cm} Martin Dominik\\
\> what do we know, and how could we \\
\> learn more? \\
12:40   \>The frequency of extrasolar planet               \> \hspace{2cm} Rieul Gendron\\
\> detections with microlensing simulations \\
\\
13:00   \> {\it Lunch Break} \\
\\
\>    \>{\bf Chair: Philippe Jetzer}\\
\\
{\bf Dark Matter Search}\>    \>\\
14:40     \>  PAndromeda - the Pan-STARRS M31          \> \hspace{2cm} Arno Riffeser\\
\> survey for dark matter \\
15:10     \>  Final OGLE-II and OGLE-III results on          \> \hspace{2cm} {\L}ukasz Wyrzykowski\\
\> microlensing towards the LMC and SMC \\
15:30     \>  Analysis of microlensing events towards the LMC  \> \hspace{2cm} Luigi Mancini \\
15:50     \>  Simulation of short time scale pixel lensing   \> \hspace{2cm} Sedighe Sajadian \\
\> towards the Virgo cluster \\
16:10     \>  M31 pixel lensing and the PLAN project  \> \hspace{2cm} Sebastiano Calchi Novati \\
\\
16:30     \>{\it Coffee Break}\\
\\
{\bf Towards the future I}\>    \>\\
17:00     \>  Microlensing with the SONG global network  \> \hspace{2cm} Uffe G. J{\o}rgensen \\
17:20     \>  Next-generation microlensing pilot planet search   \> \hspace{2cm} Dan Maoz \\
\> and the frequency of planetary systems \\
17:40     \>  Kohyama Astronomical Observatory:   \> \hspace{2cm} Atsunori Yonehara \\
\> current status \\
18:00     \>  The lucky imaging technique for microlensing   \> \hspace{2cm} Kennet Harps{\o}e\\
\> observations \\
\\
\\
\\
9:00xx\= Reports from the B factories xxxxxxxxx \=   \kill
{\large Friday, January 21, 2011}  \\
\\
\>    \> {\bf Chair: P. Yock}\\
\\
{\bf Status of current Surveys II}\>    \>\\
09:00     \>  The 2010 MicroFUN season \> \hspace{2cm} Jennifer Yee \\
09:30     \>  The RoboNet 2010 season  \> \hspace{2cm} Yiannis Tsapras \\
\\
{\bf Theoretical investigations II}\>    \>\\
10:00     \>  A semi-analytical model for gravitational \> \hspace{2cm} Denis Sullivan \\
\> microlensing events \\
10:20     \>  GPU-assisted contouring for modeling  \> \hspace{2cm} Markus Hundertmark \\
\> binary microlensing events \\
10:40     \>  Rapidly rotating lenses - repeating orbital motion \> \hspace{2cm} Matthew Penny \\
\> features in close binary microlensing \\
\\
11:00     \>{\it Coffee Break}\\
\\
{\bf Topical Speech II}\>    \>\\
11:30     \>  The dark matter –- LHC endeavour to unveil     \> \hspace{2cm} Antonio Masiero \\
\> TeV new physics \\
\\
{\bf Cosmological microlensing}\>    \>\\
12:20     \>  Dark matter determinations from Chandra         \> \hspace{2cm} David Pooley\\
\> observations of quadruply lensed quasars \\
12:40     \>  Cosmic equation of state from strong    \> \hspace{2cm} Beata Malec\\
\> gravitational lensing systems    \\
\\
13:00   \> {\it Lunch Break} \\
\\
\\
\>    \>{\bf Chair: Pascal Fouqu$\acute{\mathrm{{\bf e}}}$}\\
\\
{\bf Planetary events}\>    \>\\
14:30     \>  MOA-2009-BLG-266LB: the first cold Neptune    \> \hspace{2cm} David Bennett\\
\> with a measured mass \\
14:50     \>  Increasing the detection rate of low-mass \> \hspace{2cm} Julie Baudry\\
\> planets in high-magnification events and \\
\> MOA-2006-BLG-130 \\
15:10     \>  The complete orbital solution for  \> \hspace{2cm} Jennifer Yee\\
\> OGLE-2008-BLG-513 \\
15:30     \>  Planetary microlensing event MOA-2010-BLG-328  \> \hspace{2cm} Kei Furusawa\\
15:50     \>  Binary microlensing event OGLE-2009-BLG-020 \> \hspace{2cm} Jan Skowron\\
\> gives orbit predictions verifiable by follow-up \\
\> observations \\
\\
16:10     \>{\it Coffee Break}\\
\\
17:30     \>{\it Tour of the old town of Salerno}\\
\\
20:00     \>{\it Social dinner}\\
\\
\\
\\
\\
9:00xx\= Reports from the B factories xxxxxxxxx \=   \kill
{\large Saturday, January 22, 2011}  \\
\\
\>    \> {\bf Chair: Francesco De Paolis}\\
\\
{\bf Towards the future II}\>    \>\\
09:00     \>  IPAC's role as the science center for  \> \hspace{2cm} Kaspar von Braun \\
\> NASA's WFIRST mission \\
09:30     \>  EUCLID microlensing planet hunt  \> \hspace{2cm} Jean-Philippe Beaulieu \\
09:50     \>  Simulating the planet hunting capability of Euclid  \> \hspace{2cm} Matthew Penny \\
10:10     \>  Space-based microlensing exoplanet survey:   \> \hspace{2cm} David Bennett \\
\> WFIRST and/or Euclid \\
\\
10:40     \>{\it Coffee Break}\\
\\
{\bf Topical Speech III}\>    \>\\
11:10     \>  Characterization of planetary systems with       \> \hspace{2cm} Alessandro Sozzetti \\
\> high-precision astrometry: the Gaia Potential \\
\\
12:00     \>  Microlensing with Gaia satellite        \> \hspace{2cm} {\L}ukasz Wyrzykowski\\
\\
{\bf Theoretical investigations III}\>    \>\\
12:20     \>  Red noise effect in space-based microlensing  \> \hspace{2cm} Achille Nucita \\
\> observations \\
12:40     \>  Light curve errors introduced by limb-darkening      \> \hspace{2cm} David Heyrovsky\\
\> models   \\
\\
13:00   \> {\it Lunch Break} \\
\\
\>    \> {\bf Chair: Scott Gaudi}\\
\\
\\
14:40   \>  Isolated, stellar-mass black holes through     \> \hspace{2cm} Kailash Sahu\\
\> microlensing   \\
15:00   \>  The observability of isolated compact     \> \hspace{2cm} Nicola Sartore\\
\> remnants with microlensing   \\
15:20   \>  Gravitational microlensing by the Ellis wormhole    \> \hspace{2cm} Fumio Abe\\
15:40   \>  The deflection of light ray in strong field:    \> \hspace{2cm} Asoke Kumar Sen\\
\> a material medium approach  \\
16:00   \>  How to stop a runaway (Monte Carlo Markov)  \> \hspace{2cm} Keith Horne\\
\\
16:20     \>{\it Coffee Break}\\
\\
16:50   \>  Open session  \> \hspace{2cm} \\
\\
\\
9:00xx\= Reports from the B factories xxxxxxxxx \=   \kill
{\large \textbf{Poster Session}}  \\
\\
\>  Critical curve topology in special triple lens  \> \hspace{2cm} Kamil Danek\\
\> configurations   \\
\>  PAndromeda - a dedicated deep survey of M31   \> \hspace{2cm} Chien-Hsiu Lee\\
\> with Pan-STARRS 1 \\
\end{tabbing}
\end{small}

%\blankpage

\clearpage
%%%%%%%%%%%%%%%%%%%%%%%%%%%%%d
%   MAINMATTER --  Section by Section

\mainmatter

%%%%%%%%%%%   Section 0 %%%%%%%%%%%%%%%%%%%%%%%
\emptyheads

\begin{center}
\begin{large}
{\Huge \sffamily School on \\
Modelling Planetary Microlensing Events}
\end{large}
\end{center}
\bigskip

\begin{tabbing}
 Brief Reports from the \=B factories xxxxx \= xxxxxxxxx\= \kill
\\
\\
Introduction to microlensing \\
\>\> {\it Philippe Jetzer}\\
\\
From microlensing observations to science \\
\>\> {\it Martin Dominik}\\
\\
The theory and phenomenology of planetary microlensing \\
\>\> {\it Scott Gaudi }\\
\\
From raw images to lightcurves: how to make sense of your data \\
\>\> {\it Yiannis Tsapras}\\
\\
The efficient modeling of planetary microlensing events \\
\>\> {\it David Bennett}\\
\\
Contour integration and downhill fitting \\
\>\> {\it Valerio Bozza}\\
\\
Microlensing modelling and high performance computing \\
\>\> {\it Ian Bond}\\
\end{tabbing}

\fancyheads

\Chapter{Introduction to microlensing}
            {Introduction to microlensing}{Philippe Jetzer}
\bigskip\bigskip

\addcontentsline{toc}{chapter}{{\it Philippe Jetzer}}
\label{polingStart}

\begin{raggedright}

{\it Philippe Jetzer\index{author}{Philippe Jetzer}\\
Institute of Theoretical Physics, University of Zurich \\
Switzerland}
\bigskip\bigskip

\end{raggedright}
In the lecture I will present the basic formalism of gravitational
lensing such as the lens equation and the special case of the
Schwarzschild lens with its application in microlensing. I will
discuss the microlensing probability for different targets such as
the galactic bulge, LMC, SMC and the Andromeda galaxy as well as
give a short review of the present status of microlensing searches
conducted by the various collaborations in particular with respect
to the problem of the galactic dark matter content in form of
MACHOs.

\Chapter{From microlensing observations to science}
            {From microlensing observations to science}{Martin Dominik}
\bigskip\bigskip

\addcontentsline{toc}{chapter}{{\it Martin Dominik}}
\label{polingStart}

\begin{raggedright}

{\it Martin Dominik\index{author}{Dominik, Martin}\\
SUPA, University of St Andrews, School of Physics \& Astronomy \\
United Kingdom\\}
\bigskip\bigskip

\end{raggedright}
An elaborate integrated strategy incorporating target selection
and scheduling, data flow, assessment, and final analysis is
required to ensure that the scientific goals that we aim for are
achieved. Specific issues that need to be taken care of are
dealing with uncertainties, ambiguities, and degeneracies, as well
as having the capacity to keep track with data being acquired at
ever increasing rate. I will present you with some challenges.

\Chapter{The theory and Phenomenology of Planetary Microlensing}
            {The theory and Phenomenology of Planetary Microlensing}{Scott Gaudi}
\bigskip\bigskip

\addcontentsline{toc}{chapter}{{\it Scott Gaudi}}
\label{polingStart}

\begin{raggedright}

{\it Scott Gaudi\index{author}{Scott Gaudi}\\
Department of Astronomy, Ohio State University \\
USA\\}
\bigskip\bigskip

\end{raggedright}
I discuss the theory and phenomenology of planetary microlensing.
I begin with a review of the basic theoretical formalism: starting
with the time delay surface, and continuing with the lens
equation, I describe how critical curves, caustics, and
magnifications can be computed. I then explore the properties of
the caustic curves of planetary microlenses. I illustrate the
topology of caustic curves and how these change with the
parameters of the planetary system. In addition, I review the
generic, universal behavior of images near caustics. I then delve
into the rich phenomenology and salient observable properties of
planetary microlensing light curves, and discuss how these can be
intuitively understood based on consideration of the microlensed
images and shape of and magnification near the caustics. Finally,
I demonstrate how all of these considerations can be used to
roughly estimate the properties of a planetary system giving rise
to an observed light curve based purely on visual inspection.

\Chapter{From raw images to lightcurves: how to make sense of your
data}
            {From raw images to lightcurves: how to make sense of your data}{Yiannis Tsapras}
\bigskip\bigskip

\addcontentsline{toc}{chapter}{{\it Yiannis Tsapras}}
\label{polingStart}

\begin{raggedright}

{\it Yiannis Tsapras\index{author}{Tsapras, Yiannis}\\
Queen Mary University \\
UK}
\bigskip\bigskip

\end{raggedright}
I will present an overview of astronomical image processing with
special focus on the difference imaging technique in the context
of microlensing observations towards the galactic bulge. Starting
from how to calibrate the raw data, I will discuss the various
steps involved in the analysis, possible pitfalls, and how to
extract the photometric information from the images in order to
construct a clean lightcurve. Examples using the RoboNet DIA
software will be given.

\Chapter{The Efficient Modeling of Planetary Microlensing Events}
            {The Efficient Modeling of Planetary Microlensing Events}{David Bennett}
\bigskip\bigskip

\addcontentsline{toc}{chapter}{{\it David Bennett}}
\label{polingStart}

\begin{raggedright}

{\it David P. Bennett\index{author}{Bennett, David P.}\\
University of Notre Dame, Notre Dame, IN\\
USA\\}
\bigskip\bigskip

\end{raggedright}
I present a general method for the modeling of planetary
microlensing events, with an emphasis on the most difficult events
which involve more than two lens masses, microlensing parallax
and/or orbital motion. Finite source calculations are done with
the image centered ray-shooting method, which can be made both
highly efficient and flexible enough to model any events.

\Chapter{Contour integration and downhill fitting}
            {Contour integration and downhill fitting}{Valerio Bozza}
\bigskip\bigskip

\addcontentsline{toc}{chapter}{{\it Valerio Bozza}}
\label{polingStart}

\begin{raggedright}

{\it Valerio Bozza\index{author}{Valerio Bozza}\\
Department of Physics, University of Salerno\\
Italy\\}
\bigskip\bigskip

\end{raggedright}
By Green's theorem, the two-dimensional integration on the
microlensed images is written as a line integral on the image
boundaries. We will discuss the advantages and the shortcomings of
this method, presenting several improvements of the basic idea:
parabolic correction, error control, optimal sampling, limb
darkening. We will also review some basic downhill fitting
methods, which rapidly provide preliminary models for microlensing
events starting from suitable initial conditions.

\Chapter{Microlensing Modelling and High Performance Computing}
            {Microlensing Modelling and High Performance Computing}{Ian Bond}
\bigskip\bigskip

\addcontentsline{toc}{chapter}{{\it Ian Bond}} \label{polingStart}

\begin{raggedright}

{\it Ian Bond\index{author}{Ian Bond}\\
Massey University\\
New Zealand}
\bigskip\bigskip

\end{raggedright}
I will go over the practical aspects of the modelling and analysis
of microlensing events. I will discuss two programming
environments for high performance computing: cluster computers and
GPU (graphical processor units) platforms. I will describe, with
examples, how to design and implement software to run on these
platforms.  In particular, GPUs are emerging as a powerful tool
for scientific computation and their potential in microlensing
modelling is promising. The use of GPUs will be particularly
emphasized in this seminar.

\emptyheads

\begin{center}
\begin{large}
{\Huge \sffamily Conference topical speeches}
\end{large}
\end{center}
\bigskip

\begin{tabbing}
 Brief Reports from the \=B factories xxxxx \= xxxxxxxxx\= \kill
\\
\\
Giant planet accretion and dynamical evolution: considerations on
systems around \\small-mass stars \\
\>\> {\it Alessandro Morbidelli}\\
\\
\\
The dark matter -– LHC endeavour to unveil TeV new
physics \\
\>\> {\it Antonio Masiero}\\
\\
\\
Characterization of planetary systems with
high-precision astrometry: the Gaia \\
potential \\
\>\> {\it Alessandro Sozzetti}\\

\\
\end{tabbing}

\fancyheads

\Chapter{Giant planet accretion and dynamical evolution:
considerations on systems around small-mass stars}
            {Giant planet accretion and dynamical evolution}{Alessandro Morbidelli}
\bigskip\bigskip

\addcontentsline{toc}{chapter}{{\it Alessandro Morbidelli}}
\label{polingStart}

\begin{raggedright}

{\it Alessandro Morbidelli\index{author}{Morbidelli, Alessandro} \\
Observatoire de la Cote d'Azur, Nice \\
France \\}
\bigskip\bigskip

\end{raggedright}

\section{Introduction}

The classical model of giant planet formation envisions that
multi-Earth mass solid cores formed from the accretion of
planetesimals and that these cores then captured by gravity a
massive atmosphere from the gas in the proto-planetary disk
\cite{Pollack}. A problem in this scenario is that the solid cores
have to grow on a timescale shorter than the gas-dissipation
timescale, which observations set to be a few My only
\cite{Haish}. This is not easy, particularly in disks with small
densities, such as those around low-mass stars. For this reason it
is expected that giant planets cannot form around low-mass stars
\cite{Laughlin} or they form very rarely \cite{Alibert}. However,
micro-lensing and radial velocity detections show that giant
planets do exist around stars down to at least 0.2$M_\oplus$. This
conflict between models and observations suggests that the process
of giant planet accretion needs to be revisited

\section{The problem of core accretion}

It has been recently pointed out that our ideas on the accretion
of the cores of the giant planets were simplistic even in the case
of disks around solar-mass stars.

In fact, N-body simulations show that, once the cores become
sufficiently massive (about 1$M_\oplus$), they tend to scatter the
planetesimals away, rather than accrete them. In doing this, they
clear their neighboring region, which in turn limits their own
growth, even if initially there are a lot of solids available in
the system. If damping effects are included in the simulation, in
the hope of limiting the ability of the cores to scatter the
planetesimals away, the result is that the planetesimals end up in
stable circular orbits that are dynamically separated from the
cores (i.e. no close encounters with the cores are possible).
Effectively, the cores open radial gaps in the planetesimal
distribution \cite{Levison}. One could think that core migration
or planetesimal radial drift due to gas drag can help to break the
isolation of the cores from the planetesimals. However the
simulations show that, in these cases, most planetesimals end up
locked in resonances with the cores; resonances prevent
collisions, so radial migration is unlikely to boost accretion in
a significant way  \cite{Levison}.

\section{Hints for a new model of formation of giant planets cores}

New results on the migration of proto-planets in gas-disks show
that the concept of inward migration of Earth-mass objects (the
so-called Type I migration \cite{Goldreich}) is valid only in
idealized isothermal disks.  In more realistic radiative disks,
the migration of the proto-planets is outward in the inner part of
the disk and inward in the outer part
\cite{Paardekooper}\cite{Masset}\cite{Kley}. There are therefore
one (or sometimes two) orbital radii where migration is cancelled.
All proto-planets formed in the disk tend to migrate towards these
equilibrium radii \cite{Lyra}. This can concentrate the
proto-planets in a narrow region of space and favor further mutual
accretion, thus leading to the formation of giant planets cores. A
proof-of-concept simulation will be presented at the conference,
leading to the formation of a 12$M_\oplus$ body from a system of
1$M_\oplus$ planetary embryos.

If this idea is correct, the formation of the cores of the giant
planets is not due to the local runaway accretion of small
planetesimals, but to the concentration of all planetary embryos
that formed throughout the disk and their subsequent mutual
collisions. This is a big change of emphasis. The concept of the
{\it local} surface density is not relevant any more because the
material is migrated to the site of growth of the core from a
large range of distances. This gives hope that the formation of
the cores is less sensitive to the surface density of the disk
than previously expected, and that therefore giant planet
formation is more likely also around low-mass stars.

\section{Dynamical evolution of giant planets}

Once giant planets are formed, they open gaps in the gas
distribution in the proto-planetary disk. Once this happens, giant
planets are expected to undergo Type II migration towards the
central star \cite{Lin}.

Also Type-II migration, though, is an idealized concept. It is
valid in the case where planets are much less massive than disks
and they open extremely deep gaps that gas cannot pass through. In
the other cases, the planets do not migrate at the Type-II rate;
they can even move outward and, instead of gaps, they can open
cavities in the inner part of the disk, removing most the gas
between the star and the orbit of the planet \cite{Crida}.

In case of two (or more) planets, the migration pattern is even
more complex. For two planets in resonance, migration is inward if
the outer planet has a mass comparable to the inner one, or
larger. If the outer planet has a mass that is a sizeable fraction
of the inner one (as in the Jupiter-Saturn case), migration is
outward. If the outer planet has a negligible mass (say Jupiter
and Neptune), migration is inward again \cite{Morby-Crida}. Thus,
for our solar system we can envision that Jupiter migrated inwards
while Saturn was growing and then, once Saturn reached a mass
close to its current one, they migrated outward. This kind of
evolution may be relevant for the system around the star
OGLE-06-109L, discovered by microlensing, which looks like a
Sun-Jupiter-Saturn system slightly scaled down in mass.

\Chapter{The Dark Matter - LHC Endeavour to Unveil TeV New
Physics}
            {Dark matter - LHC and TeV new physics}{Antonio Masiero}
\bigskip\bigskip

\addcontentsline{toc}{chapter}{{\it Antonio Masiero}}
\label{polingStart}

\begin{raggedright}

{\it Antonio Masiero\index{author}{Masiero, Antonio} \\
INFN - University of Padova \\
Italy \\}
\bigskip\bigskip

\end{raggedright}

After more than four decades of relentless tests of the Standard
Model (SM) of particle physics, one can safely state that it
correctly describes the fundamental interactions all the way up to
the energy scale of 100 GeV. Yet, the observational evidences that
neutrinos are massive and that a large amount of non-baryonic dark
matter exists corroborate the theoretical demand for the presence
of new non-SM physics at the TeV scale. Interestingly enough, the
main theoretical motivation for new physics (NP) at the
electroweak scale (i.e., the presence of an ultraviolet SM
completion to enforce the stability of the electroweak breaking
scale) nicely joins the need for some form of cold matter: indeed,
most theoretically dictated extensions of the SM (low-energy
supersymmetry, extra-dimensions, etc.) entail the existence of
some new stable particle which can play the role of dark matter
candidate. I'll discuss the interplay between the searches for TeV
NP which are going on at the LHC and the searches for dark matter
related to TeV NP both in direct and indirect DM probes. It is
exciting that the coming decade has the potentiality to witness
the simultaneous success of the high-energy (LHC) and
astroparticle (DM) roads in our endeavour to unveil the presence
of NP at the electroweak scale.

\Chapter{Characterization of Planetary Systems with High-Precision
Astrometry: The Gaia Potential}
            {The Gaia potential}{Alessandro Sozzetti}
\bigskip\bigskip

\addcontentsline{toc}{chapter}{{\it Alessandro Sozzetti}}
\label{polingStart}

\begin{raggedright}

{\it Alessandro Sozzetti\index{author}{Sozzetti, Alessandro} \\
INAF - Osservatorio Astronomico di Torino \\
Italy \\}
\bigskip\bigskip

\end{raggedright}

\section{Introduction}

In its all-sky survey, the ESA global astrometry mission Gaia, due
to launch in late 2012, will perform high-precision astrometry and
photometry for 1 billion stars down to V = 20 mag. The data
collected in the Gaia catalogue, to be published by the end of the
decade, will likely revolutionize our understanding of many
aspects of stellar and Galactic astrophysics. One of the relevant
areas in which the Gaia observations will have great impact is the
astrophysics of planetary systems. There are many complex
technical problems related to and challenges inherent in correctly
modelling the signals of planetary systems present in measurements
collected with an observatory poised to carry out precision
astrometry at the micro-arcsecond ($\mu$as) level. Provided these
are well understood, Gaia $\mu$as astrometry has great potential
for important contributions to the astrophysics of planetary
systems, particularly when seen in synergy with other indirect and
direct methods for the detection and characterization of planetary
systems.

\section{Astrometric Modeling of Planetary Systems}

The problem of the correct determination of the astrometric orbits
of planetary systems using Gaia data (highly non-linear orbital
fitting procedures, large numbers of model parameters) will
present many difficulties. For example, it will be necessary to
assess the relative robustness and reliability of different
procedures for orbital fits, with a detailed understanding of the
statistical properties of the uncertainties associated with the
model parameters. For multiple systems, a trade-off will have to
be found between accuracy in the determination of the mutual
inclination angles between pairs of planetary orbits,
single-measurement precision and redundancy in the number of
observations with respect to the number of estimated model
parameters. It will be challenging to correctly identify signals
with amplitude close to the measurement uncertainties,
particularly in the presence of larger signals induced by other
companions and/or sources of astrophysical noise of comparable
magnitude. Finally, for systems where dynamical interactions are
important (a situation experienced already by Doppler surveys),
fully dynamical fits involving an n-body code might have to be
used to properly model the Gaia astrometric data and to ensure the
dynamical stability of the solution (see Sozzetti 2005). All the
above issues could significantly impact Gaia's planet detection
and characterization capabilities. For these reasons, a
Development Unit (DU), within the pipeline of Coordination Unit 4
(object processing) of the Gaia Data Processing and Analysis
Consortium, has been specifically devoted to the modelling of the
astrometric signals produced by planetary systems. The DU is
composed of several tasks, which implement multiple robust
procedures for (single and multiple) astrometric orbit fitting
(such as Markov Chain Monte Carlo algorithms) and the
determination of the degree of dynamical stability of multi-planet
systems.

\begin{table}[tbh]
\begin{minipage}{0.5\linewidth}
\begin{center}
   \renewcommand{\arraystretch}{1.4}
   \setlength\tabcolsep{7pt}
{\tiny
      \begin{tabular}{|c|c|c|c|c|c|c|}
       \hline\noalign{\smallskip}
       $\Delta d$ & $N_\star$  & $\Delta a$ & $\Delta M_p$ &  $N_{\rm d}$ &  $N_{\rm m}$ \\
       (pc) & & (AU) & ($M_J$) & & \\
              \noalign{\smallskip}
       \hline
       \noalign{\smallskip}
0-50 & $1\times10^4$ & 1.0 - 4.0 & 1.0 - 13.0 & $1400$ & $ 700$\\
\hline 50-100 & $5\times10^4$ & 1.0 - 4.0 & 1.5 - 13.0 & $2500$ &
$ 1750$\\ \hline 100-150 & $1\times10^5$ & 1.5 - 3.8 & 2.0 - 13.0&
$2600$ & $ 1300$\\\hline 150-200 & $3\times10^5$ & 1.4 - 3.4 & 3.0
- 13.0& $2150$ & $ 1050$\\\hline

   \end{tabular}
}
\end{center}
\centering\large (a)
\end{minipage}
\hspace{0.5cm}
\begin{minipage}{0.5\linewidth}
\begin{center}
   \renewcommand{\arraystretch}{1.4}
   \setlength\tabcolsep{7pt}
{\tiny
      \begin{tabular}{|l|c|}
       \hline\noalign{\smallskip}
        Case & N. of systems \\
              \noalign{\smallskip}
       \hline
       \noalign{\smallskip}
Detection & $\sim 1000$\\ \hline
Orbits and masses to  & \\
$<15-20\%$ accuracy & $\sim 400-500$ \\ \hline
Successful  & \\
coplanarity tests & $\sim 150$\\ \hline
   \end{tabular}
}
\end{center}
\centering\large (b)
\end{minipage}
\caption{Left: Number of giant planets of given ranges of mass
($\Delta M_p$) and orbital separation ($\Delta a$) that could be
detected ($N_d$) and measured ($N_m$) by Gaia, as a function of
increasing distance ($\Delta d$) and stellar sample ($N_\star$).
Right: Number of planetary systems that Gaia could potentially
detect, measure, and for which coplanarity tests could be carried
out successfully. See Casertano et al. (2008) for details.}
\label{nsyst}
\end{table}

\subsection{The Gaia Legacy}

Gaia's main contribution to exoplanet science will be its unbiased
census of thousands of planetary systems (see Table~\ref{nsyst})
orbiting hundreds of thousands nearby ($d< 200$ pc), relatively
bright ($V \leq 13$) stars across all spectral types, screened
with constant astrometric sensitivity. As a result, the actual
impact of Gaia measurements in exoplanets science is broad, and
rather structured. The Gaia data have the potential to: a)
significantly refine our understanding of the statistical
properties of extrasolar planets; b) help crucially test
theoretical models of gas giant planet formation and migration; c)
achieve key improvements in our comprehension of important aspects
of the formation and dynamical evolution of multiple-planet
systems; d) aid in the understanding of direct detections of giant
extrasolar planets; e) provide important supplementary data for
the optimization of the target selection for future observatories
aiming at the direct detection and spectral characterization of
habitable terrestrial planets. Finally, ongoing studies are now
focusing on the detailed understanding of the planet discovery
potential of Gaia as far as low-mass stars (Sozzetti et al., in
preparation) and post-main-sequence objects (Silvotti, Sozzetti,
\& Lattanzi 2011) and are concerned.

In conclusion, the Gaia mission is now set to establish the
European leadership in high-precision astrometry for the next
decade. The largest compilation of high-accuracy astrometric
orbits of giant planets, unbiased across all spectral types up to
$d\simeq200$ pc, will allow Gaia to crucially contribute to
several aspects of planetary systems astrophysics (formation
theories, dynamical evolution), in combination with present-day
and future extrasolar planet search programs.

%%%%%%%%%%%%%%%%%%%%%%%%%%%%%%%%%%%%%%%%%%%%%%%%%%%%%%%%%%%%%%%%%%%%%%%%%
%%
%%   use this format to include an .eps figure into your paper
%%
%\begin{figure}[htb]
%\begin{center}
%\epsfig{file=rgb.eps,height=1.5in}
%\caption{Plan of the magnet used in the Mesmeric studies.}
%\label{fig:magnet}
%\end{center}
%\end{figure}
%%%%%%%%%%%%%%%%%%%%%%%%%%%%%%%%%%%%%%%%%%%%%%%%%%%%%%%%%%%%%%%%%%%%%%%%%%%

%%%%%%%%%%%%%%%%%%%%%%%%%%%%%%%%%%%%%%%%%%%%%%%%%%%%%%%%%%%%%%%%%%%%%%%%%
%%
%%   use this format to include a LaTeX table  into your paper
%%
%\begin{table}[b]
%\begin{center}
%\begin{tabular}{l|ccc}
%Patient &  Initial level($\mu$g/cc) &  w. Magnet &
%w. Magnet and Sound \\ \hline
% Guglielmo B.  &   0.12     &     0.10      &     0.001  \\
% Ferrando di N. &  0.15     &     0.11      &  $< 0.0005$ \\ \hline
%\end{tabular}
%\caption{Blood cyanide levels for the two patients.}
%\label{tab:blood}
%\end{center}
%\end{table}
%%%%%%%%%%%%%%%%%%%%%%%%%%%%%%%%%%%%%%%%%%%%%%%%%%%%%%%%%%%%%%%%%%%%%%%%%%%

%\bigskip
%I am grateful to Don Alfonso d'Alba for certain services essential to
%this investigation.

%%%%%%%%%%%   Section 1 %%%%%%%%%%%%%%%%%%%%%%%
\emptyheads

%\begin{center}

%\begin{large}
%{\Huge \sffamily Cosmological Microlensing}
%\end{large}
%\end{center}

%\newpage

\begin{center}
\begin{large}
{\Huge \sffamily Status of current surveys}
\end{large}
\end{center}
\bigskip

\begin{tabbing}
 Brief Reports from the \=B factories xxxxx \= xxxxxxxxx\= \kill
%\>{\bf Session Chair:}   \>\> Andrew Gould\\
%\>{\bf Scientific Secretary:}   \>\> Sibylle Petrak \\
\\
\\
Status of the OGLE-IV Survey \>\> {\it Andrzej Udalski}\\
\\
MOA-II observation in 2010 season \>\> {\it Takahiro Sumi}\\
\\
\\
\\
\>{\bf Session Chair:}  \> \> Martin Perl\\
%\>{\bf Scientific Secretaries:} \>  \> Ian Adam\\
%\>                                 \>\> Yuval Grossmann \\
\\
\\
The deflection of light ray
    \>      \>{\it Asoke Kumar Sen}\\
\textit{in strong field: a material medium approach}   \\
\end{tabbing}

\fancyheads

\Chapter{Status of the OGLE-IV Survey}
            {Status of the OGLE-IV Survey}{Andrzej Udalski}
\bigskip\bigskip

\addcontentsline{toc}{chapter}{{\it Andrzej Udalski}}
\label{polingStart}

\begin{raggedright}

{\it Andrzej Udalski\index{author}{Udalski, Andrzej}\\
Warsaw University Observatory\\
Poland}
\bigskip\bigskip
\end{raggedright}

On the night of March 4/5, 2010, after ten months long break, the
OGLE survey resumed regular observations entering the OGLE-IV
phase. With the new 32 chip mosaic camera covering 1.4 square
degrees on the sky with the resolution of 0.26 arcsec/pixel and
fast reading time of the entire array of 20 seconds the observing
capabilities of the OGLE survey increased by an order of magnitude
compared to the previous  OGLE-III phase. With this observing
set-up the OGLE project continues its 18 years long tradition of
being one of the largest sky surveys  worldwide with the current
outcome of about 40 Terabytes of raw data per year.

During the 2010 Galactic bulge observing season OGLE-IV regularly
monitored large fraction of the Galactic bulge conducting pilot
observations for the second generation microlensing survey. About
4.5 square degrees of the Galactic bulge regions with the highest
microlensing rate were observed with the cadence of 20 minutes
while additional 8.5 square degrees with the cadence of 1 hour.
Moreover, large area of the Galactic bulge was observed once or
twice a day.

The automatic data pipeline of OGLE-IV images reductions has been
recently finished. Its full implementation at the telescope,
expected in early 2011, will allow to restart on-line (real time)
reductions of the huge data-stream coming from the OGLE telescope
and restart the data analysis systems like the OGLE Early Warning
System.

%\begin{thebibliography}{99}

%\end{thebibliography}

%%

\Chapter{MOA-II observation in 2010 season}
            {MOA-II observation in 2010 season}{Takahiro Sumi}
\bigskip\bigskip

\addcontentsline{toc}{chapter}{{\it Takahiro Sumi}}
\label{polingStart}

\begin{raggedright}

{\it Takahiro Sumi\index{author}{Sumi, Takahiro}\\
Solar-Terrestrial Environment Laboratory \\
Nagoya University\\
Japan}
\bigskip\bigskip
\end{raggedright}

\section{Introduction}
Since the first discovery of exoplanets orbiting main-sequence
stars in 1995 \cite{may95}, more than 500 exoplanets have been
discovered via the radial velocity method and more than 50 have
been detected via their transits. Several planetary candidates
have also been detected via direct imaging, and astrometry.

The gravitational microlensing was proposed as a unique method to
search exoplanets \cite{Liebes64,mao91}. The planet's gravity
induces small caustics, which can generate small deviations in
standard \cite{pac86} single-lens microlensing light curves.
Compared to other techniques, microlensing is sensitive to smaller
planets, down to an Earth mass \cite{bennett96}, and in wider
orbits of 1-6 AU. Because microlensing observability does not
depend on the light from the lens host star, it is sensitive to
planets orbiting faint host stars like M-dwarfs and even brown
dwarfs. Furthermore, it is sensitive to distant host stars at
several kpc from the Sun, which allows the Galactic distribution
of planetary systems to be studied.

In 2003, the gravitational microlensing method yielded its first
definitive exoplanet discovery \cite{bon04}. So far ten planetary
systems with eleven planets have been found by this technique
%\cite{uda05,bea06,gou06,gau08,bennett08,dong09b,Janczak10,sumi2010,Miyake2011},
\cite{Miyake2011}, which have very distinct properties from those
detected by other techniques.

Although the radial velocity and transit discoveries are more
numerous, microlensing is uniquely sensitive to the cold Neptunes
outside of the snow-line, and the microlensing results to date
indicate that this class of planets may be the most common type of
exoplanet yet discovered \cite{sumi2010}.

\section{Observations}

The MOA-II carries out survey observations toward the Galactic
Bulge (GB) to find exoplanets and toward the large and small
Magellanic clouds (LMC and SMC) for searching MACHOs via the
gravitational microlensing using a 1.8m telescope, equipped with a
mosaic camera with 10 chips of 2kx4k-pixel CCD\cite{sako2008}, at
Mt. John Observatory in New Zealand.  We observe our target fields
very frequently (every 15-90 min) and analyze data in real-time to
issue alerts. This high cadence is specifically designed to find
the short timescale planetary signature.  Our real-time anomaly
alert system searches for planetary signatures in ongoing
microlensing events in which new data points are available on the
light curves within 5 min after exposure. In 2010 season, we
continued this observational strategy.

\section{Results}

In 2010 season, we detected 607 microlensing events and issued the
alerts toward the GB. Among these, 5 events show planetary or
brown dwarf signatures in their light curves thanks to the data by
OGLE-IV survey and intensive follow-up observations by $\mu$FUN,
PLANET, RoboNet, MiNDSTEp.

In 2010, we found one event toward the LMC that shows a clear
asymmetry due to the parallax. This indicates that the lens object
is likely a foreground disk star.

\bigskip
We are grateful to OGLE, $\mu$FUN, PLANET, RoboNet, MiNDSTEP
collaborations for close cooperation.

\Chapter{The RoboNet 2010 season}
            {The RoboNet 2010 season}{Yiannis Tsapras}
\bigskip\bigskip

\addcontentsline{toc}{chapter}{{\it Yiannis Tsapras}}
\label{polingStart}

\begin{raggedright}

{\it Yiannis Tsapras\index{author}{Tsapras, Yiannis}\\
Queen Mary University \\
UK}
\bigskip\bigskip
\end{raggedright}

\section{The Las Cumbres Observatory Global Telescope Network}
The Las Cumbres Observatory (LCOGT) \cite{lcogt} is a privately
funded institute which owns and operates the 2m Faulkes Telescopes
in Hawaii and Australia.  LCOGT is in the process of developing a
global network of 1m and 0.4m telescopes capable of robotic
around-the-clock observations. These will be used for science as
well as education. The 0.4m and 1m prototypes are currently being
tested and deployment around the world will begin in 2011. The
first ones will be deployed in Chile and South Africa, followed by
Tenerife, Australia and China. The complete network will comprise
of twelve to fifteen 1m and twenty-four 0.4m telescopes and is
expected to be complete by the end of 2014. The final numbers
depend on the relative costs. The SUPA-II Planet Hunter bid from
the University of St Andrews will finance the construction of an
extra three 1m telescopes with exactly the same specifications.

The 0.4m telescopes will be primarily used for educational
purposes. Schools from around the world will have access to these
telescopes through web interfaces and will be able to make
observations during daylight hours in Europe by using the
telescopes in Chile, Hawaii and Australia. Furthermore, they will
have the option to get involved in certain science projects, like
microlensing or transits, and contribute observations. The LCOGT
education pages provide descriptions of possible projects
available to the students. The 1m telescopes will be used for
science observations. Each site will eventually host two to four
telescopes which can be operated individually or in parallel,
allowing simultaneous observations of targets in different
pass-bands. Each telescope will be fitted with a fast readout CCD
camera that has a field of view of about half a degree and filters
in the UBVRI and Pan-STARRS Z and Y bands. Each cluster of
telescopes will have an optical fiber-feed to a shared
medium-resolution spectrograph.

The observation requests sent out to the telescopes will be
scheduled by a highly dynamic scheduling algorithm that will
constantly farm the database of TAC-approved science projects for
the most appropriate observations to perform at any given time
given the current conditions and the scientific priority.

\section{RoboNet: A microlensing search for cold planets}

In order to draw conclusions about planetary populations, it is
not enough to just detect planets, it is necessary to understand
the selection bias of the surveys. This requires the adoption of
an observing strategy that is not based on human intervention,
since this cannot be simulated. RoboNet's unique advantage is that
our entire system is completely automated, which makes it easier
to understand the selection biases of our strategy. As the system
is designed for a network with continuous Bulge coverage, we
expect to address this question with the help of the full LCOGT
network in future seasons.

The methodology used by the RoboNet team is described in
\cite{Tsapras09}. Our pilot campaign (RoboNet-I) contributed to 5
of the first 6 microlensing planet discoveries and provided the
impetus to develop our system further. We are continuing this
development in preparation for the deployment of the LCOGT network
of 1m and 0.4m telescopes.

RoboNet has pioneered the technique of adaptive scheduling of
microlensing targets through the use of a prioritisation algorithm
that continuously shuffles the observable microlensing targets,
giving higher priority to the ones that are more likely to reveal
a planetary signal \cite{Horne09}. It calculates the optimal
frequency at which each ongoing microlensing event needs to be
sampled at in order to maximize the planet detection probability.
Once an observation is obtained for a specific event, it's
priority is adjusted upwards in the case that it deviates from the
expected single-lens case, or downwards if it doesn't. The
relative priorities are reassessed every few minutes and active
events that have not been observed recently will start to appear
higher and higher in the list until an observation is obtained. At
that point their relative priority will be readjusted according to
the previous scheme. In order to be sensitive to "cold Earths", a
sampling interval of ~15-45 mins is required, which the extended
robotic network will be able to deliver.

RoboNet makes use of difference image analysis software
\cite{Bramich08} to extract accurate brightness measurements from
the incoming images which are used to construct the event
lightcurves. One image is selected as a template and all other
images are geometrically and photometrically aligned to and
subtracted from it. Any stars that show variability are
immediately obvious on the resulting subtracted images, while
stars of constant brightness leave no residuals.

The resulting lightcurves are immediately shared with the extended
microlensing community via automated rsync transfers but are also
available on the project website for downloading. The lightcurves
are continuously assessed by our SIGNALMEN software, which uses
robust methods to automatically detect the onset of anomalous
features. If such a feature is detected, an override request is
sent to the network to obtain prompt observations that will
confirm or disprove the anomaly. While most observations are
performed in queued observing mode, we sometimes use our override
capability in order to respond to anomaly alerts. Once more
telescopes are included in our network, we will be able to operate
in a fully automated mode throughout the microlensing season.

%%%%%%%%%%%%  Section 2   %%%%%%%%%%%%%%%%%%%%%%%%%%%%
\emptyheads

\begin{center}
\begin{large}
{\Huge \sffamily Cosmological Microlensing}
\end{large}
\end{center}
\bigskip

\begin{tabbing}
 Brief Reports from the \=B factories xxxxx \= xxxxxxxxx\= \kill
\\
\\
Dark matter determinations from Chandra observations of quadruply lensed quasars\\
\>\> {\it David Pooley}\\
\\
Cosmic equation of state from strong gravitational lensing systems \\
\>\> {\it Marek Biesiada \& Beata Malec}\\
\\

\end{tabbing}

\fancyheads

\Chapter{Dark Matter Determinations from \textit{Chandra}
Observations of Quadruply Lensed Quasars}
            {Dark matter determinations from quadruply lensed quasars}{David Pooley}
\bigskip\bigskip

\addcontentsline{toc}{chapter}{{\it David Pooley}}
\label{polingStart}

\begin{raggedright}

{\it David Pooley\index{author}{Pooley, David}\\
Eureka Scientific\\
USA}
\bigskip\bigskip
\end{raggedright}

I present all publicly available \textit{Chandra} observations of
14 X-ray bright quadruply lensed quasars.  The X-ray data reveal
flux ratio anomalies which are more extreme than those seen at
optical wavelengths, confirming the microlensing origin of the
anomalies originally seen in the optical data.  The X-ray emitting
regions are essentially point sources and therefore give a
microlensing signal unencumbered by source size considerations.
Building on our previous work, we have completed a thorough
investigation of the best way to analyze the X-ray data, resulting
in improved determinations of the X-ray flux ratios.  We have
constructed custom microlensing magnification maps for a range of
stellar fractions for the four images of each quasar, and we have
implemented a more sophisticated Bayesian analysis of the data to
determine the most likely dark matter fraction in the lensing
galaxies.

\Chapter{Cosmic Equation of state from Strong Gravitational
Lensing Systems}
            {Cosmic equation of state from SGL
systems}{Biesiada \& Malec}
\bigskip\bigskip

\addcontentsline{toc}{chapter}{{\it Biesiada \& Malec}}
\label{polingStart}

\begin{raggedright}

{\it Marek Biesiada\index{author}{Biesiada, Marek}\\
Department of Astrophysics and Cosmology, Institute of Physics \\
University of Silesia
Poland \\}
\bigskip
{\it Beata Malec\index{author}{Malec, Beata}\\
Copernicus Center for Interdisciplinary Studies \\
Cracow
Poland\\}

\bigskip\bigskip
\end{raggedright}

\section{Introduction}
Accelerating expansion of the Universe is a great challenge for
both physics and cosmology. In light of lacking the convincing
theoretical explanation, an effective description of this
phenomenon in terms of cosmic equation of state turns out useful.

The strength of modern cosmology lies in consistency across
independent, often unrelated pieces of evidence. Therefore, every
alternative method of restricting cosmic equation of state is
important. Strongly gravitationally lensed quasar-galaxy systems
create such new opportunity by combining stellar kinematics
(central velocity dispersion measurements) with lensing geometry
(Einstein radius determination form position of images).

\section{Idea and methods}

Strong gravitationally lensed systems create opportunity to test
cosmological models of dark energy in a way alternative to Hubble
diagrams (from SNIa or GRBs), CMBR or LSS. The idea is that
formula for the Einstein radius in a SIS lens (or its SIE
equivalent)
$$
\theta_E = 4 \pi \frac{\sigma_{SIS}^2}{c^2} \frac{D_{ls}}{D_s}
$$
depends on the cosmological model through the ratio:
(angular-diameter) distance between lens and source to the
distance between observer and lens. Provided one has reliable
knowledge about lensing system: the Einstein radius $\theta_E$
(from image astrometry) and stellar velocity dispersion
$\sigma_{SIS}$ (central velocity dispersion inferred from
spectroscopy) one can used such well studied systems to test
background cosmology. This method is independent on the Hubble
constant value ($H_0$ gets cancelled in the distance ratio) and is
not affected by dust absorption or source evolutionary effects. It
depends, however, on the reliability of lens modelling (e.g. SIS
or SIE assumption). The method was proposed by
Biesiada~\cite{biesiada} and also discussed in details by Grillo
et al. (2008)~\cite{grillo}.

Here we apply such method to a combined data sets from SLACS and
LSD surveys of gravitational lenses~\cite{grillo}. In result we
obtain the cosmic equation of state parameters, which generally
agree with results already known in the literature. This
demonstrates that the method can be further used on larger samples
obtained in the future~\cite{biesiadMNras}. We have also performed
joint analysis taking into accout standard rulers (combining
lensing system data with CMB acoustic peak location and BAO data)
and standard candles (SNIa data - we used Union08 compilation by
Kowalski et al. (2008)~\cite{kowalski} ). The observables we used
had different parameter degeneracies and so different restrictive
power in the parameter spaces of cosmological models. It can be
best seen in figures.

\begin{figure}[htb]
\begin{center}
\epsfig{file=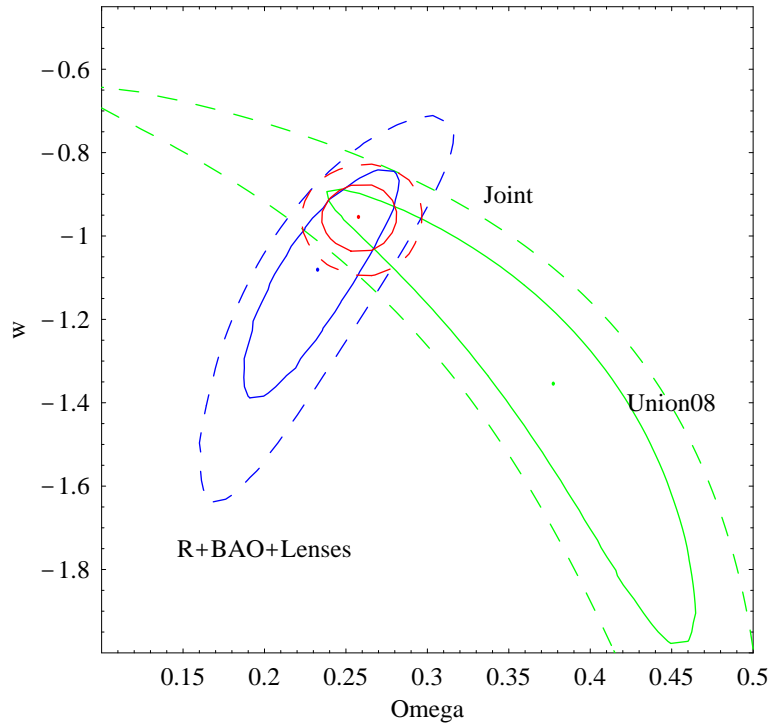,height=4.0in} \caption{Best fits (dots) and
(68 \%, 95\%) confidence regions in $(\Omega, w)$ plane for
quintessence model. Confidence regions displayed separately for
standard rulers, standard candles and joint analysis.}
\label{fig:quint}
\end{center}
\end{figure}

We considered five cosmological scenarios of dark energy, widely
discussed in current literature. These are $\Lambda$CDM,
Quintessence, Chevalier-Polarski-Linder model, Chaplygin gas and
Braneworld scenario. Then we performed $\chi^2$ fits to different
cosmological scenarios on described samples. The probes were
combined by calculating joint likelihoods.

Because standard rulers and standard candles probe distance
measures based on different concepts (angular diameter distance
and luminosity distance), one step before making a full joint fit
we performed fits based on rulers and candles separately.

\begin{figure}[htb]
\begin{center}
\epsfig{file=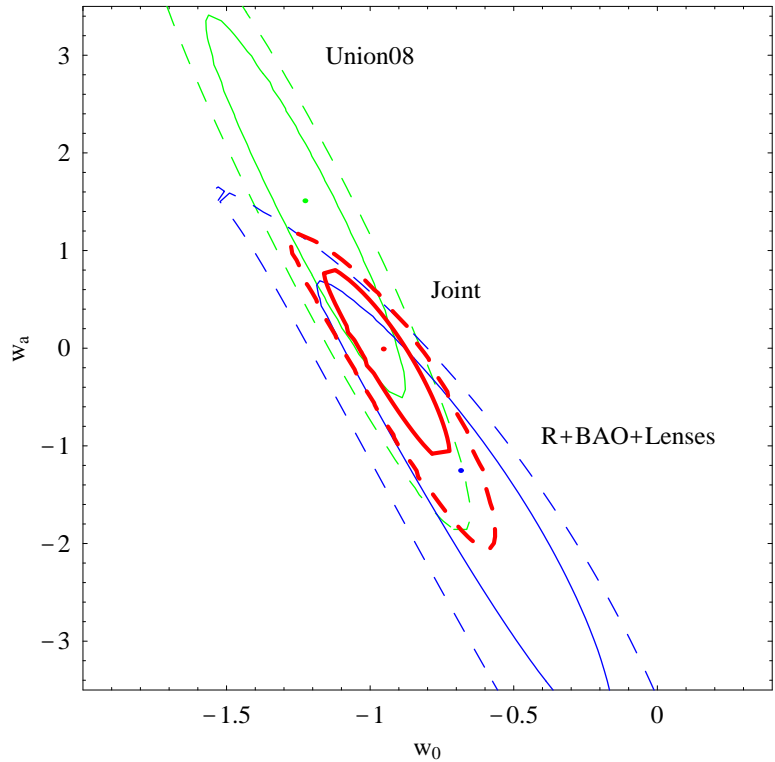,height=4.0in} \caption{Best fits (dots) and
(68 \%, 95\%) confidence regions in $(w_0, w_a)$ plane for
Chevalier-Linder-Polarski
  model. Confidence regions displayed separately for standard rulers, standard candles and joint analysis.} \label{fig:quint}
\end{center}
\end{figure}

\section{Conclusions}

The best fits we obtained for the model parameters in joint
analysis turned out to prefer cases effectively equivalent to
$\Lambda$CDM model. They are also in agreement with other combined
studies performed by other authors on different sets of diagnostic
probes. As comparison between models in terms of chi-square values
does not account for the relative structural complexity of the
models we also made use of information theoretic methods.
According to Akaike criterion (AIC) $\Lambda$CDM it is only
slightly preferred over the quintessence and both models with a
dynamical equation of state $w(z)$ (CPL parametrization) and
Chaplygin gas scenario get considerably less support from the
data. Odds against the brane-world scenario are so high that it
can be considered as ruled out by the data. According to the
Schwartz Bayesian Information criterion (BIC) $\Lambda$CDM wins,
the quintessence model is considerably less supported by data, and
the other ones are ruled out.

\bigskip
This work was supported by the Polish Ministry of Science Grant
no. N N203 390034.

%%%%%%%%%%%%  Section 3   %%%%%%%%%%%%%%%%%%%%%%%%%%%%
\emptyheads

\begin{center}
\begin{large}
{\Huge \sffamily Galactic Microlensing: \\the dark matter search}
\end{large}
\end{center}
\bigskip

\begin{tabbing}
 Brief Reports from the \=B factories xxxxx \= xxxxxxxxx\= \kill
\\
\\
PAndromeda - the Pan-STARRS M31 survey for dark matter\\
\hspace{2.0cm}{\it Arno Riffeser, Stella
Seitz, Ralf Bender, C.-H. Lee, Johannes Koppenhoefer}\\
\\
Final OGLE-II and OGLE-III results on microlensing towards
the LMC and SMC \\
\>\> {\it {\L}ukasz Wyrzykowski}\\
\\
Analysis of microlensing events towards the LMC \\
\>\> {\it Luigi Mancini \& Sebastiano Calchi Novati}\\
\\
Simulation of short time scale pixel lensing towards the Virgo cluster \\
\>\> {\it Sedighe Sajadian \& Sohrab Rahvar}\\
\\
M31 pixel lensing and the PLAN project \\
\>\> {\it Sebastiano Calchi Novati}\\
\\
\end{tabbing}

\fancyheads

\Chapter{PAndromeda - A Dedicated Deep Survey of M31 with
Pan-STARRS 1}
            {PAndromeda - Deep Survey of M31}{Riffeser et al.}
\bigskip\bigskip

\addcontentsline{toc}{chapter}{{\it Riffeser et al.}}
\label{polingStart}

\begin{raggedright}

{\it Arno Riffeser\index{author}{Riffeser, Arno}, Stella
Seitz\index{author}{Seitz, Stella}, Ralf
Bender\index{author}{Bender, Ralf}, Chien-Hsiu Lee\index{author}{Lee, Chien-Hsiu},
Johannes Koppenhoefer\index{author}{Koppenhoefer, Johannes }\\
Max Planck Institute for Extraterrestrial Physics, Garching \\
Germany\\}

\bigskip

\end{raggedright}
{\it Pan-STARRS 1 Science Consortium University of Hawaii,
Pan-STARRS Project Office, Max Planck Institute for Astronomy, Max
Planck Institute for Extraterrestrial Physics, Johns Hopkins
University, University of Durham, University of Edinburgh, Queens
University of Belfast, Harvard-Smithsonian Center for
Astrophysics, Los Cumbres Observatory Global Telescope Network,
National Central University of Taiwan\\}

The 1.8~m Panoramic Survey Telescope and Rapid Response System
(Pan-STARRS) on Haleakala, Maui, began its regular survey in
spring 2010.  With its $\sim$7~deg$^2$ field of view and the use
of orthogonal transfer array CCDs \cite{2007SPIE.6501..007B} it
represents the system with the highest data-flow today. In
addition to the large area 3Pi Survey some fields are exposed
deeper, and visited more frequently.  One of these so called
medium deep fields (MD) targets the Andromeda galaxy. PAndromeda
monitors M31 for 2\% of the overall PS1 time. This means that M31
is observed for 0.5 h per night during a period of 5 months per
year.  PAndromeda is designed to identify gravitational
microlensing events, caused by bulge and disk stars (self-lensing)
and by compact matter in the halos of M31 and the MW (halo
lensing, or lensing by MACHOs).  The main science goals of
PAndromeda are measuring the masses and mass-fraction of compact
objects in the M31 and MW halos, and constraining the M31 bulge
mass function at the low mass end.  As a side product PAndromeda
is also able to search for microlensing events towards M32 and NGC
205.

The interpretation of the microlensing events (see
\cite{2006ApJS..163..225R}) requires i) understanding the mix of
stellar ages and metalicities in the bulge, disk, and halo of M31
as obtained from resolved stellar populations (census of
supergiants, OB-associations, analysis of CMD diagrams as a
function of location) variability studies (from Cepheids to LPVs),
and color gradients in the light profiles, ii) deriving improved 3
dimensional models for the density and velocity distributions of
the bulge and disk, iii) including improved constraints on the
extinction in M31 \cite{2009A&A...507..283M}.  All these
informations can directly extracted from the PAndromeda data
itself.

The continuous monitoring in the $r'$ and $i'$ bands with
PAndromeda allows to confirm the achromaticity of events; the main
filter (detection filter) is the $r'$-band to optimize the number
of source stars accessible for measurable lensing events, which
depends on the brightness of the stars, the photon noise (mostly
by M31), and the sensitivity of filters in each band.  During the
first season we focussed on the integration depth in the filters
r' and i'.  We split the $r'$-band imaging into two integrations
(separated by 4-6 hours) per night, for the whole season.  This
allows to measure light curves with shorter time scales than 1 day
more accurately, since a large fraction of events is predicted
(and found in previous surveys) with short time scales.
PAndromeda monitored M31 from 07/23/2010 till 12/27/2010 on 91
nights (58\%). In total 1782 images were exposed, 1179 in r' (90
nights, 70740 sec) and 603 in i' band (66 nights, 36180 sec).  The
total amount of reduced data is 14 TB.  Depending on our observing
strategy in 2011 the remaining filters and integration times for
PAndromeda will be chosen such, that in turn these observations
can be used to study many other aspects of M31 with the aim to
improve the predictions of the microlensing event rates.  These
microlensing predictions depend on the mass function of stars,
their luminosities and sizes as determined by their ages and
metallicities, on their density and velocities distributions, and
on the extinction.

The PS1 Gigapixel Camera (GPC1) produces images consisting of 3840
CCDs ($60\times8\times8$) with roughly $580\times590$ pixels each.
This results in $1.3\times 10^{9}$ pixels which are exposed and
read-out.  The GPC1 data are de-biased, flat-fielded, and
astrometrically registered with the Pan-STARRS Image Processing
Pipeline (by Magnier E.).  A reduced GPC1 exposure requires a disk
storage of 8.0 GB consisting of frame, weight and mask.  After
this basic reduction the data are further processed by our own
image processing software (MUPIPE, \cite{2002A&A...381.1095G}).
For the PAndromeda difference imaging analysis we adapted the
MUPIPE pipeline to the specifications of PS1 data and data flow
and implemented it into the Astro-WISE system.

From the 2010 season we analyzed the central field of M31
($21'\times 21'$). This is to test the detection process in the
field where we expect the highest lensing rate because of self
lensing.  So far we detected 3 high quality microlensing
light-curves. The third one is very bright with 19 mag in r'. Note
that high flux excess events are more difficult to reconcile with
self-lensing than with halo-lensing \cite{2008ApJ...684.1093R}.
The full data set is currently analyzed.

\Chapter{Final OGLE-II and OGLE-III results on microlensing
towards the LMC and SMC}
            {OGLE results towards the LMC and SMC}{{L}ukasz Wyrzykowski}
\bigskip\bigskip

\addcontentsline{toc}{chapter}{{\it {L}ukasz Wyrzykowski}}
\label{polingStart}

\begin{raggedright}

{\it {L}ukasz Wyrzykowski\index{author}{Wyrzykowski, {L}ukasz}\\
Institute of Astronomy, University of Cambridge \\
UK}
\bigskip\bigskip

\end{raggedright}

\section{Introduction}

For almost two decades compact halo objects (MACHOs) have been one
of the best candidate for Dark Matter (DM). The main channel of
detecting them is through the microlensing phenomenon, however the
results from two microlensing surveys, MACHO and EROS, are in
contradiction as to the abundance of MACHOs in the Galaxy.

We present an independent and the most comprehensive result so far
from the OGLE microlensing survey of the Large and Small
Magellanic Clouds running continuously for over 13 years. In both
Clouds, we report the detection of the total of 8 events with 2
more plausible candidates. All but one of them are consistent with
the expected signal from lensing by Clouds own stars (self-
lensing), therefore there is no need of introducing dark
microlenses.

\section{LMC: 2+2 events and 2 candidates}
There were 4 events found in the OGLE-II and OGLE-III data.
Another 2 more were flagged as potential candidates in OGLE-III
data. The optical depth derived for these events was
$\tau_{\mathrm{LMC-O2}} = 0.43\pm0.33\times10^{-7}$ and
$\tau_{\mathrm{LMC-O3}} = 0.16\pm0.12\times10^{-7}$ for OGLE-II
and OGLE-III, respectively (\cite{Wyrzykowski2009},
\cite{Wyrzykowski2011}).

Time-scales, locations and positions of these events on the
colour-magnitude diagram indicate all of them are very likely to
be solely caused by self-lensing. Source stars in all events seem
to belong to the LMC and events occurred in the areas where the
self-lensing contributes the most to the overall microlensing
optical depth.

Small number of events seen by OGLE indicates the excess of events
seen by MACHO group is probably not caused by microlensing. OGLE
data ruled out as microlensing one of the MACHO events (\#7),
which showed another bump after many years.

\section{SMC: 1+3 events}
Previously there were 2 events found in the SMC by MACHO and EROS
groups and both were confirmed to be due self-lensing.

In the OGLE data there were 4 events detected giving
$\tau_{\mathrm{SMC-O2}} = 1.55\pm1.55\times10^{-7}$
\cite{Wyrzykowski2010} and $\tau_{\mathrm{SMC-O3}} =
1.27\pm0.96\times10^{-7}$. Events 03 and 04 are good candidates
for self-lensing due to their time-scales and locations. Event 01
is a rather weak candidate and could be also due to a variable
star.

In the most likely microlensing model for the event OGLE-SMC-02
the lens is a 10 $M_{\odot}$ binary black hole from the Galactic
halo \cite{Dong2007}. Such events, however, are not seen towards
the LMC, which calls for further observations and theoretical
studies.

\begin{figure}[htb]
\begin{center}
\epsfig{file=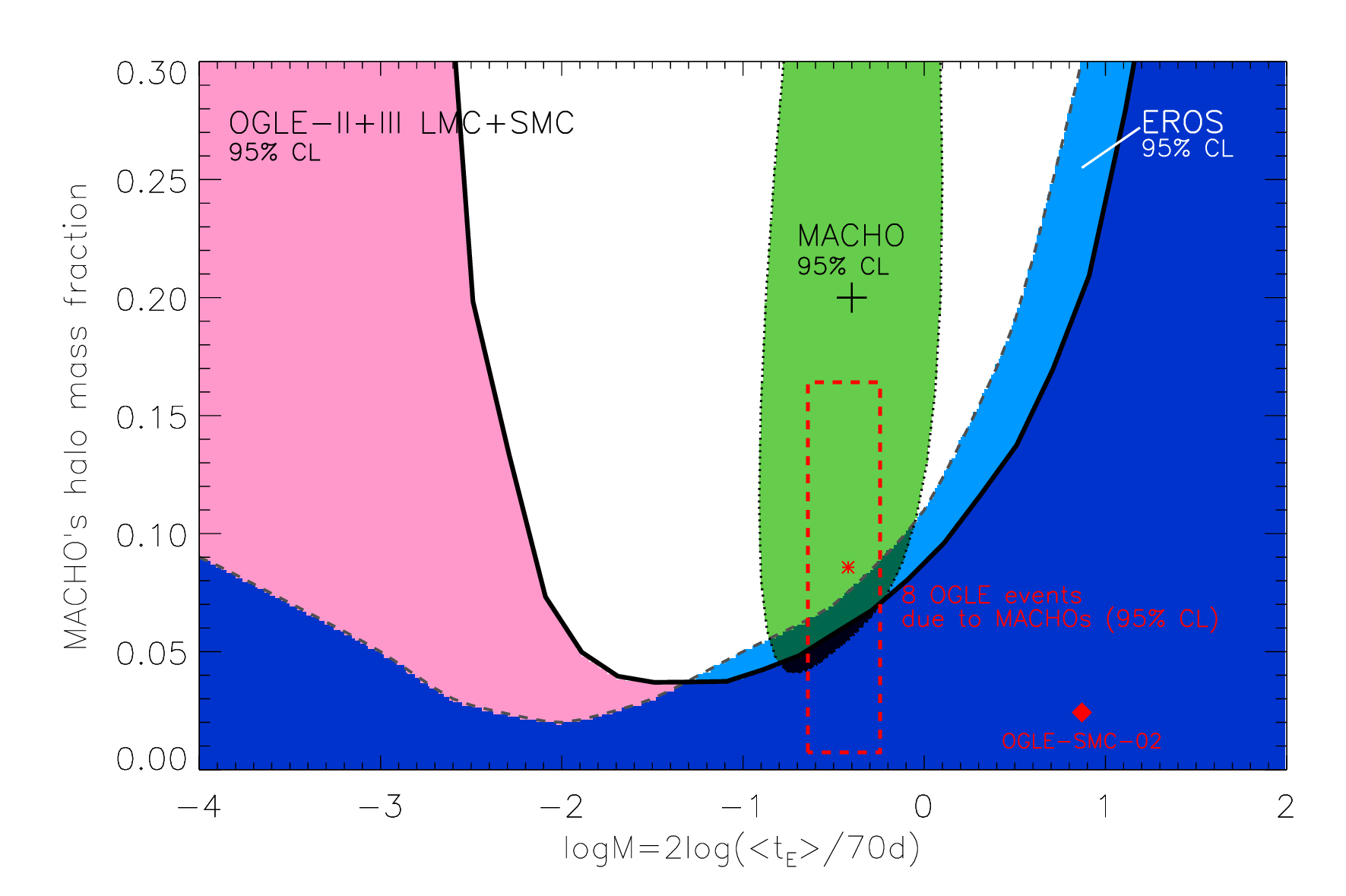,height=4.0in} \caption{MACHOs contribution
to the total halo mass as measured by MACHO group and upper limits
from EROS and OGLE data.} \label{fig:taulim}
\end{center}
\end{figure}

\section{Conclusions}
Based on our detection efficiency, we derived new constrains on
the MACHOs presence in the Galactic Halo of 6\% for M=0.1-0.4
$M_{\odot}$ and below 4\% for masses in range 0.01-0.1 $M_{\odot}$
(Fig. \ref{fig:taulim}). For MACHOs with mass of 1 $M_{\odot}$ the
upper limit is $f<9$\% and $f<20$\% for $M< 6 M_{\odot}$.

Our result indicates that baryonic DM in the form of relics of
stars and very faint objects in the sub-solar-mass range are
unlikely to inhabit the Milky Way's dark matter halo in any
significant numbers. Presence of a black-hole lens candidate
towards the SMC agrees with expected no more than 2\% contribution
of black-holes to the mass of the Galactic halo.

\Chapter{Analysis of microlensing events towards the LMC}
            {Analysis of microlensing events towards the LMC}{Mancini \& Calchi Novati}
\bigskip\bigskip

\addcontentsline{toc}{chapter}{{\it Mancini \& Calchi Novati}}
\label{polingStart}

\begin{raggedright}

{\it Luigi Mancini\index{author}{Mancini, Luigi} \&
Sebastiano Calchi Novati\index{author}{Calchi Novati, Sebastiano}\\
Department of Physics, University of Salerno, Italy
\\
IIASS, Vietri Sul Mare\\
Italy}
\bigskip\bigskip

\end{raggedright}

The nature of the observed microlensing events towards the
Magellanic Clouds is still an open issue. Indeed, in order to draw
meaningful insights into the contribution of dark matter objects
in form of MACHOs, it is essential first to estimate accurately
all the possible contributions due to (luminous) lenses belonging
to known populations located along the line of sight (to which we
refer too as ``self lensing''). A possibly non exhaustive list
includes lenses belonging to the luminous components of the LMC
(the disc and the bar) that can act as sources , the disc of the
Milky Way and the somewhat elusive stellar halo for both the Milky
Way and the LMC.

The results reported so far by the collaborations who carried out
observational campaigns towards the LMC as for the MACHO
contribution to the Galactic halo are in agreement to exclude
MACHOs as a viable dark matter candidate for masses below
$(10^{-1}-10^{-2})~\mathrm{M}_\odot$. However, a relevant
discrepancy still exists as for compact halo object in the mass
range $(0.1-1)~\mathrm{M}_\odot$. The MACHO collaboration claimed
for an mass halo fraction in form of MACHOs of about $f\sim 20\%$
out of observations towards the LMC (Alcock et~al., 2000), a
result more recently confirmed by Bennett (2005). On the other
hand, the EROS (Tisserand et~al., 2007) and the OGLE (Wyrzykowski
et~al., 2009, 2010) collaborations, out of observations towards
both the LMC and SMC, concluded that the expected self-lensing
rate is sufficient to explain the observed rate also in this mass
range. It is therefore important to address the issue of the
nature of the observed events, either to be attributed to MACHO
lensing or to self lensing.

In previous analyses we have considered the set of events reported
by the MACHO collaboration and shown that, on the basis of both
their number number and their characteristics (event duration and
spatial distribution), they cannot all be attributed to self
lensing (Mancini et~al., 2004). In Calchi~Novati et~al. (2006) we
have considered the possible role played by the LMC dark matter
halo, in particular suggested that  the halo fraction in form of
MACHOs for the Milky Way and the LMC might not be equal. Finally,
in Calchi~Novati et~al. (2009) we have discussed the results of
the OGLE-II campaigns towards the LMC.

Here we report on a detailed analysis of the recent results of the
OGLE-III campaiagn towards the LMC (Wyrzykowski et~al, 2010). For
all the possible lens populations (both luminous and dark), we
present maps of the optical depth and a study of the expected
characteristics (duration, spatial distribution and number), to be
compared with the observed events. This is done through an
evaluation of the microlensing rate towards all the observed
fields. Finally, we evaluate the probability distribution for the
mass halo fraction in form of MACHO via a likelihood analysis.
Overall, we find the observed rate to be compatible with the
expected lensing signal by known luminous populations.

\Chapter{Simulation of pixel lensing of M87 by HST}
            {Simulation of pixel lensing of M87 by HST}{Sajadian \& Rahvar}
\bigskip\bigskip

\addcontentsline{toc}{chapter}{{\it Sajadian \& Rahvar}}
\label{polingStart}

\begin{raggedright}

{\it Sedighe Sajadian\index{author}{Sajadian, Sedighe} \& Sohrab Rahvar\index{author}{Rahvar, Sohrab}\\
Department of Physics, Sharif University of Technology, Tehran \\
Iran\\}

\bigskip\bigskip

\end{raggedright}

In this work we propose a new strategy of pixel lensing
observation of the unresolved stars of M87 in the Virgo cluster by
HST. We show that in contrast to the previous observational
strategy in Baltz \index{Baltz} et al. (2004) with one observation
per night for the duration of one month, a few days intensive
observation with taking one image per one HST orbit, we can
substantially increase the number of events more than one order of
magnitude. In this observational strategy, high magnification
microlensing events is predicted to be observed with the rate of
$\sim 4$ event per day with a typical transit time scale of $\sim
19$ hours. We examine the possible detection of dark matter
mini-halos with this observational strategy.

\Chapter{M31 pixel lensing and the PLAN project}
            {M31 pixel lensing and the PLAN project}{Sebastiano Calchi Novati}
\bigskip\bigskip

\addcontentsline{toc}{chapter}{{\it Sebastiano Calchi Novati}}
\label{polingStart}

\begin{raggedright}

{\it Sebastiano Calchi Novati\index{author}{Calchi Novati, Sebastiano}\\
Department of Physics, University of Salerno
Italy\\
IIASS, Vietri Sul Mare\\
Italy}
\bigskip\bigskip

\end{raggedright}

\section{Introduction: M31 Pixel Lensing and the issue of MACHO versus self lensing}

Pixel lensing is a unique tool for the study of dark matter
objects in form of MACHOs up to distant galaxies as M31. More than
20 pixel lensing events towards M31 have been reported so far from
several different campaigns, but no firm conclusions on the MACHO
issue have been reached yet (Calchi~Novati, 2010). A fundamental
problem in the analysis is the difficulty to disentangle
microlensing signal due to dark matter MACHOs from ``self
lensing'', microlensing signal due to lenses belonging to known
luminous populations. This is complicated first by the additional
degeneracy in the microlensing event parameter space
characteristics of \emph{pixel} lensing, where one studies flux
variations of \emph{unresolved} objects. Second, by the rather
large expected self-lensing signal, as compared to the MACHO
lensing one, at least in the inner M31 region, where in any case
most of the events, either MACHO or self lensing, are expected. In
this respect, as for many others, M31 pixel lensing is peculiar
with respect to microlensing towards the Magellanic Clouds, LMC
and SMC (Moniez, 2010). Still, besides the underlying technique,
also the fundamental physical issue one addresses to, that of
MACHOs, is the same. The recent analyses of OGLE-II/SMC
(Wyrzykowski et~al., 2010) and OGLE-III/LMC (Wyrzykowski et~al.,
2010) both indicates that the observed rate of events is
compatible with the expected self-lensing signal. This is in
agreement with the previous OGLE-II/LMC (Wyrzykowski et~al., 2009)
and the EROS results (Tisserand et~al. 2007). On the other hand,
the MACHO collaboration claimed for a MACHO signal of about
$0.5~\mathrm{M}_\odot$ for a mass halo fraction in form of MACHO
$f\sim 20\%$ (Alcock et~al., 2000), a result more recently
confirmed by the further analyses of Bennett (2005). To address
the issue of MACHOs along a different line of sight is therefore
important. Besides, looking towards M31 one can map its own full
dark matter halo, which is not possible for the Galaxy one.

To explore the MACHO versus self-lensing issue, and more in
general to address the problem of the nature of the observed
events, the careful and through analysis of \emph{single events}
have shown to be a very efficient tool even if it can only give
partial indications. Relevant results have been reported for the
M31 microlensing event PA-N1 (Auri\`ere et~al., 2001), PA-S3/GL1
(Riffeser et~al., 2008), and by the more recent analysis of OAB-N2
by the PLAN collaboration (Calchi~Novati et~al., 2010).
Remarkably, though with different level of confidence, all these
analyses conclude that MACHO lensing is to be preferred to self
lensing for the events studied. This approach runs in parallel
with the  analysis of full data set completed by a careful
comparison of the observed  and the expected rate (an approach
which still has to deal with the limitation given by the rather
small statistics involved). The more relevant results have been
reported by the POINT-AGAPE (Calchi~Novati et~al., 2005) and the
MEGA (de~Jong et~al., 2006) collaborations, working on the same
data set. POINT-AGAPE claimed for an evidence of a MACHO signal,
in the same mass range indicated by the MACHO LMC analysis,
whereas MEGA concluded that self lensing was sufficient to explain
the detected rate. An extremely promising new project for M31
pixel lensing observation is that of PAndromeda (Riffeser et~al,
2011).

\section{The PLAN project: an update}

Within this still open framework we discuss some of the results we
have obtained with the PLAN collaboration which is carrying out a
long term monitoring project of M31 (with observational campaigns
at the 1.5m OAB telescope from 2006 to 2010 and a pilot campaign
at the 2m HCT telescope in October 2010). In Calchi~Novati et~al.
(2009) as a result of a fully automated pipeline we have reported
2 microlensing candidate events out of the 2007 campaign (OAB-N1
and OAB-N2). The observed rate turned out to be relatively large,
still, compatible with the expected self-lensing one. The
available statistics was however too small to allow us to draw
stringent conclusions on the MACHO issue. Since then we have
followed both the indicated approaches to address the MACHO versus
self-lensing issue.

First, some additional data made available by the WeCAPP
collaboration along the OAB-N2 light curve and a more through
analysis of KPNO and HST archive images used to better constrain
the possible event source characteristics, have allowed us a much
more detailed analysis of OAB-N2 (Calchi~Novati et~al., 2010). In
particular, through a study of the lens proper motion, we have
shown that this event is more likely to be attributed to MACHO
lensing than to self lensing.

As for our full data set, more statistics is badly needed. Here we
report some preliminary results of the 2008 and 2009 campaigns. As
an output, no news microlensing candidate events have been
detected. In particular we discuss the crucial role played by the
selected bump unicity analysis we carry out on archive POINT-AGAPE
light curves. We compare the observed rate to the expected one
evaluated through a Monte~Carlo simulation completed by an
analysis of the efficiency of the pipeline. The observed rate is
compatible with the expected self-lensing one but still does not
allow us to constrain MACHO lensing (Calchi~Novati et~al, 2011).

%%%%%%%%%%%%  Section 4   %%%%%%%%%%%%%%%%%%%%%%%%%%%%
\emptyheads

\begin{center}
\begin{large}
{\Huge \sffamily Planetary events}
\end{large}
\end{center}
\bigskip

\begin{tabbing}
 Brief Reports from the \=B factories xxxxx \= xxxxxxxxx\= \kill
\\
\\
OA-2009-BLG-266LB: the first cold Neptune with a measured
mass\\
\>\> {\it David Bennett}\\
\\
Increasing the detection rate of low-mass planets in
high-magnification events and \\MOA-2006-BLG-130 \\
\>\> {\it Julie Baudry \& Philip Yock}\\
\\
The complete orbital solution for OGLE-2008-BLG-513 \\
\>\> {\it Jennifer Yee}\\
\\
Planetary microlensing event MOA-2010-BLG-328 \\
\>\> {\it Kei Furusawa}\\
\\
Binary microlensing event OGLE-2009-BLG-020 gives orbit
predictions verifiable by \\follow-up observations \\
\>\> {\it Jan Skowron}\\
\\
\end{tabbing}

\fancyheads

\Chapter{MOA-2009-BLG-266LB: The First Cold Neptune with a
Measured Mass}
            {MOA-2009-BLG-266LB: The First Cold Neptune with a Measured Mass}{David P. Bennett}
\bigskip\bigskip

\addcontentsline{toc}{chapter}{{\it David P. Bennett}}
\label{polingStart}

\begin{raggedright}

{\it David P. Bennett\index{author}{Bennett, David P.}\\
University of Notre Dame, Notre Dame, IN\\
USA\\}
\bigskip
{\it for the MOA, $\mu$FUN, RoboNet, PLANET, MiNDSTEp, and OGLE
Collaborations\\}
\bigskip\bigskip
\end{raggedright}

The MOA-II survey detected beginning of the planetary signal in
microlensing event MOA-2009-BLG-266, so MOA's prompt anomaly alert
led to nearly complete sampling of the planetary anomaly. Good
light curve coverage from MOA and a number of the follow-up
telescopes after the anomaly enabled the measurement of a strong
microlensing parallax signal that allows the mass to be measured.

MOA-2009-BLG-266 is also the first planetary microlensing event to
be observed from a telescope in Heliocentric orbit, as it was
observed for $\sim 2\,$days by the High Resolution Instrument
(HRI) of the Deep Impact (DI) spacecraft. The short duration and
unfortunate timing of the space-based observations imply that
these observations have only a weak constraint on the microlensing
parallax measurement. But this data also indicates that future
observations with the DI/HRI instrument in 2011-2014 will allow
mass measurements of many microlens planets and their host stars,
including the first mass measurements of planets in the Galactic
bulge.

\Chapter{Increasing the detection rate of low-mass planets and
MOA--2006-BLG-130}
            {Increasing the detection rate of low-mass planets}{Julie Baudry}
\bigskip\bigskip

\addcontentsline{toc}{chapter}{{\it Julie Baudry}}
\label{polingStart}

\begin{raggedright}

{\it Julie Baudry\index{author}{Baudry, Julie}\\
University of Orsay, Paris Sud XI \\
France\\}
\bigskip
{\it Philip Yock\index{author}{Yock, P.}\\
Department of Physics, University of Auckland \\
New Zealand\\}

\bigskip\bigskip

\end{raggedright}

\section{Introduction}

Extra-solar planets detection is one of the applications of the
gravitational microlensing phenomenon: a perturbation on the light
curve can betray the presence of a planet orbiting the lens
star~\cite{GouldLoeb,BennettRhie}. Most microlensing planets have
been found in events of high magnification. Events of highest
magnification occur when the lens star transits the source star,
but these events provide relatively poor sensitivity to planets.
Previous research concentrated on highest magnification events,
but, on purely geometrical grounds, near-perfect alignement is
relatively rare and there must be more events at low magnification
than at high magnification. It will be presented in a first part
the interest of monitoring events of lower magnification, in order
to increase the detection rate of low-mass planets.

MOA-2006-BLG-130 (OGLE-2006-BLG-437) is a microlensing event that
occurred in August 2006 which exhibited a slight perturbation on
the light curve. In a second part, we will present the analysis of
this event, in order to  see whether or not it is a binary lens.

\section{Planetary perturbation detectability and magnification}

\begin{figure}[htb]
\begin{center}
\epsfig{file=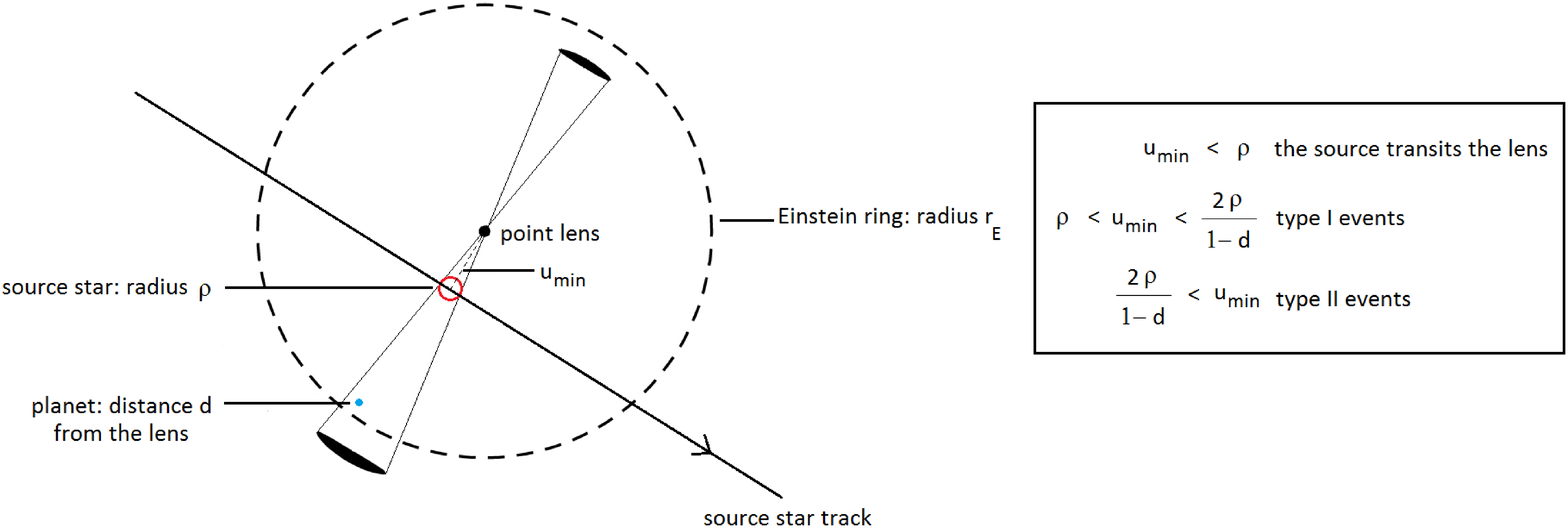,height=3.0in} \caption{Different
types of events, depending on the geometrical construction of the
lens.} \label{fig:events_zones}
\end{center}
\end{figure}

Events of lower magnification may be subdivided into two classes :
those in which the planet lies closer to the Einstein ring than
the length of either Einstein arc (type I), and those in which the
planet lies further away (type II), Figure~\ref{fig:events_zones}.
Both type I and type II events enjoy good sensitivity to planets.
In type II events, the planet is closer to the lens, and the
planetary perturbation is greater when the magnification is
higher. In type I events, the Einstein arc is affected by the
proximity of the planet, and it will be shown that the sensitivity
of these events to low-mass planets is almost independent of
magnification. Using simulation programs of the MOA@UoA
project~\cite{MOAcode}, producing theoritical light curves, it
will be demonstrated that the planetary perturbation is reasonably
detectable in type I events, even at low magnification. It follows
that monitoring further events of this type could increase the
detection rate of low-mass planets.

%%%%%%%%%%%%%%%%%%%%%%%%%%%%%%%%%%%%%%%%%%%%%%%%%%%%%%%%%%%%%%%%%%%%%%%%%
%%
%%   use this format to include an .eps figure into your paper
%%
%%%%%%%%%%%%%%%%%%%%%%%%%%%%%%%%%%%%%%%%%%%%%%%%%%%%%%%%%%%%%%%%%%%%%%%%%%%

\section{MOA-2006-BLG-130}

\begin{figure}[htb]
\begin{center}
\epsfig{file=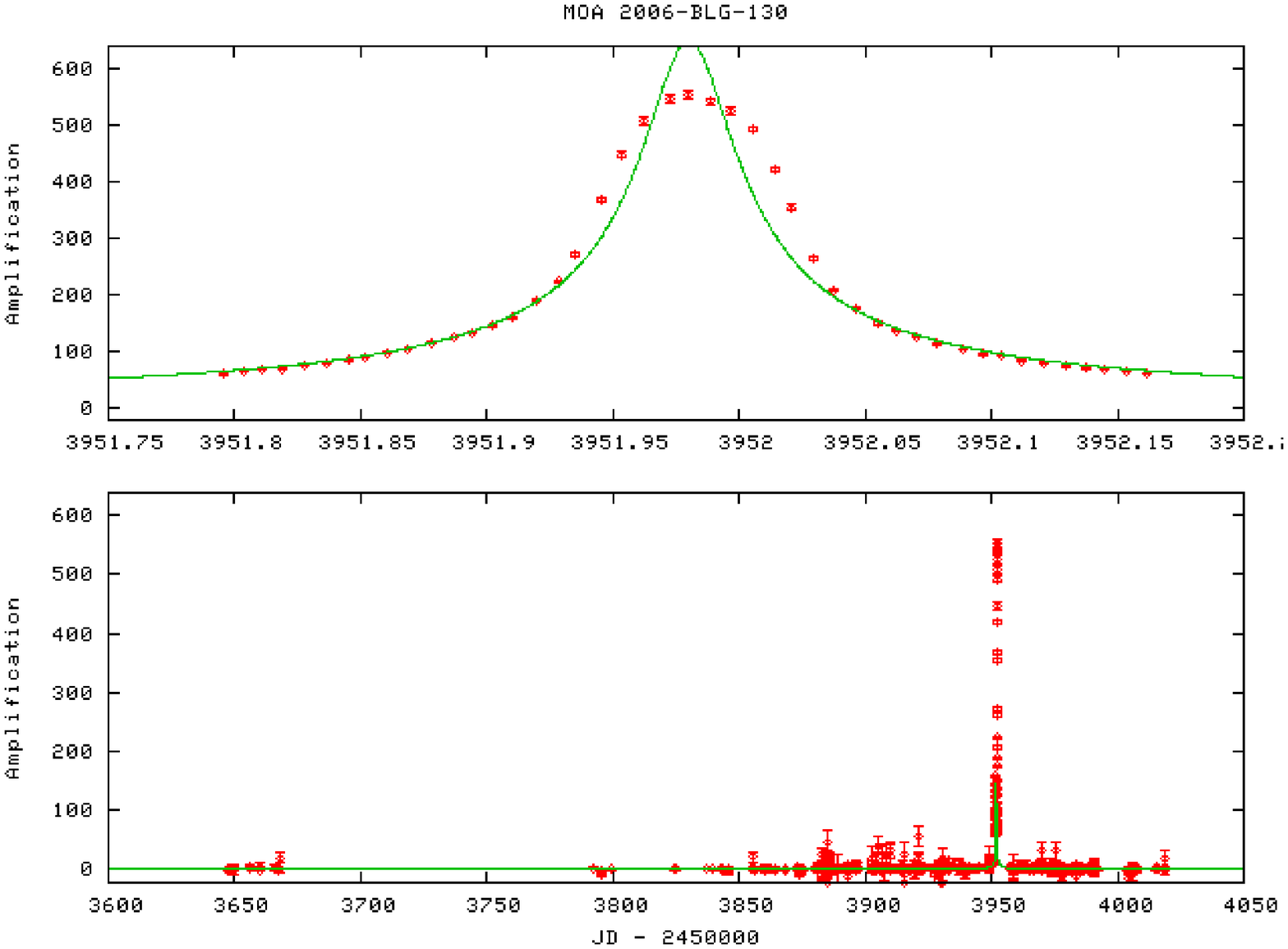,height=4.0in} \caption{Data
light curve for MOA-2006-BLG-130, from the MOA microlensing
alerts~\cite{MOAwebsite}.} \label{fig:MOA-2006-BLG-130}
\end{center}
\end{figure}

This event, for which the data light curve is given in
Figure~\ref{fig:MOA-2006-BLG-130}, was tested with the MOA@UoA
simulation code. We used a $\chi^2$ marginalisation method, to
test 27 binary lens, operation being repeated for each for 17
values of the source track angle. Results and analysis method will
be presented. However, the coverage of this event was really
sparse, so other phenomena were considered as possible cause of
the perturbation.

%%%%%%%%%%%%%%%%%%%%%%%%%%%%%%%%%%%%%%%%%%%%%%%%%%%%%%%%%%%%%%%%%%%%%%%%%
%%
%%   use this format to include an .eps figure into your paper
%%
%%%%%%%%%%%%%%%%%%%%%%%%%%%%%%%%%%%%%%%%%%%%%%%%%%%%%%%%%%%%%%%%%%%%%%%%%%%

%%

\Chapter{The Keplerian Orbit of
OGLE-2008-BLG-513/MOA-2008-BLG-401}
            {Orbit of OGLE-2008-BLG-513/MOA-2008-BLG-401}{Jennifer Yee}
\bigskip\bigskip

\addcontentsline{toc}{chapter}{{\it Jennifer Yee}}
\label{polingStart}

\begin{raggedright}

{\it Jennifer Yee\index{author}{Yee, Jennifer}\\
Ohio State University \\
USA\\}

\bigskip\bigskip

\end{raggedright}

The light curve of OGLE-2008-BLG-513/MOA-2008-BLG-401 is
characterized by two strong peaks due to companion to the lens
with a mass ratio of $q=0.027$. The timescale of this event is
fairly long ($t_E$ = 32 days) and the span of the 2-body
perturbation is $>12$ days because the caustic is resonant (the
separation between the lens star and its companion in Einstein
radii is $\sim 1$). Because of the timescales involved we were
able to measure both the parallax effect and the orbital motion of
the lens. Using information from the parallax and orbital motion
effects, we find that the companion is a 7.8 $M_{\mathrm{Jup}}$
giant planet orbiting an 0.3 $M_{\odot}$ host with a semi-major
axis of 2.2 AU.

We initially fit the data with a static, 2-body lens model
including parallax effects but were unable to find a model that
satisfactorily data. We are able to fit the data successfully by
allowing orbital motion in the 2-body lens. We use two different
models: the first uses only the lens velocity projected onto the
plane of the sky and the second uses a full Keplerian description
for the lens motion. We find that the full Keplerian model is
slightly preferred ($\chi^2$ improvement of ~10). This preference
for full Keplerian motion indicates that we are measuring the
curvature of the orbit during the event and allows us to place
constraints on most of the orbital parameters.

The source star in this event is unusually blue compared to the
other stars in the field: it is 0.95 mag bluer than the red clump.
Given its blue color, one possibility for the source is that it is
an A dwarf in the bulge or behind the the bulge in the far disk.
An alternative hypothesis is that the source is well in the
foreground, in front of most of the dust. In this case, it would
be a low mass dwarf that appears blue because it experiences much
less reddening than the other stars in the field. Various
inconclusive arguments can be made supporting each hypothesis,
drawing on the observed relative proper motion, astrometry, source
limb-darkening, and arguments about the observed color and
magnitude of the source. However, in our analysis of the light
curve, we find that in order for the orbit of the lens planet to
be bound, the source must be a nearby M dwarf. If the source star
is a nearby, metal-poor thick disk star, this explanation accounts
for almost all of the information we have on the source. In order
to clarify the nature of the source, we have applied for and been
granted time on the VLT to get a spectrum of the source. Given our
orbital motion observations, we predict that the source will be a
low-mass star.

Regardless of the nature of the source, the mass ratio between the
companion and the lens star is robust. A mass ratio of 0.027 is
very rare among binary/brown dwarf/planetary systems regardless of
the specific nature of the companion. If the spectrum confirms our
interpretation of the event, this will be yet another example of a
massive planet orbiting an M dwarf. Such planets are extremely
difficult to form using present theories of planet formation. The
core accretion theory predicts that such planets should be
extremely rare because M dwarfs have less massive disks and short
orbital timescales at the snow line where giant planets should
form. The gravitational instability theory is able to form giant
planets around M dwarfs but only at much larger distances than we
observe in this system. This planet, and others like it,
increasingly form a challenge to the accepted theories of planet
formation.

%%
%%    Changes to this file from the original article.tex format:
%%
%%       1.  The article.tex file has been stripped of its header material.
%%       2.  The definition of \Discussion near the end of the article
%%               has also been stripped out
%%       3.  The line \addtocontents and the following line have been
%%                     uncommented
%%       4.  The labels {Heavy Quark Decay}{Roling A. Poling} have been added
%%                    below the title; these are for the running heads.
%%       5.  Index items markers \index{name} have been changed to
%%               \index{subject}{name}, since this volume has
%%               both a subject and an author index.
%%       6.  EPS figures have been placed in directory poling, and references
%%               to these figures have been modified to refer to this
%%               directory.

\Chapter{Planetary microlensing event MOA-2010-BLG-328}
            {Planetary microlensing event MOA-2010-BLG-328}{Kei Furusawa}
\bigskip\bigskip

\addcontentsline{toc}{chapter}{{\it Kei Furusawa}}
\label{polingStart}

\begin{raggedright}
{\it Kei Furusawa\index{author}{Furusawa, Kei}\\
Nagoya University\\
Japan}
\bigskip\bigskip
\end{raggedright}

\section{Observation}
Microlensing Observations in Astrophysics (MOA) monitors $\sim$ 40
degree${^2}$ of the Galactic bulge to find Microlensing events by
1.8m MOA-II telescope in New Zealand. Our real-time anomaly alert
system checks new data points of each of microlensing events after
five minutes of observation to find deviations due to planets. MOA
has discovered 607 microlensing events and some anomaly events by
the real-time anomaly alert system toward the Galactic bulge in
2010.

\bigskip

The microlensing event MOA-2010-BLG-328 was detected and alerted
by MOA on 16 June 2010. Around UT 12:30 27 July, we found the
deviations from single lens light curve of the event by anomaly
alert system and issued the anomaly alert. Two days after the
anomaly alert was issued, the deviations could be seen more
clearly and David Bennett sent his preliminary planetary model.
Following his e-mail, follow-up groups, $\mu$FUN: CTIO (Chile),
Farm Cove Observatory (New Zealand),Palomar Observatory (CA, USA),
PLANET: Canopus (Australia), RoboNet: Faulkes Telescope North
(Hawaii, USA), Faulkes Telescope South (Australia), Liverpool
Telescope (Spain), MiNDSTEp: SAAO (South Africa), Danish (Chile)
conducted intensive observation. Fortunately, the event occurred
in OGLE-IV monitored fields so we got also OGLE data.

\section{Analysis}

The light curve is fitted by Markov Chain Monte Carlo (MCMC).
Initial parameters for best fit model search are used over the
range of the planet-star mass ratio ($q$) $-5 \leq$ log$q \leq 0$
and planet-star separation ($s$) $-1 \leq$ log$s \leq 1$, the
total number of initial parameters is 858.

We estimate the dereddened source color and magnitude from fitting
and color magnitude diagram (CMD). Then the estimated source star
is G type star and we adopt Limb Darkening parameter of G type
star.

Currently, as the result of the detail analysis of the light
curve, the mass ratio $q$ and the separation $s$ are yielded $9.7
\times 10^{-4}$, $1.3$ respectively. It remains possible that the
light curve also shows the microlensing parallax which is caused
by the orbital motion of the Earth around the Sun during the
event. The parallax effect enables us to estimate the physical
parameters of the planet.

\section{Conclusion}
We analyzed planetary microlensing event MOA-2010-BLG-328. As the
result of this analysis, the lens has a companion with planet mass
ratio. This result is preliminary because the parallax signal is
very sensitive to systematic errors of data. The data are
gradually improving, so we continue evaluating parallax in detail

\Chapter{Binary microlensing event OGLE-2009-BLG-020 gives orbit
predictions verifiable by follow-up observations}
            {Binary microlensing event OGLE-2009-BLG-020}{Jan Skowron}
\bigskip\bigskip

\addcontentsline{toc}{chapter}{{\it Jan Skowron}}
\label{polingStart}

\begin{raggedright}

{\it Jan Skowron\index{author}{Skowron, Jan}\\
Department of Astronomy, Ohio State University\\
USA\\}

\bigskip\bigskip

\end{raggedright}

\section{Introduction}
We present the first example of binary microlensing for which the
parameter measurements can be verified (or contradicted) by future
Doppler observations. This test is made possible by a confluence
of two relatively unusual circumstances.  First, the binary lens
is bright enough ($I=15.6$) to permit Doppler measurements.
Second, we measure not only the usual 7 binary-lens parameters,
but also the ``microlens parallax'' (which yields the binary mass)
and two components of the instantaneous orbital velocity. Thus we
measure, effectively, 6 'Kepler+1' parameters (two instantaneous
positions, two instantaneous velocities, the binary total mass,
and the mass ratio). Since Doppler observations of the brighter
binary component determine 5 Kepler parameters (period, velocity
amplitude, eccentricity, phase, and position of periastron), while
the same spectroscopy yields the mass of the primary, the combined
Doppler + microlensing observations would be overconstrained by $6
+ (5 + 1) - (7 + 1) = 4$ degrees of freedom. This makes possible
an extremely strong test of the microlensing solution.

We also introduce the uniform microlensing notation for single and
binary lenses, define conventions, summarize all known
microlensing degeneracies and extend set of parameters to describe
full Keplerian motion of the binary lenses (see Appendix A in the
main paper Skowron et al. 2011, in prep.).

\section{Observational data}

On 15 February 2009, heliocentric Julian Date (HJD) $\sim$
2454878, the Optical Gravitational Lensing Experiment (OGLE) team
announced ongoing microlensing event OGLE-2009-BLG-020 detected by
Early Warning System (EWS) and observed on the 1.3m Warsaw
Telescope in Las Campanas Observatory in Chile. The event was also
monitored by the Microlensing Observations in Astrophysics (MOA)
1.8m telescope on Mt. John in New Zealand. On HJD' $\sim$ 4915
($HJD' = HJD - 2450000$) it could be seen that light curve was
deviating from the standard Paczy{\'n}ski model, and follow-up
observations by other telescopes began. Intense monitoring of the
event was performed until HJD' $\sim$ 4920. During this period
source crossed caustic twice and both crossings where observed.

\begin{figure}[htb]
\begin{center}
\epsfig{file=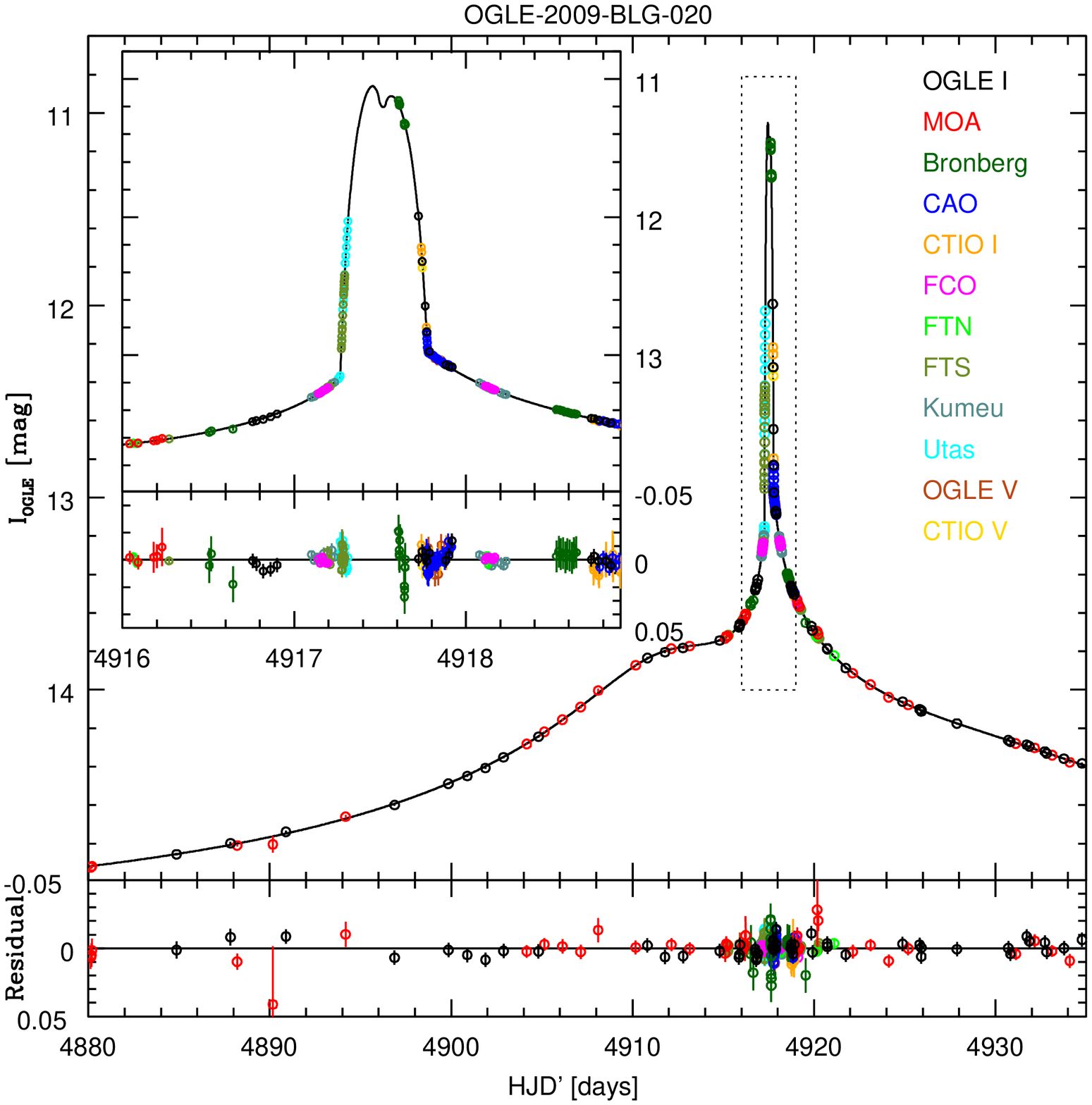,width=4in} \caption{Light
curve and best-fit model of OGLE-2009-BLG-020.} \label{f:lc}
\end{center}
\end{figure}

In this work, beside the data obtained by survey telescopes, we
use follow-up data from 8 additional observatories: the 2.0m
Faulkes South (FTN), the 2.0m Faulkes North (FTN), the 36cm
telescope at Kumeu Observatory, the 36cm telescope at Bronberg
Observatory, the 1m telescope on Mt. Canopus (Utas), the 36cm
telescope in Farm Cove Observatory (FCO), the SMARTS 1.3m Cerro
Tololo Inter-American Observatory (CTIO), and the 40cm telescope
in Campo Catino Austral Observatory (CAO). Light curve of the
event is shown on Figure~\ref{f:lc}.

%%%%%%%%%%%%%%%%%%%%%%%%%%%%%%%%%%%%%%%%%%%%%%%%%%%%%%%%%%%%%%%%%%%%%%%%%
%%
%%   use this format to include an .eps figure into your paper
%%
%%%%%%%%%%%%%%%%%%%%%%%%%%%%%%%%%%%%%%%%%%%%%%%%%%%%%%%%%%%%%%%%%%%%%%%%%%%

The whole light curve consists of 9 years of data with 121 days
during the course of the visible magnification, of which 5 days
constitute intense follow-up observations. The OGLE telescope
performed observations in $V$ and $I$ bands, and CTIO telescope in
$V$, $I$ and $H$ bands, which permit measurements of the color of
the magnified source star. Together we have 5333 data points in
the light curve with 2247 during the course of the magnification
event.

\section{Microlensing fit with full Keplerian orbit}

To find a microlensing model we use the method developed by S.
Dong and described in Dong et al. (2006) and its modification in
Dong et al. (2009, Section 3). Although we introduce a few changes
in parametrization of the model. In order to allow comparison with
the radial velocity (RV) measurements it is profitable to use the
full Keplerian orbit parametrization. In addition to being more
accurate, the additional advantage of this approach is to avoid
all unbound orbital solutions (with eccentricity $\ge 1$) and to
enable introduction of priors on the values of orbital parameters
directly into MCMC calculations, if one decides to adopt that
approach.

\begin{figure}[htb]
\begin{center}
\epsfig{file=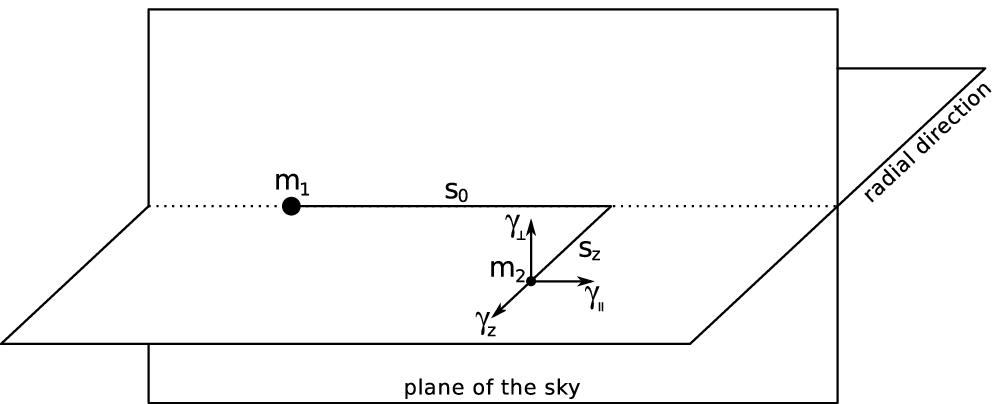,width=4in} \caption{Definition of the
phase-space parameters $(s_0, 0, s_z, \gamma_\parallel,
\gamma_\perp, \gamma_z)$ at $t_{0,kep}$, describing motion of the
secondary binary lens component ($m_2$) relative to the primary
($m_1$). Vertical plane is the plane of the sky.} \label{f:gammas}
\end{center}
\end{figure}

We describe the orbit of the secondary component relative to the
primary by giving 3 cartesian positions and 3 velocities at one
arbitrarily chosen time ($t_{0,kep}$). See Figure~\ref{f:gammas}.
Our extended microlensing model (with parallax parameters) carries
information about the mass ratio, the total mass of the lens, and
the physical scale in the lens plane, so together with the six
instantaneous phase-space coordinates it comprises a complete set
of system parameters (except systemic radial velocity). This, for
example, allows us to calculate the relative RV at any given time.
We can use this for comparison with the RV curve from follow-up
observations.

%%%%%%%%%%%%%%%%%%%%%%%%%%%%%%%%%%%%%%%%%%%%%%%%%%%%%%%%%%%%%%%%%%%%%%%%%
%%
%%   use this format to include an .eps figure into your paper
%%
%%%%%%%%%%%%%%%%%%%%%%%%%%%%%%%%%%%%%%%%%%%%%%%%%%%%%%%%%%%%%%%%%%%%%%%%%%%

\begin{figure}[htb]
\begin{center}
\epsfig{file=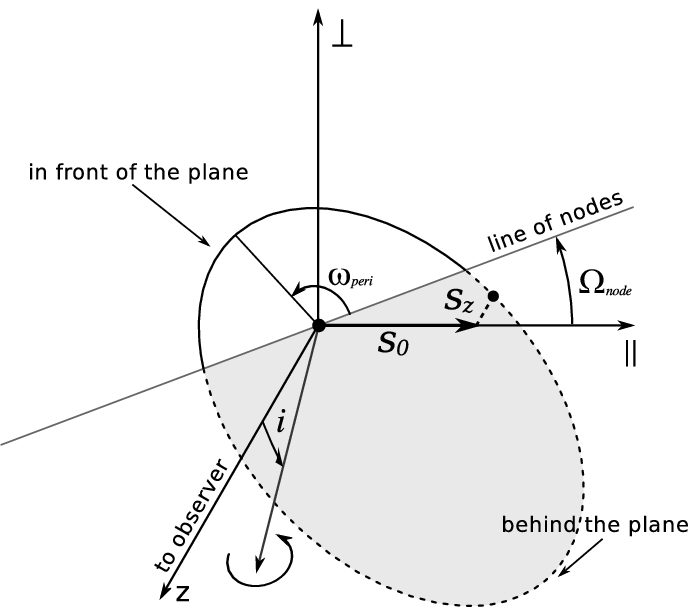,width=3in} \caption{Figure shows an
example of the relative binary orbit which is rotated by three
Euler angles. The binary components at the time $t_{0,kep}$ are
marked with 2 dots. The axis $z$ points toward the observer; axes
marked with symbols $\parallel$ and $\perp$ define the plane that
is parallel to the plane of the sky and crosses the primary
component of the binary. The portions of the line that lie behind
the plane are dashed. The base coordinate system is related to the
microlensing event such that at the time $t_{0,kep}$ the first
axis coincides with the binary axis projected onto plane of the
sky.} \label{f:euler}
\end{center}
\end{figure}

It is possible to calculate all properties in physical units as
well as standard Kepler parameters of the orbit -- i.e.,
eccentricity ($e$), time of periastron ($t_{peri}$), semi-major
axis ($a$) and 3 Euler angles: longitude of ascending node
($\Omega_{node}$), inclination ($i$) and argument of periapsis
($\omega_{peri}$).
%Having those we can find the exact position of the lens components at
%any given time, which in turn specifies the lens geometry $(s,q)$,
%and the projected position of the source relative to this
%geometry. We use this in magnification calculations.
(See Figure~\ref{f:euler} for illustration of the orbital
parameters and they relation to the phase-space parameters
specified at the fixed time $t_{0,kep}$.)

%%%%%%%%%%%%%%%%%%%%%%%%%%%%%%%%%%%%%%%%%%%%%%%%%%%%%%%%%%%%%%%%%%%%%%%%%
%%
%%   use this format to include an .eps figure into your paper
%%
%%%%%%%%%%%%%%%%%%%%%%%%%%%%%%%%%%%%%%%%%%%%%%%%%%%%%%%%%%%%%%%%%%%%%%%%%%%

Likelihoods obtained from MCMC procedure suffer from degeneracy
between one of the parallax parameters ($\pi_{E,\perp}$ --
perpendicular to the Earth acceleration) and angular velocity of
the lens in the plane of the sky. We break this degeneracy by
choosing subset of solutions which yields observed brightness of
the lens consistent with the theoretical main-sequence isochrone
(we assume that all blended light is coming from the lens, and,
with mass ratio $q=0.27$, we expect that the secondary component
of the lens does not contribute to the light). This leads to
estimations of the lens distance and the mass of its primary
component:
\begin{equation}
D_l = 1.1 \pm 0.1 \, {\rm kpc} \quad M_1 = 0.84 \pm 0.03 \,
M_{\odot} \quad
\end{equation}
The observed magnitudes of the blend and the source are:
\begin{equation}
\left((V-I), I, (I-H) \right)_b = \left(1.316 \pm 0.01,\; 15.680
\pm 0.01,\;  1.409 \pm 0.05 \right)
\end{equation}
\begin{equation}
\left( (V-I), I \right)_{s} = \left(  1.929 \pm 0.002, \; 16.43
\pm 0.04 \right)
\end{equation}

\section{Test with radial velocity}

Every link in the MCMC chain consist a complete set of parameters
of the binary lens. It yields not only the mass, distance and
separation in physical units, but also all Keplerian parameters of
the orbit. Thus we can calculate the RV curve for any period of
the binary.

Assuming the observations of the RV curve of the primary are
taken, we can assign to every link the likelihood that it is
consistent with those data. We derive the radial velocity for any
time the data were taken and, allowing for systemic velocity of
the center of mass of the binary to vary, we calculate the
likelihoods for every point. In this way we construct new set of
links weighted by, both, the microlensing light curve and radial
velocity curve. This new set of solutions yields new, most
probable, values of all lens parameters, which may or may not
coincide with the values derived from the microlensing only
solution. This would be a test of microlensing solution.

If new values of parameters will lie outside 3-$\sigma$ limits of
microlensing solution we can say that our method failed. By
contrast, if solutions, which are consistent with the RV curve,
lie near the best-fit values we obtained from microlensing we can
not only believe our solution, but we can also read all parameters
of the binary, which are not given by one or the other method
alone. For example, radial velocity curve will give us period and
systemic radial velocity, which we cannot read from the
microlensing light curve alone. However, microlensing will yield
the inclination, orientation on the sky and the 2-d velocity of
the binary projected on the plane of the sky.

\subsection{Useful information for RV observations}
The selective power of the RV curve holds when the observation are
taken through at least one period of the binary. We propose a 2
year span of observations since the period we derive is between
200 and 700 days (2-$\sigma$ limit).

The binary coordinates are ($18^h 04^m 20^s.99$, $-29^\circ 31'
08''.6$, $J2000.0$). The finding chart can be found on the OGLE
EWS
webpage\footnote{http://ogle.astrouw.edu.pl/ogle3/ews/2009/blg-020.html}.

The binary is blended with the ``microlensing source'' which, as
its position on the CMD suggests, is a Galactic bulge giant. The
binary has a mass ratio of $0.272$, so assuming both components
are the main-sequence stars, majority of the light is coming from
the primary. The binary is $1.3$ magnitude brighter in $V$ and
$0.8$ mag brighter in $I$ than the giant, however they are of
similar brightness in $H$.

Since the binary is located in the Galactic disk, we anticipate it
will be clearly separated in the velocity space from the blended
Bulge giant. The radial velocity amplitude of the primary is
expected to be of order of a few ${\rm km}\,{\rm s}^{-1}$.

\section{Conclusions}

The binary star that manifested itself in the microlensing event
OGLE-2009-BLG-020 is the first case of a lens that is close enough
and bright enough to allow ground-based spectroscopic follow-up
observations. This makes it a unique tool to test the microlensing
solution.

We derive lens parameters using the same method by which majority
of planetary candidates discovered by microlensing are analyzed.
We detect a signal from the orbital motion of the lens in the
microlensing light curve. This signal, as well as our measurements
of the orbital parameters of the binary lens, can be confirmed or
contradicted by the future observations.

Combining the microlensing solution with the radial velocity curve
will yield a complete set of system parameters including 3-d
Galactic velocity of the binary and all Kelperian orbit elements.

This work undertakes an effort to establish a uniform microlensing
notation, extending work of Gould (2000) by including full set of
orbital elements of the binary lens (see Appendix A in the paper).
We also summarize all known microlensing degeneracies.

The method of deriving orbital elements from the 6 phase-space
coordinates, used to parametrize microlensing event, is described
in Appendix B in the paper.

%%%%%%%%%%%%  Section 5   %%%%%%%%%%%%%%%%%%%%%%%%%%%%
\emptyheads

\begin{center}
\begin{large}
{\Huge \sffamily Theoretical investigations}
\end{large}
\end{center}
\bigskip

\begin{tabbing}
 Brief Reports from the \=B factories xxxxx \= xxxxxxxxx\= \kill
\\
\\
Microlensing and planet populations - What do we know, and how could we learn more\\
\>\> {\it Martin Dominik}\\
\\
The frequency of extrasolar planet detections with microlensing
simulations \\
\>\> {\it Rieul Gendron \& Shude Mao}\\
\\
A semi-analytical model for gravitational microlensing events \\
\>\> {\it Denis Sullivan, Paul Chote, Michael Miller}\\
\\
GPU-assisted contouring for modeling binary microlensing events \\
\> {\it Markus Hundertmark, Frederic V. Hessman, Stefan Dreizler }\\
\\
Red noise effect in space-based microlensing observations \\
\hspace{1.0cm}{\it Achille Nucita, Daniele Vetrugno, Francesco De
Paolis, Gabriele Ingrosso,}\\
\hspace{1.0cm}{\it Berlinda M. T. Maiolo, Stefania Carpano}\\
\\
Light curve errors introduced by limb-darkening models \\
\>\> {\it David Heyrovsky}\\
\\
Isolated, stellar-mass black holes through microlensing \\
\hspace{1.0cm}{\it Kailash Sahu, Howard E. Bond, Jay Anderson, Martin Dominik,}\\
\hspace{1.0cm}{\it Andrzej Udalski, Philip Yock}\\
\\
The observability of isolated compact remnants with microlensing \\
\>\> {\it Nicola Sartore \& Aldo Treves}\\
\\
Gravitational microlensing by the Ellis wormhole \\
\>\> {\it Fumio Abe}\\
\\
The deflection of light ray in strong field: a material medium approach \\
\>\> {\it Asoke Kumar Sen}\\
%\\
%How to stop a runaway (Monte Carlo Markov) chain \\
%\>\> {\it Keith Horne}\\
\\
Rapidly rotating lenses - repeating orbital motion features in close binary microlensing \\
\>\> {\it Matthew Penny, Eamonn Kerins, Shude Mao}\\
\\
\end{tabbing}

\fancyheads

\Chapter{Microlensing and planet populations - What do we know,
and how could we learn more?}
            {Microlensing and planet populations}{Martin Dominik}
\bigskip\bigskip

\addcontentsline{toc}{chapter}{{\it Martin Dominik}}
\label{polingStart}

\begin{raggedright}

{\it Martin Dominik\index{author}{Dominik, Martin}\\
SUPA, University of St Andrews, School of Physics \& Astronomy \\
United Kingdom\\}

\bigskip\bigskip

\end{raggedright}
\section{Gravitational microlensing campaigns}
Following a concrete suggestion brought up by Bohdan Paczy\'{n}ski
\cite{Pac86}, the first dedicated gravitational microlensing
surveys \cite{EROS,MACHO} were designed to measure a potential
clumpy dark matter content in the galactic halo by means of
observations towards the Magellanic Clouds. Observations towards
the Galactic bulge have then been proposed for robustly testing
whether the experimental set-ups are able to detect signals
\cite{Pac91,KP94}, given that a possible negative result towards
the Magellanic Clouds would remain ambiguous. Wider-reaching
astrophysical applications of gravitational microlensing simply
became add-ons to existing opportunities, and have subsequently
shaped the evolutionary process of experiments. Exploiting the
discovery potential of planets orbiting stars other than the Sun
\cite{MP91} by means of follow-up observations on ongoing
microlensing surveys \cite{GL92} became possible with real-time
processing and public dissemination of data
\cite{OGLE:EWS,MACHO:alert}. The explicit goal to detect planets
by microlensing was in particular engraved in the acronym of the
Probing Lensing Anomalies NETwork (PLANET) \cite{PLANET1}.

\section{Efficiency and planet population statistics}

Planetary abundance statistics cannot be extracted simply from
counting claimed detections. We need to understand what the
characteristics of the monitoring campaigns are, which might have
changed over time. In fact, the observed sample is a probabilistic
realisation of the product of the underlying planet population and
the detection efficiency of the experiment(s). This however means
that the considered sample and the detection efficiency need to
refer to the same detection criterion and monitoring strategy. Not
only does the detection efficiency need to be evaluated in a
statistical meaningful way, but also do the "detections" need to
be counted accordingly \cite{GRGreview}.

Moreover, not suprisingly, with different approaches being
followed by the various surveys and follow-up monitoring
campaigns, their detection efficiencies are substantially
different functions of the planet mass $m_\mathrm{p}$ and orbital
separation $r_\mathrm{p}$. For the most massive planets, there is
not much gain with less than daily sampling, so that survey
experiments with their larger number of observed events have the
largest sensitivity \cite{OGLE:Tsapras, OGLE:Snodgrass}, whereas
less massive planets with their shorter signals become
undetectable by falling into the data gaps. With regular
quasi-continuous high-precision round-the-clock monitoring, as
carried out by PLANET \cite{PLANET:EGS}, RoboNet
\cite{RoboNet-1.0,RoboNet-II}, or MiNDSTEp \cite{MiNDSTEp0}, the
planet detection efficiency drops roughly as
$\sqrt{m_\mathrm{p}}$, until the finite size of the source stars
at a few Earth masses causes a faster drop-off and subsequently
undetectability \cite{fiveshort,fivelong}. If, like $\mu$FUN, one
focuses on highly-magnified peaks, probing the central caustic
close to the lens star, the detection efficiency is close to
100\,\% for a wide range of planetary masses, with just the range
of orbital separations shrinking towards lower masses, and it then
falls to zero rather suddenly \cite{Gould:abundance}.

The history of claimed microlensing planet detections seems to
follow a pattern that matches the characteristics of the different
campaigns: A first super-Jupiter from the OGLE and MOA surveys
\cite{OB03235}, the 5-Earth-mass planet OGLE-2005-BLG-390Lb from
PLANET observations \cite{OB05390}, and planets from
high-magnification peaks monitored by $\mu$FUN
\cite{OB05169,MB08310}. Moreover, a larger number of detections in
the region between Neptune and Saturn \cite{OB07368,KB09319} arose
in the more recent years due to increased survey cadence.

However, rough planet abundance estimates give rather puzzling
results. A well-defined sample of events with peak magnifications
$A_0 \geq 200$, densely covered mostly by $\mu$FUN data
\cite{Gould:abundance} indicates an abundance for massive gas
giants that is about 10 times larger than what one would estimate
from PLANET observations and the corresponding detections
\cite{planetMF}. Despite the fact that both estimates are
statistically compatible within their large uncertainties, these
findings are somewhat surprising given that the PLANET
observations are {\em more} powerful in this mass range, but have
provided less detections. In fact, the 4-year $\mu$FUN sample
contains just 13 events, which means that the adopted strategy is
not powerful enough for providing a planetary mass function in
foreseeable time; increased efforts of monitoring a larger number
of events at more moderate magnifications, as follow-up efforts or
high-cadence surveys, are required \cite{planetMF}. The detection
of a planetary signal in event MOA-2009-BLG-266, resembling the
path that led to OGLE-2005-BLG-390Lb, clearly demonstrates the
power of the regular follow-up approach (which in fact is just
copied by the MOA high-cadence monitoring).

With the current small number of planetary detections,
microlensing observations are still in their very infancy on the
path to determining planet abundances, and any kind of planetary
mass function is a long way ahead. A particular complication
arises from the fact that the planet abundances strongly depend on
the properties of the parent star, and microlensing events come
with a probabilistic mixture of host stars. Should more massive
stars come with more or less planets of a certain kind, what
planetary mass function are we talking about? Distinguishing
between masses of planet host stars requires a much larger sample
for drawing meaningful conclusions. Stellar metallicity is a
further relevant parameter that is difficult to control, and the
fact that microlensing probes both Galactic disk and bulge stars
is a challenge as well as an opportunity.

While planetary microlensing naturally depends on the
dimensionless separation parameter $d$ and the planet-to-star mass
ratio $q$, planet formation does not obey such a scaling law,
regardless of whether planet formation theories give us good
guidance or not. Moreover, the adoption of any specific functional
form, just for its simplicity, that is not clearly supported by
the data, may not be adequate to probe certain features. In
particular, smooth monotonic functions are ill-suited to describe
planetary deserts or cut-offs.

%[OGLE-2005-BLG-390: PLANET to MiNDSTEp, SIGNALMEN\

\section{Moving forward}
Maximizing the number of detections without adopting well-defined
monitoring procedures might lead to further spectacular findings,
but will be useless for studying planet populations. High-cadence
surveys have already proven their extended sensitivity to planets
with smaller masses. In order to obtain meaningful results on
planet population statistics, deviations from a deterministic
survey programme must be made in a deterministic way only as well.
Otherwise, claimed detections cannot be used to gather statistical
information. It has also been demonstrated that the automated
detection of anomalies \cite{SIGNALMEN} increases the detection
efficiency, and in particular allows for the realistic detection
of planets of Earth mass and below, while keeping the
deterministic process chain intact. The real-time identification
of potential deviations from ordinary light curves is essential
for the current ability to push further down in mass, but if we
want to determine the abundance of such planets as well, this
identification must not be left to human judgement. If planets of
Earth mass and below are common, recent survey data give rise to
speculating that opportunities have been missed for claiming such
a detection. If we want to make an impact {\em now}, the real-time
provision of photometric data within minutes (not within hours) is
likely to make the crucial difference. Beyond that, and in the
longer term, precision photometry on fainter stars would allow for
far better statistics on low-mass planets.

\Chapter{The frequency of extrasolar planet detections with
microlensing simulations}
            {The frequency of extrasolar planet detections with microlensing simulations}{Rieul Gendron}
\bigskip\bigskip

\addcontentsline{toc}{chapter}{{\it Rieul Gendron}}
\label{polingStart}

\begin{raggedright}

{\it Rieul Gendron\index{author}{Gendron, Rieul}\\
Jodrell Bank Centre for Astrophysics, University of Manchester \\
United Kingdom\\}

\bigskip
{\it Shude Mao\index{author}{Mao, Shude}\\
Jodrell Bank Centre for Astrophysics, University of Manchester,
UK\\
National Astronomical Observatories, Chinese Academy of Sciences \\
China\\}

\bigskip\bigskip

\end{raggedright}

It was first proposed by~\cite{shude_paper} that gravitational
microlensing could be used to detect extrasolar planets. To date,
over 500 have been discovered, with microlensing planets, although
relatively few in number, probing a region of the mass versus
semi-major axis plane that is currently out of reach of the other
methods, with the sensitivity peaking just beyond $\sim 1$ AU.
Microlensing planets are therefore very useful for placing
constraints on planet formation models.

We simulate light curves for 1000 extrasolar planet systems around
host stars of mass $0.25 M_{\odot}$ drawn from the Ida \& Lin core
accretion models of planet formation~\cite{idalin}. The simulated
data is first fitted with a single-lens model. If the fit is poor,
combinations of the planets most likely to be causing
perturbations are placed around the star and new light curves are
generated to attempt to reproduce the original light curve.

We found that the majority of the light curves were well-fitted by
single-lens models. However, 26 were not and hence were determined
to contain pertubations due to at least one planet. Of the 26
interesting light curves, it was found that 16 could be explained
by single planets. A greater number of planets were necessary to
explain the remaining 10 cases. These results place an upper limit
on the number of planets that would be deduced by a fitting
procedure since it might be possible to produce the original light
curve with a fewer number of planets.

\Chapter{A Semi-Analytical Model for Gravitational Microlensing
Events}
            {A semi-analytical model for gravitational ML events}{Sullivan et al.}
\bigskip\bigskip

\addcontentsline{toc}{chapter}{{\it Sullivan et al.}}
\label{polingStart}

\begin{raggedright}

{\it Denis Sullivan\index{author}{Sullivan, Denis} \& Paul Chote\index{author}{Chote, Paul},
Michael Miller\index{author}{Miller, Michael}\\
School of Chemical \& Physical Sciences, Victoria University of Wellington \\
New Zealand\\}

\bigskip

\end{raggedright}

\section{Introduction}

This paper describes the latest version of the software we have
developed for modelling gravitational microlensing events that
involve multiple lensing masses.  Of particular interest is the
modelling of planetary microlensing events requiring two or more
lensing masses.

The basic thin lens equation relating a point source image
position, $\mathbf s$, to the positions, $\mathbf r$, of the
multiple images created by the $N$ lensing masses is simple
enough,
\begin{displaymath}
 {\mathbf s} = {\mathbf r}
             - \sum_{j=1}^{\mathrm N}\epsilon_j\frac{\mathbf{r} - \mathbf{r}_j}
                                           {|\mathbf{r} - \mathbf{r}_j|^2}
\hspace{5mm} \mbox{, where $\mathbf{r}_j$ and $\epsilon_j$ are the
lens positions and
        mass fractions,}
\end{displaymath}
The positions in this expression are in units of the Einstein ring
radius corresponding to the total lens mass.  Except for the
single lens case, inverting the equation in order to determine
each image position directly is not a trivial exercise.  In fact,
the inversion procedure is relatively complicated for even the
two-mass lens and it rapidly becomes unmanageable when the number
of lensing massses increases much beyond two.

\section{Inverse Ray-tracing}

The above inversion difficulty has led to wide-spread use of the
``brute force'' approach called inverse ray tracing: ``all''
potential positions in the image plane are checked in order to
determine which ones are consistent with a given source position.
Each determination is straight forward since it is clear that a
simple substitution of an image position in the above lens
equation yields a unique source position origin.

Due to the potentially large number of calculations required,
access to significant computing power makes this technique a
practical proposition, but it can also be improved by use of
efficient image plane search techniques or better still recording
all image plane, source plane correspondences in order to create a
so-called \emph{magnification map}.  This map can be used to
generate a microlensing light curve, since the source
magnification at a given position can be directly estimated from
the area density of points (inverse rays) in the source plane
relative to the density in the image plane.  A particular light
curve can be rapidly computed by calculating the changing ``ray''
density ratio for any chosen source track.  This method includes
finite source size effects directly along with source limb
darkening if required, by weighting ray densities in the source
disk.  Observer/source parallax effects can also be included via
appropriately curved source tracks.  However, internal lens
configuration changes will alter the magnification map and
accuracy depends directly on the density of rays covering the
image plane, which of course has computational implications.

Modelling code using this methodology was developed as part of a
PhD programme at VUW~\cite{Korpela07} and it has been used to
model a range of planetary microlensing events.  Recently, we
decided to pursue the approach of directly inverting the lensing
equation.  One particular motivation for this work was to enable
satisfactory modelling of lens motion effects, but the overall
approach is also more efficient, at least for systems with no more
than four lenses.

\section{The Polynomial Method}

The algebra is greatly simplified by representing the two
dimensional plane positions using complex coordinates as first
introduced by Witt~\cite{Witt90}. The lens equation relating
source position ($\mathbf{s} \rightarrow \omega = u+iv$) to image
positions ($\mathbf{r} \rightarrow z = x+iy$) for $N$ lenses
($\mathbf{r}_{j} \rightarrow z_{j}=x_{j}+iy_{j}$) then becomes:
\begin{displaymath}
\omega = z - \sum_{j=1}^{N}\frac{\epsilon_{j}}{\bar{z}- \bar{z}_{j}}
\hspace{10mm} \mbox{or,} \hspace{5mm}
\bar{\omega} = \bar{z} - \sum_{j=1}^{N}\frac{\epsilon_{j}}{z - z_{j}}
\hspace{5mm} \mbox{(in complex conjugate form)}
\end{displaymath}

Combining both forms of this lensing equation into one equation
for image position $z$ for a two mass lens system produces a
complex fifth order polynomial, whose five roots are potential
image positions. Back substitution is required in order determine
if a root corresponds to a real image position, as there can be
three or five images depending on the geometry.  In general, the
complex fifth order polynomial can only be solved numerically but
there are algorithms to accomplish this.  We use the method of
Jenkins and Traub~\cite{Jenkins72} in combination with root
polishing using the Laguerre method~\cite{Riley06} to obtain
accurate root values.

Given that the light bending in a gravitational field does not
change the intrinsic source brightness, the point source
magnification at each image position, $\mathbf z=x+iy$, is given
by the ratio of the infinitesimal area element $\delta A_{xy} =
dxdy$ in the image plane to the corresponding elemental area
$\delta A_{uv}$ in the source plane at the position $\mathbf
w=u+iv$. This quantity is given by the determinant of the Jacobian
for the transformation of the point $\mathbf z(x,y)$ in the image
plane to the point $\mathbf w(u,v)$ in the source plane, as
governed by the above lensing equation.  But, it is the `Jacobian'
($J$) of the inverse transformation from the source position to
the image position that is readily determined and, given the
geometrical nature of the quantity, the absolute value of the
inverse provides the value of interest.  We have:
\begin{displaymath}
 J = \frac{\partial u}{\partial x}\frac{\partial v}{\partial y}
   - \frac{\partial u}{\partial y}\frac{\partial v}{\partial x}
\Longrightarrow
     \frac{\partial \omega}{\partial z}
     \frac{\partial \bar{\omega}}{\partial \bar{z}}
   - \frac{\partial\omega}{\partial\bar{z}}
     \frac{\partial \bar{\omega}}{\partial z }
  = 1 - \left |\frac{\partial\bar{\omega}}{\partial{z}}\right |^2
  = 1 - \left |\frac{\partial\omega}{\partial\bar{z}}\right |^2
\end{displaymath}

The point source magnification of each image formed at the
position $\mathbf z$ is therefore $1/|J(\mathbf z)|$ and the
combined magnification for all unresolved images is simply the sum
over these quantities.

The critical curves (and the corresponding caustics) are the locii
of points where the point source magnification formally diverges
and are thus determined from the requirement that this inverse
Jacobian is zero; this leads to a set of fourth order complex
polynomials for the two mass lens system.

Assembling the expressions that give the coefficients of the
complex polynomials is rather tedious~\cite{Rhie02} and a symbolic
algebra computer package such as Maple is of considerable
assistance. However, we have found that (substantial) further
manual manipulation can lead to more compact and efficient
expressions.  This is particularly the case for lensing systems
with \emph{three} and even \emph{four} masses.  And, we have
included both these cases in our code.  The three-mass case yields
a complex $10^{\rm th}$ order polynomial \cite{Rhie02}, while the
four-mass system leads to a $17^{\rm th}$ order polynomial
requiring the determination of no less than 18 complex
coefficients.  The critical curve polynomials for these systems
are of $6^{\rm th}$ and $8^{\rm th}$ order respectively.

Finite source effects become important when the source disk is
near a caustic curve and are important in planetary events.  We
have included these in the software package by using a combination
of two methods: a hexadecapole expansion and a numerical
integration procedure we call the polygon method.  The
hexadecapole expansion method~\cite{Gould08} employs multiple
point source magnification values to estimate the magnification
resulting from a finite source, including limb darkening effects
if required.  The polygon method is based on the work of Gould and
Gaucherel~\cite{Gould97} and uses polygons to represent the source
disk and image figures and thereby estimate the areas.  Limb
darkening effects are included in this method by using a network
of ``concentric'' polygons.

A method promoted by Bennett~\cite{Bennett10} for finite source
effects first employs the polynomial approach to locate the image
centres and then adopts the inverse ray-tracing method in the
vicinity of each image to estimate the limb-darkening integrals
over the images.  We are looking at implementing this approach as
well.

\section{Conclusion}

In spite of the expression complexity in calculating the lensing
polynomial coefficients, we have found our new code to be
competitive speed-wise with the inverse ray-tracing approach.  The
rapidly rising level of complexity and the large numbers of terms
in the coefficient expressions makes the polynomial method
impractical for more than four lensing masses.  For $n$ lenses the
order of the polynomial is $n^2 + 1$, so for $n=5$ lenses we have
27 coefficients to determine. Furthermore, we need to employ
numerical methods to extract 26 complex root values from the
$26^{\rm th}$ order polynomial and determine which ones correspond
to real image positions. It is reasonably clear that the inverse
ray-tracing approach will be more efficient for these cases, at
least for examples with no lens motion.

When lens motion is involved, the polynomial method is both more
efficient and faster than inverse ray tracing.  A three-mass
lensing system consisting of a star and two planets has already
been discovered and successfully modelled~\cite{Gaudi08}, and it
could be the case that a system of a star and three planets is
found and usefully modelled.  But it is probably the case that
endeavouring to model a microlensing light curve with five lensing
masses or more (and perhaps even four) is very unlikely to produce
useful physical information.

%%  bibliographic items can be constructed using the LaTeX format in SPIRES:
%%    see    http://www.slac.stanford.edu/spires/hep/latex.html

%%

\Chapter{GPU-assisted contouring for modeling binary microlensing
events}
            {GPU-assisted contouring for modeling binary ML events}{Hundertmark et al.}
\bigskip\bigskip

\addcontentsline{toc}{chapter}{{\it Hundertmark et al.}}
\label{polingStart}

\begin{raggedright}

{\it Markus Hundertmark\index{author}{Hundertmark, Markus},
Frederic V. Hessman\index{author}{Hessman, Frederic V.},
Stefan Dreizler\index{author}{Dreizler, Stefan}\\
Institut f\"ur Astrophysik, Georg-August-Universit\"at G\"ottingen \\
Germany}

\bigskip\bigskip

\end{raggedright}

\section{Introduction}

Due to the continuously improving computer power available in
dedicated hardware, clusters and computing grids, real-time binary
modeling is now being tested and used by different microlensing
groups. In addition to CPU-based computing, one can exploit the
possibility of speeding-up calculations using graphic cards which
are available at reasonable prices and have a reduced carbon
footprint. These Graphics Processing Units (GPUs) offer a
increased capability concerning the number of floating point
operations per second but at the cost of careful flow control
management. Naturally, these GPUs are well suited for ray-tracing
but in this work a pointwise approach for simulating lightcurves
on the basis of contouring techniques is also possible, despite
some system dependent drawbacks concerning the accuracy.

\section{GPU-contouring}

The use of GPUs for gravitational microlensing simulations seems
natural as their design is driven by the needs of intensive
ray-tracing computer games applications. The advantages and
disadvantages when used for scientific ray-tracing have been
shown, e.g. by \cite{bate2010},\cite{thompson2010}.  One
bottle-neck for a fast fitting environment is the limited data
transfer capacity between CPU and GPU. In order to ease the burden
of transfer back and forth to the GPU a large fraction of the
analysis must be executed on the GPU and needs to be optimized for
these machines, since they offer the highest performance for
standard single-precision floating point numbers.

As an alternative, we have implemented the contouring technique on
GPUs aiming to preserve the pointwise calculation of lightcurves
by parallelizing the root-finding and subsequent integration of
the image contours as by \cite{gould1996}.

%%%%%%%%%%%%%%%%%%%%%%%%%%%%%%%%%%%%%%%%%%%%%%%%%%%%%%%%%%%%%%%%%%%%%%%%%
%%
%%   use this format to include an .eps figure into your paper
%%
\begin{figure}[htb]
\begin{center}
\epsfig{file=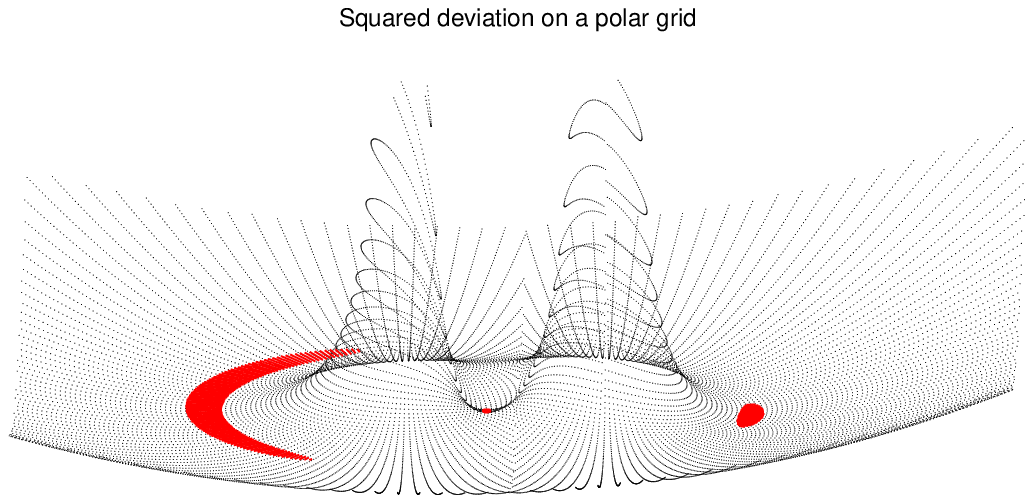,height=3.0in} \caption{Two-grided
polar parameterization of the lens plane for root-finding.}
\label{fig:polargrid}
\end{center}
\end{figure}
%%%%%%%%%%%%%%%%%%%%%%%%%%%%%%%%%%%%%%%%%%%%%%%%%%%%%%%%%%%%%%%%%%%%%%%%%%%

Due to the technical limitations of a computing system with many
processing units, the number of required computations and
variables has to be kept low.  This comes at the price of careful
flow control management. For this purpose, the squared deviation
between lens and source position as introduced by
\cite{schramm1987} has been used. A point-wise parallelisation of
the root-finding was achieved by solving the lens equation on a
polar grid (cf. \cite{bennett2010}), where the radial search for
potential roots was carried out as one thread, as illustrated in
Fig.~\ref{fig:polargrid}. Afterwards, the elements are integrated
with a simple Gaussian quadrature. This approach is particularly
efficient for planetary events with modest angular separations and
source positions $\left|\beta\right| < 1.5\ \theta_{\rm{E}}$. A
test implementation on a system with 240 streaming
processors\footnote{NVIDIA\textregistered\ Tesla\texttrademark\
C1060} is able to simulate a lightcurve consisting of 1000 points
in one second. Root-finding threads cannot communicate in this
approach and thus it would be more difficult to implement a
hierarchical approach as by \cite{dominik2007}.

In contrast to existing CPU-based solutions, the demanded accuracy
cannot always be achieved, as most commonly used GPUs exclusively
compute with single-precision\footnote{IEEE 754 single-precision
(32 bits)} floating numbers or require a substantial overhead for
operating at double-precision. On single-precision machines the
root-finding accuracy is limited by $5 \cdot 10^{-7}\
\theta_{\rm{E}}$ in the radial direction, which directly affects
the maximal achievable accuracy and accessible sources star radii
$r_{\star}$. Estimating the magnification as an integration of
evenly sized squares in $128^2$ search directions per grid gives
at zeroth order an uncertainty of
\begin{equation}
\sigma_{\mu} \approx 10 ^{-6} r^{-1}_{\star}.
\end{equation}
Models for ground-based microlensing lightcurves require a model
accuracy of better than $10^{-3}$ and thus the applicability of
the model is limited to source star radii~$\geq10^{-3}\
\theta_{\rm{E}}$. Beyond that limit, only relative precision can
be guaranteed.  The effect of missing solutions at the edges is
$\propto 1/N$ and thus the aforementioned number of $128^2$
elements is a compromise between runtime, single-precision
round-off error, and numerical uncertainty of the integration. The
benefits of refining the grids for keeping the requested accuracy
is shown in Fig.~\ref{fig:contouring lightcurves from GPUs}.
Comparing current simulations with magnification maps
\cite{kayser1986},\cite{wambsganss1997}, indicate that the
numerical accuracy stays below $10^{-3}$ at one sigma level as
demanded.

%%%%%%%%%%%%%%%%%%%%%%%%%%%%%%%%%%%%%%%%%%%%%%%%%%%%%%%%%%%%%%%%%%%%%%%%%
%%
%%   use this format to include an .eps figure into your paper
%%
\begin{figure}[htb]
\begin{center}
\epsfig{file=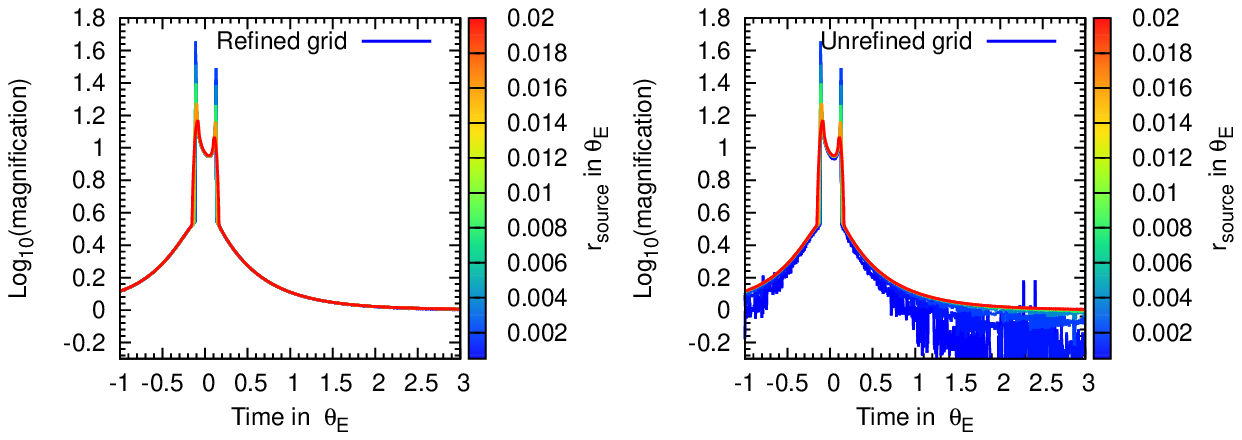,height=2.0in}
\caption{Lightcurves simulated with GPU-assisted contouring are
shown for different source star radii with a constant number of
integration elements and for a number of integration elements
adapted to a predefined accuracy limit.} \label{fig:contouring
lightcurves from GPUs}
\end{center}
\end{figure}
%%%%%%%%%%%%%%%%%%%%%%%%%%%%%%%%%%%%%%%%%%%%%%%%%%%%%%%%%%%%%%%%%%%%%%%%%%%

\section{Summary and further work}

A first point-wise implementation of the contouring technique on
GPUs has been achieved. The rapidly increasing number of streaming
processors and the recently introduced double-precision
capability\footnote{NVIDIA\textregistered\ Fermi\texttrademark}
supports the development of these systems as the existing software
can be scaled to a new state-of-the-art equipment just by
recompiling. In addition, this approach can be used with existing
sophisticated fitting codes and the simulation can even be mixed
with other pointwise contouring techniques which are more suitable
for highly accurate simulations, such as the system introduced by
\cite{bozza2010}.

\bigskip
M.H. would like to acknowledge the support from the DFG and the
DFG Research Training Group GrK - 1351 ``Extrasolar Planets and
their host stars''.

\Chapter{Red noise effect in space-based microlensing
observations}
            {Red noise effect in space-based microlensing observations}{Nucita et al.}
\bigskip\bigskip

\addcontentsline{toc}{chapter}{{\it Nucita et al.}}
\label{polingStart}

\begin{raggedright}

{\it Achille A. Nucita\index{author}{Nucita, Achille A.}, Daniele
Vetrugno\index{author}{Vetrugno, Daniele}, Francesco De Paolis\index{author}{De Paolis, Francesco},
Gabriele Ingrosso\index{author}{Ingrosso, Gabriele}, Berlinda M. T. Maiolo\index{author}{Maiolo, Berlinda M. T.}\\
Department of Physics, University of Salento, Lecce, Italy \\
INFN, Sez. di Lecce, Lecce, Italy\\}
\bigskip
{\it Stefania Carpano\index{author}{Carpano, Stefania}\\
ESA, Research and Scientific Support Department, ESTEC, Noordwijk \\
The Netherlands\\}

\bigskip\bigskip

\end{raggedright}
\section{Introduction}

Gravitational microlensing, first developed to investigate the
distribution of Galactic dark matter \cite{1986ApJ...304....1P},
represents now a powerful and important tool to explore many other
astrophysical contexts. For example, in the last years, the
possibility to investigate the distribution of exo-planetary
systems by means of the microlensing has been pointed out to the
scientific community. The advantages for using it are huge and
span from the high sensitivity to relatively low mass planets and
large observer-lens distances to the unique possibility of
detecting free-floating planets. Gravitational microlensing
surveys may also give the possibility of a planetary census
throughout the Galaxy \cite{2005NewAR..49..424G}. Another
technique used to search for exo-planets is that of detecting
transit events by means of the differential photometry analysis of
the sources.

Both transits and microlensing techniques are very close from the
observational point of view, since both are photometric-based and
the observer has to search for deviations from the baseline flux.
In the transit case one seeks for a decrease in the signal flux
while for microlensing events an amplification is present. In both
cases, the analysis is to be made on light curves.

To this aim, we decided to study and characterize the noise
(typically red or pink noise which could mimic a microlensing
feature) of a sample of light curves from a space-based
experiment. In this prospective we have considered a sample of
light curves taken from the CoRoT\footnote{More detailed
information are avaible at the web site
http://smsc.cnes.fr/COROT/} satellite, a space-based experiment
projected by the ``Centre National d'Etudes Spatials'' (CNES) in
collaboration with the French laboratories and other international
partners (Europe, Brasil) launched in December $2006$.

% %%%%%%%%%%%%%%%%%%%%%%%%%%%%%%%%%%%%%%%%%%%%%%%%%%%%%%%%%%%%%%%%%%%%%%%%%
% %%
% %%   use this format to include an .eps figure into your paper
% %%
% \begin{figure}[htb]
% \begin{center}
% \epsfig{file=rgb.eps,height=1.5in}
% \caption{Plan of the magnet used in the Mesmeric studies.}
% \label{fig:magnet}
% \end{center}
% \end{figure}
% %%%%%%%%%%%%%%%%%%%%%%%%%%%%%%%%%%%%%%%%%%%%%%%%%%%%%%%%%%%%%%%%%%%%%%%%%%%
%
%
% %%%%%%%%%%%%%%%%%%%%%%%%%%%%%%%%%%%%%%%%%%%%%%%%%%%%%%%%%%%%%%%%%%%%%%%%%
% %%
% %%   use this format to include a LaTeX table  into your paper
% %%
% \begin{table}[b]
% \begin{center}
% \begin{tabular}{l|ccc}
% Patient &  Initial level($\mu$g/cc) &  w. Magnet &
% w. Magnet and Sound \\ \hline
%  Guglielmo B.  &   0.12     &     0.10      &     0.001  \\
%  Ferrando di N. &  0.15     &     0.11      &  $< 0.0005$ \\ \hline
% \end{tabular}
% \caption{Blood cyanide levels for the two patients.}
% \label{tab:blood}
% \end{center}
% \end{table}
% %%%%%%%%%%%%%%%%%%%%%%%%%%%%%%%%%%%%%%%%%%%%%%%%%%%%%%%%%%%%%%%%%%%%%%%%%%%

\bigskip
D.Vetrugno acknowledges support from the Faculty of the European
Space Astronomy Centre (ESAC) and for the kind hospitality. The
CoRoT space mission, launched on 2006 December 27, was developed
and is operated by the CNES, with participation of the Science
Programs of ESA, ESA's RSSD, Austria, Belgium, Brazil, Germany and
Spain.

\Chapter{Light curve errors introduced by limb-darkening models}
            {Light curve errors introduced by limb-darkening models}
\bigskip\bigskip

\addcontentsline{toc}{chapter}{{\it David Heyrovsk\'y}}
\label{polingStart}

\begin{raggedright}

{\it David Heyrovsk\'y\index{author}{Heyrovsk\'y, David}\\
Institute of Theoretical Physics, Charles University in Prague \\
Czech Republic}

\bigskip\bigskip

\end{raggedright}

During the analysis of caustic-crossing microlensing events the
source star is typically modeled by analytical limb-darkening laws
- most often linear, in a few cases higher-order laws
(square-root, quadratic). An alternative option is to use a
non-analytical limb-darkening model based on principal component
analysis of model atmospheres \cite{Heyrovsky2008,Zub2011}, which
gives a more accurate description of the source-star intensity
profile \cite{Heyrovsky2003}. To guide the choice of
limb-darkening model during event analysis we need to know the
photometric error introduced by assuming a particular model. In
order to answer this question, we compute single-lens
caustic-crossing light curves for stellar model atmospheres from
Kurucz's ATLAS9 grid in a representative set of photometric bands,
and compare them with light curves of sources with different
approximations of the underlying limb darkening. We demonstrate
both the amplitude of the introduced photometric error and the
corresponding residual pattern. We discuss the implications of our
results for the analysis of single and two-point-mass
caustic-crossing events.

\Chapter{Detecting Isolated Stellar-Mass Black Holes through
Astrometric Microlensing using HST}
            {Detecting isolated stellar-mass black holes using HST}{Sahu et al.}
\bigskip\bigskip

\addcontentsline{toc}{chapter}{{\it Sahu et al.}}
\label{polingStart}

\begin{raggedright}

{\it Kailash C. Sahu\index{author}{Sahu, Kailash C.}, Howard E.
Bond\index{author}{Bond, Howard E.}, Jay
Anderson\index{author}{Anderson,
Jay}\\
Space Telescope Science Institute, Baltimore, USA\\}
\bigskip
{\it Martin Dominik\index{author}{Dominik, Martin}\\
School of Physics \& Astronomy, University of St. Andrews, St.
Andrews, UK
\\}
\bigskip
{\it Andrzej Udalski\index{author}{Udalski, Andrzej}\\
Warsaw University Observatory, Warszawa, Poland\\}
\bigskip
{\it Philip Yock\index{author}{Yock, Philip}\\
Department of Physics, University of Auckland, Auckland, New
Zealand\\}

\bigskip\bigskip

\end{raggedright}

\section{Introduction}

Stars with masses greater than $\sim 20$ $M_{\odot}$ are thought
to end their lives as stellar-mass black holes (BHs) (e.g., Fryer
\& Kalogera 2001). Yet, there has never been an unambiguous
detection of an isolated black hole, since they emit no radiation.
The only detected BHs to date are through their X-ray emissions in
binary systems, where their masses have been measured through
radial velocity measurements. Solitary BHs are detectable with
current microlensing searches like OGLE and MOA. These campaigns
have already detected several BH candidates. Bennett et al. (2002)
have described two events: MACHO 96-BLG-5 and MACHO 98-BLG-6, with
mass estimates of  $>6$ solar mass. Mao et al. (2002) concluded
that MACHO 99-BLG-22 is a BH candidate with a minimum mass of
$10.5$ $M_{\odot}$ .  However, all these claims are statistical in
nature, since determination of the lens mass  from the
microlensing light curve alone suffers from a degeneracy between
the distance to the lens, the mass of the lens, and the
source-lens relative proper motion.

\section{Black Hole Detections through Microlensing }

Microlensing survey programs typically detect $>1000$ microlensing
events towards the Galactic bulge per year. A few of them have
long durations, and these long-duration events ($t_{\mathrm{E}}
> 150$ days) could potentially be arising from lenses with masses
larger than 10 $M_{\odot}$. For most of them, the light from the
lens itself is negligible. So it has been suggested that a
majority of them are due to massive, non-luminous stellar
remnants. However, a long-duration microlensing event can be also
by a low-mass lens passing in front of the source with a small
relative motion.

%%   use this format to include an .eps figure into your paper
%%
\begin{figure}[htb]
\begin{center}
\epsfig{file=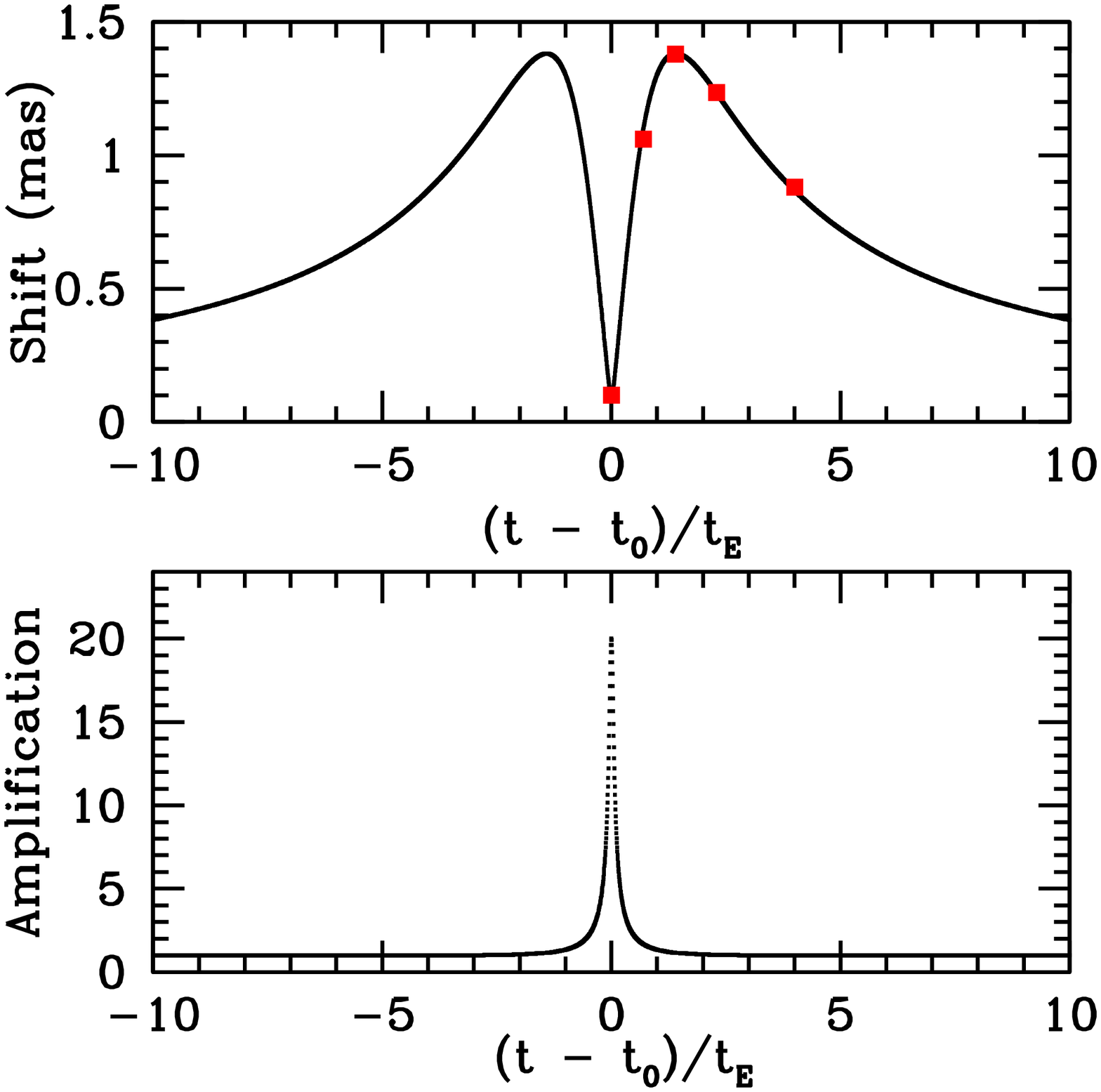,height=4in} \caption{ The lower panel
shows a photometric microlensing light curve with minimum impact
parameter u=0.05, and the upper panel shows the corresponding
expected astrometric shifts. The astrometric shift peaks at
u$\sim$1.4, which is about 1.4 milliarcsec for a BH lens with a
mass of 5 $M_{\odot}$ in the Sagittarius arm.  The timing of our 5
proposed observations is shown by red dots in the upper panel.
These 5 measurements will unambiguously separate the shifts caused
by microlensing from the shifts caused by blending and proper
motion of the source. } \label{fig:magnet}
\end{center}
\end{figure}
%%%%%%%%%%%%%%%%%%%%%%%%%%%%%%%%%%%%%%%%%%%%%%%%%%%%%%%%%%%%%%%%%%%%%%%%%%%

\section{Astrometry Resolves the Degeneracy}
Microlensing not only causes a photometric amplification, but it
also causes a small astrometric shift in the position of the
source (see Fig. 1 for illustrations). This astrometric shift is
generally undetectable from ground-based observations. But, for
lenses with masses of $\sim 5$ $M_{\odot}$, the astrometric shift
is expected to be $\sim 1.4$ mas, which is easily detectable with
HST. This astrometric shift, coupled with the parallax effect
generally observed for the long-duration events, allows a
determination of the mass of the lens unambiguously, thus
resolving the degeneracy between the low-mass and the high-mass
lenses (Sahu \& Dominik 2001).

\section{Mass Determinations of the BH Candidates and Implications}

We expect to achieve an astrometric precision of $\sim$0.2~mas at
each epoch of HST observations. Thus a shift of 1.4~mas as
expected from a  5 $M_{\odot}$ BH would be detected at 7 sigma.
The 0.2~mas error translates to an error of about 0.75 $M_{\odot}$
in mass measurement. Thus we can determine masses with about
$15\%$ precision at 5 $M_{\odot}$ or $8\%$ at 10 $M_{\odot}$.

Thus our program has the potential of making the first unequivocal
detections of isolated black holes, and the first direct mass
measurements for isolated stellar-mass BHs through any technique.
We expect to observe 5 promising long-duration events with
WFC3/HST, possibly leading to detection of some confirmed BHs.
Detection of these BHs will provide the very first clues on the
frequency of isolated BHs in the Galaxy, which have strong
implications on the slope of the IMF at high masses.

\bigskip
This project uses the NASA/ESA Hubble Space Telescope, operated by
AURA for NASA.

\Chapter{The observability of isolated compact remnants with
microlensing}
            {The observability of isolated compact remnants with ML}{Sartore \& Treves}
\bigskip\bigskip

\addcontentsline{toc}{chapter}{{\it Nicola Sartore}}
\label{polingStart}

\begin{raggedright}

{\it Nicola Sartore\index{author}{Sartore, Nicola}\\
INAF - Istituto di Astrofica Spaziale e Fisica Cosmica, Milano \\
Italy
\\}
\bigskip

{\it Aldo Treves\index{author}{Treves, Aldo}\\
Dipartimento di Fisica e Matematica, Universit\'a dell'Insubria, Como \\
Italy}

\bigskip

\end{raggedright}

\section{Introduction}

The observability of old isolated compact remnants is a long
standing problem. To date, blind searches for neutron stars and
stellar-mass black holes (NSs and BHs herafter) likely accreting
from the interstellar medium (ISM) have failed to reveal any
promising candidates.

Dark compact objects may still manifest themselfs through the
effect of their gravitational field on the light of background
sources, i.e. as microlensing events, in particular towards the
Galactic bulge, where the density sources is larger. The
contribution of NSs and BHs to the microlensing rate towards the
bulge has been estimated by e.g. \cite{G00}, \cite{WM05} and
\cite{CN08}. Here we present a quantitative analisys of this
contribution, under the hypothesis that both NSs and BHs are born
with large kick velocities (see e.g. \cite{H05}) and compare this
contribution with that of ``normal'' stars, that is brown dwarfs,
main sequence stars and white dwarfs.

\section{Model and Results}
We built a model of the distribution of normal stars in the bulge
and disk of the Galaxy (see \cite{ST10} for a detailed description
of the model and procedure adopted) and followed the orbits of
synthetic bulge-born and disk-born isolated compact objects in the
Galactic potential. Following \cite{G00}, the normalization of the
phase space distribution of NSs and BHs was estimated from the
stellar initial mass function assuming that all stars with initial
mass greater than $1\,M_\odot$ are in the remnant phase. In
particular, stars with inital mass between 8 and 40 solar masses
are now NSs with $M\,=\,1.4\,M_\odot$, while stars with mass
$>\,40\,M_\odot$ are now BHs with $M\,=\,10\,M_\odot$.

Then, we calculated the optical depth and the distribution of
event time scales for normal stars, NSs and BHs towards different
lines of sight (l.o.s. hereafter). The optical depth obtained was
lower than the case in which NSs and BHs are born without kicks.
This can be easily explained since objects with large velocity
have a larger scale height. Thus, the spatial density along the
l.o.s. is lower. On the other hand, a fast moving object is
expected to give a higher rate of events with respect to an object
with lower velocity. In fact, we found that the fraction of
microlensing events due to isolated NSs and BHs is larger than
that expected in the case of non kicks. In particular, for events
longer than  100 days, we estimated that about $50\%$ of the
events are likely due to NSs and BHs.

\section{Conclusions and Future Prospects}

Cosidering the large number of microlensing events observed by
several microlensing surveys, our results suggest that a
non-negligible fraction of these events is likely related to NSs
and BHs, in particular among long duration events. Indeed, an
excess of long duration events has been reported in the
literature. Thus, microlensing surveys are a suitable tool to
probe the statistical properties of old isolated compact remnants.
However, the nature of single lenses has to be assessed as further
confirmation. We are performing a cross-check of archival X-ray
data to find counterparts of long duration events. The detection
of X-ray emission would be a strong indication that the lens is a
compact object.

\Chapter{Gravitational microlensing by the Ellis wormhole}
            {Gravitational microlensing by the Ellis wormhole}{Fumio Abe}
\bigskip\bigskip

\addcontentsline{toc}{chapter}{{\it Fumio Abe}}
\label{polingStart}

\begin{raggedright}

{\it Fumio Abe\index{author}{Abe, Fumio}\\
Solar-Terrestrial Environment Laboratory\\
Nagoya University\\
Japan}
\bigskip\bigskip
\end{raggedright}

\section{abstract} A method to calculate light curves of the
gravitational microlensing of the Ellis wormhole is derived in the
weak-field limit. In this limit, lensing by the wormhole produces
one image outside the Einstein ring and one other image inside.
The weak-field hypothesis is a good approximation in Galactic
lensing if the throat radius is less than $10^{11}$ km. The light
curves calculated have gutters of approximately 4\% immediately
outside the Einstein ring crossing times. The magnification of the
Ellis wormhole lensing is generally less than that of
Schwarzschild lensing. The optical depths and event rates are
calculated for the Galactic bulge and Large Magellanic Cloud
fields according to bound and unbound hypotheses. If the wormholes
have throat radii between $100$ and $10^7$ km, are bound to the
galaxy, and have a number density that is approximately that of
ordinary stars, detection can be achieved by reanalyzing past
data. If the wormholes are unbound, detection using past data is
impossible.

\section{Introduction}
A solution of the Einstein equation that connects distant points
of space--time was introduced by \cite{ein35}. This
``Einstein--Rosen bridge'' was the first solution to later be
referred to as a wormhole. Initially, this type of solution was
just a trivial or teaching example of mathematical physics.
However, \cite{morr88} proved that some wormholes are
``traversable''; i.e., space and time travel can be achieved by
passing through the wormholes. They also showed that the existence
of a wormhole requires exotic matter that violates the null energy
condition. Although they are very exotic, the existence of
wormholes has not been ruled out in theory. Inspired by the
Morris--Thorne paper, there have been a number of theoretical
works (see \cite{vis95, lob09} and references therein) on
wormholes. The curious natures of wormholes, such as time travel,
energy conditions, space--time foams, and growth of a wormhole in
an accelerating universe have been studied. Although there have
been enthusiastic theoretical studies, studies searching for real
evidence of the existence of wormholes are scarce. Only a few
attempts have been made to show the existence or nonexistence of
wormholes.

A possible observational method that has been proposed to detect
or exclude the existence of wormholes is the application of
optical gravitational lensing. The gravitational lensing of
wormholes was pioneered by \cite{cram95}, who inferred that some
wormholes show ``negative mass'' lensing. They showed that the
light curve of the negative-mass lensing event of a distant star
has singular double peaks. Several authors subsequently conducted
theoretical studies on detectability \cite{saf01, bog08}. Another
gravitational lensing method employing gamma rays was proposed by
\cite{tor98}, who postulated that the singular negative-mass
lensing of distant active galactic nuclei causes a sharp spike of
gamma rays and may be observed as double-peaked gamma-ray bursts.
They analyzed BASTE data and set a limit for the density of the
negative-mass objects.

There have been several recent works \cite{sh04, per04, nan06,
rah07, dey08} on the gravitational lensing of wormholes as
structures of space--time. Such studies are expected to unveil
lensing properties directly from the space--time structure. One
study \cite{dey08} calculated the deflection angle of light due to
the Ellis wormhole, whose asymptotic mass at infinity is zero. The
massless wormhole is particularly interesting because it is
expected to have unique gravitational lensing effects. The Ellis
wormhole is expressed by the line element
\begin{equation}
ds^2 = dt^2 - dr^2 - (r^2 + a^2) (d\theta^2 + sin^2(\theta)
d\phi^2),
\end{equation}
where $a$ is the throat radius of the wormhole. This type of
wormhole was first introduced by \cite{ell73} as a massless scalar
field. Later, \cite{morr88} studied this wormhole and proved it to
be traversable. The dynamical feature was studied by \cite{shi02},
who showed that Gaussian perturbation causes either explode to an
inflationary universe or collapse to a black hole. \cite{das05}
showed that the tchyon condensate can be a source for the Ellis
geometry.

In this paper, we derive the light curve of lensing by the Ellis
wormhole and discuss its detectability. In Section 2, we discuss
gravitational lensing by the Ellis wormhole in the weak-field
limit. The light curves of wormhole events are discussed in
Section 3. The validity of the weak-field limit is discussed in
Section 4.  The optical depth and event rate are discussed in
Section 5. The results are summarized in Section 6.

\section{Gravitational lensing}
Magnification of the apparent brightness of a distant star by the
gravitational lensing effect of another star was predicted by
\cite{ein36}. This kind of lensing effect is called
``microlensing'' because the images produced by the gravitational
lensing are very close to each other and are difficult for the
observer to resolve. The observable effect is the changing
apparent brightness of the source star only. This effect was
discovered in 1993 \cite{uda93, alc93, aub93} and has been used to
detect astronomical objects that do not emit observable signals
(such as visible light, radio waves, and X rays) or are too faint
to observe. Microlensing has successfully been applied to detect
extrasolar planets \cite{bon04} and brown dwarfs \cite{nov09,
gou09}. Microlensing is also used to search for unseen black holes
\cite{alc01, ben02, poi05} and massive compact halo objects
\cite{alc00, tis07, wyr09}, a candidate for dark matter.

The gravity of a star is well expressed by the Schwarzschild
metric. The gravitational microlensing of the Schwarzschild metric
\cite{ref64, lieb64, Pacz86} has been studied in the weak-field
limit. In this section, we simply follow the method used for
Schwarzschild lensing. Figure \ref{fig1} shows the relation
between the source star, the lens (wormhole), and the observer.
The Ellis wormhole is known to be a massless wormhole, which means
that the asymptotic mass at infinity is zero. However, this
wormhole deflects light by gravitational lensing
\cite{cle84,che84,nan06,dey08} because of its curved space--time
structure. The deflection angle $\alpha(r)$ of the Ellis wormhole
was derived by \cite{dey08} to be
\begin{equation}
\alpha(r) = \pi \left\{\sqrt \frac{2 (r^2 + a^2)}{2 r^2 + a^2} -1
\right\},
\end{equation}
where $r$ is the closest approach of the light. In the weak-field
limit ($r \rightarrow \infty$), the deflection angle becomes
\begin{equation}
\alpha(r) \rightarrow \frac{\pi}{4}\frac{a^2}{r^2} - \frac{5
\pi}{32} \frac{a^4}{r^4} + o\left(\frac{a}{r}\right)^6.
\label{eqn:defl}
\end{equation}

The angle between the lens (wormhole) and the source $\beta$ can
then be written as
\begin{equation}
\beta  = \frac{1}{D_L} b - \frac{D_{LS}}{D_S}\alpha(r),
\end{equation}
where $D_L$, $D_S$, $D_{LS}$, and $b$ are the distances from the
observer to the lens, from the observer to the source, and from
the lens to the source, and the impact parameter of the light,
respectively. In the asymptotic limit, Schwarzschild lensing and
massive Janis--Newman--Winnicour (JNW) wormhole lensing
\cite{dey08} have the same leading term of $o\left(1/r\right)$.
Therefore, the lensing property of the JNW wormhole is
approximately the same as that of Schwarzschild lensing and is
difficult to distinguish. As shown in Equation (\ref{eqn:defl}),
the deflection angle of the Ellis wormhole does not have the term
of $o\left(1/r\right)$ and starts from $o\left(1/r^2\right)$. This
is due to the massless nature of the Ellis wormhole and indicates
the possibility of observational discrimination from the ordinary
gravitational lensing effect. In the weak-field limit, $b$ is
approximately equal to the closest approach $r$. For the Ellis
wormhole, $b = \sqrt{r^2 + a^2} \rightarrow r (r \rightarrow
\infty)$. We thus obtain
\begin{equation}
\beta = \frac{r}{D_L} - \frac{\pi}{4}
\frac{D_{LS}}{D_S}\frac{a^2}{r^2}\hspace{1cm}(r > 0).
\label{eqn:proj}
\end{equation}
The light passing through the other side of the lens may also form
images. However, Equation (\ref{eqn:proj}) represents deflection
in the wrong direction at $r < 0$. Thus, we must change the sign
of the deflection angle:
\begin{equation}
\beta = \frac{r}{D_L} + \frac{\pi}{4}
\frac{D_{LS}}{D_S}\frac{a^2}{r^2}\hspace{1cm}(r < 0).
\label{eqn:proj2}
\end{equation}
It would be useful to note that a single equation is suitable both
for $r > 0$ and $r < 0$ images in the Schwarzschild lensing.
However, such treatment is applicable only when the deflection
angle is an odd function of $r$.

If the source and lens are completely aligned along the line of
sight, the image is expected to be circular (an Einstein ring).
The Einstein radius $R_E$, which is defined as the radius of the
circular image on the lens plane, is obtained from Equation
(\ref{eqn:proj}) with $\beta = 0$ as
 \begin{equation}
 R_E = \sqrt[3]{\frac{\pi}{4}\frac{D_L D_{LS}}{D_S} a^2}. \label{eqn:re}
 \end{equation}

The image positions can then be calculated from
\begin{equation}
\beta = \theta - \frac{\theta_E^3}{\theta^2} \hspace{1cm} (\theta
> 0)  \label{eqn:imgeq}
\end{equation}
and
\begin{equation}
\beta = \theta + \frac{\theta_E^3}{\theta^2} \hspace{1cm} (\theta
< 0),  \label{eqn:imgeq2}
\end{equation}
where $\theta = b / D_L  \approx r / D_L$ is the angle between the
image and lens, and $\theta_E = R_E / D_L$ is the angular Einstein
radius. Using reduced parameters $\hat{\beta} = \beta / \theta_E$
and $\hat{\theta} = \theta / \theta_E$, Equations
(\ref{eqn:imgeq}) and (\ref{eqn:imgeq2}) become simple cubic
formulas:
\begin{equation}
\hat{\theta}^3 - \hat{\beta} \hat{\theta}^2 -1 = 0 \hspace{1cm}
(\hat{\theta} > 0) \label{eqn:poly}
\end{equation}
and
\begin{equation}
\hat{\theta}^3 - \hat{\beta} \hat{\theta}^2 +1 = 0 \hspace{1cm}
(\hat{\theta} < 0). \label{eqn:poly2}
\end{equation}
 As the discriminant of Equation (\ref{eqn:poly}) is $-4\hat{\beta}^3 - 27 < 0$, Equation (\ref{eqn:poly}) has two conjugate complex solutions and a real solution:
 \begin{equation}
 \hat{\theta} = \frac{\hat{\beta}}{3} + U_{1+} + U_{1-} ,
 \end{equation}
 with,
 \begin{equation}
 U_{1\pm} = \sqrt[3]{\frac{\hat{\beta}^3}{27} + \frac{1}{2} \pm \sqrt{\frac{1}{4}\left(1 + \frac{2 \hat{\beta}^3}{27}\right)^2 - \frac{\hat{\beta}^6}{27^2}}}.
 \end{equation}
The real positive solution corresponds to the physical image.

The discriminant of Equation (\ref{eqn:poly2}) is $4\hat{\beta}^3
- 27$. Thus it has a real solution if $\hat{\beta} <
\sqrt[3]{27/4}$:
\begin{equation}
\hat{\theta} = \frac{\hat{\beta}}{3} + U_{2+} + U_{2-},
\end{equation}
where,
\begin{equation}
U_{2\pm} = \omega \sqrt[3]{\frac{\hat{\beta}^3}{27} - \frac{1}{2}
\pm \sqrt{\frac{1}{4}\left(1 - \frac{2 \hat{\beta}^3}{27}\right)^2
- \frac{\hat{\beta}^6}{27^2}}} ,
\end{equation}
with $\omega = e^{(2\pi/3)i}$. This solution corresponds to a
physical image inside the Einstein ring. For $\hat{\beta} >
\sqrt[3]{27/4}$, Equation (\ref{eqn:poly2}) has three real
solutions. However, two of them are not physical because they do
not satisfy $\hat{\theta} < 0$. Only the solution
\begin{equation}
\hat{\theta} = \frac{\hat{\beta}}{3} +  \omega U_{2+} + U_{2-}
\end{equation}
corresponds to a physical image inside the Einstein ring.

Figure \ref{fig2} shows the calculated images for source stars at
various positions on a straight line (source trajectory). The
motion of the images are similar to those of the Schwarzschild
lensing. Table \ref{tbl-1} shows the Einstein radii and angular
Einstein radii for a bulge star ($D_S = 8$ kpc and $D_L = 4$ kpc
are assumed) and a star in the Large Magellanic Cloud (LMC, $D_S =
50$ kps and $D_L = 25$ kpc are assumed) for various throat radii.
The detection of a lens for which the Einstein radius is smaller
than the star radius ($\approx 10^6$ km) is very difficult because
most of the features of the gravitational lensing are smeared out
by the finite-source effect. Thus, detecting a wormhole with a
throat radius less than 1 km from the Galactic gravitational
lensing of a star is very difficult.

\section{Light curves}
The light curve of Schwarzschild lensing was derived by
\cite{Pacz86}. The same method of derivation can be used for
wormholes. The magnification of the brightness $A$ is
 \begin{eqnarray}
 A  = A_1 + A_2
      &=& \left|\frac{\hat{\theta}_1}{\hat{\beta}} \frac{d\hat{\theta}_1}{d\hat{\beta}}\right|
      + \left|\frac{\hat{\theta}_2}{\hat{\beta}} \frac{d\hat{\theta}_2}{d\hat{\beta}}\right|
      =  \left|\frac{\hat{\theta}_1}{\hat{\beta} \left(1 + \frac{2}{\hat{\theta}_1^3}\right)}\right|
      + \left|\frac{\hat{\theta}_2}{\hat{\beta} \left(1 - \frac{2}{\hat{\theta}_2^3}\right)}\right|, \\
      &=& \left|\frac{1}{\left(1 - \frac{1}{\hat{\theta}_1^3}\right) \left(1 + \frac{2}{\hat{\theta}_1^3}\right)}\right|
               + \left|\frac{1}{\left(1 + \frac{1}{\hat{\theta}_2^3}\right) \left(1 - \frac{2}{\hat{\theta}_2^3}\right)}\right|,
      \label{eqn:a}
 \end{eqnarray}
 where $A_1$ and $A_2$ are magnification of the outer and inner images,  $\hat{\theta}_1$ and $\hat{\theta}_2$ correspond to outer and inner images, respectively.
The relation between the lens and source trajectory in the sky is
shown in Figure \ref{fig3}. The time dependence of $\hat{\beta}$
is
\begin{equation}
\hat{\beta}(t) = \sqrt{\hat{\beta}_0^2 + {(t -t_0)^2/t_E}^2},
\label{eqn:hbeta}
\end{equation}
where $\hat{\beta}_0$ is the impact parameter of the source
trajectory and $t_0$ is the time of closest approach. $t_E$ is the
Einstein radius crossing time given by
\begin{equation}
t_E = R_E / v_T,  \label{eqn:te}
\end{equation}
where $v_T$ is the transverse velocity of the lens relative to the
source and observer. The light curves obtained from Equations
(\ref{eqn:a}) and (\ref{eqn:hbeta}) are shown as thick red lines
in Figure \ref{fig4}. The light curves corresponding to
Schwarzschild lensing are shown as thin green lines for
comparison. The magnifications by the Ellis wormhole are generally
less than those of Schwarzschild lensing. The light curve of the
Ellis wormhole for $\hat{\beta_o} < 1.0$ shows characteristic
gutters on both sides of the peak immediately outside the Einstein
ring crossing times ($t = t_0 \pm t_E$). The depth of the gutters
is about 4\% from the baseline. Amazingly, the star becomes
fainter than normal in terms of apparent brightness in the
gutters. This means that the Ellis wormhole lensing has off-center
divergence. In conventional gravitational lensing theory
\cite{sch92}, the convergence of light is expressed by a
convolution of the surface mass density. Thus, we need to
introduce negative mass to describe divergent lensing by the Ellis
wormhole. However, negative mass is not a physical enitity. As the
lensing by the Ellis wormhole is convergent at the center, lensing
at some other place must be divergent because the wormhole has
zero asymptotic mass. For $\hat{\beta_o} > 1.0$, the light curve
of the wormhole has a basin at $t_0$ and no peak. Using these
features, discrimination from Schwarzschild lensing can be
achieved. Equations (\ref{eqn:re}) and (\ref{eqn:te}) indicate
that physical parameters ($D_L$, $a$, and $v_T$) are degenerate in
$t_E$ and cannot be derived by fitting the light-curve data. This
situation is the same as that for Schwarzschild lensing. To obtain
or constrain these values, observations of the finite-source
effect \cite{nem94} or parallax \cite{alc95} are necessary.

The detectability of the magnification of the star brightness
depends on the timescale. The Einstein radius crossing time $t_E$
depends on the transverse velocity $v_T$. There is no reliable
estimate of $v_T$ for wormholes. Here we assume that the velocity
of the wormhole is approximately equal to the rotation velocity of
stars ($v_T = 220$ km/s) if it is bound to the Galaxy. If the
wormhole is not bound to our Galaxy, the transverse velocity would
be much higher. We assume $v_T = 5000$ km/s \cite{saf01} for the
unbound wormhole. Table 2 shows the Einstein radius crossing times
of the Ellis wormhole lensings for the Galactic bulge and LMC in
both bound and unbound scenarios. As the frequencies of current
microlensing observations are limited to once every few hours, an
event for which the timescale is less than one day is difficult to
detect. To find very long timescale events ($t_E \geq 1000$ days),
long-term monitoring of events is necessary. The realistic period
of observation is $\leq 10$ years. Thus, the realistic size of the
throat that we can search for is limited to $100$ km $\le a \le
10^7$ km both for the Galactic bulge and LMC if wormholes are
bound to our Galaxy. If wormholes are unbound, the detection is
limited to $ 10^5$ km $\le a \le 10^9$ km.

\section{Validity of the weak-field hypothesis}
First, we consider the outer image. In the previous section, we
applied the weak-field approximation to the impact parameter $b$
and the deflection angle $\alpha(r)$. As previously mentioned, the
impact parameter $b$ is written as
 \begin{equation}
 b = \sqrt{r^2 + b^2} \approx r (1 + \frac{1}{2} \frac{a^2}{r^2}).
 \end{equation}
The condition to neglect the second term is $a \ll \sqrt{2} r$. As
the image is always outside the Einstein ring,
 \begin{equation}
 a \ll \sqrt{2} R_E.
 \end{equation}
From the deflection angle, we obtain a similar relation from
Equation (\ref{eqn:defl}):
 \begin{equation}
 a \ll \sqrt{\frac{8}{5}} R_E.
 \end{equation}
The values of $a$ and $R_E$ in Table \ref{tbl-1} show that the
weak-field approximation is suitable for $a \ll 10^{11}$ km in the
Galactic microlensing. More generally, $R_E \approx D_S^{1/3}
a^{2/3}$ is derived from Equation (\ref{eqn:re}) for $D_L \approx
D_S/2$. This means that $R_E$ is much greater than $a$ if $a \ll
D_S$. Thus, the weak-field approximation is suitable if the throat
radius is negligibly small compared with the source distance. For
the inner image, the higher-order effect is expected to be greater
than that for the outer image. However, the contribution of the
inner image to the total brightness is small and decreases quickly
with $\hat{\beta}$ ($A_2/A = 0.034$ for $\hat{\beta} = 2$ and
$0.013$ for $\hat{\beta} = 3$). On the other hand, the absolute
value of the corresponding $\hat{\theta}$ does not decrease as
quickly ($\hat{\theta} = -0.618$ for $\beta = 2$ and $-0.532$ for
$\beta = 3$). Thus the contribution of the higher order effect of
the second image to the total brightness is expected to be small.

Another possibility of deviation from the weak-field approximation
is the contribution of relativistic images. Recently,
gravitational lensing in the strong-field limit \cite{vir00} has
been studied for lensing by black holes. In this limit, light rays
are strongly bent and wound close to the photon sphere. As a
result, a number of relativistic images appear around the photon
sphere. However, it has been shown that there is no photon sphere
\cite{dey08} in Ellis wormhole lensing. Therefore, there is no
contribution of relativistic images to the magnification in Ellis
wormhole lensing. We thus conclude that the weak-field hypothesis
is a good approximation unless the throat radius is comparable to
the galactic distance.

\section{Optical depth and event rate}
The probability of a microlensing event to occur for a star is
expressed by the optical depth $\tau$:
\begin{equation}
\tau = \pi \int_0^{D_S} n(D_L) R_E^2 d D_L,
\end{equation}
where $n(D_L)$ is the number density of wormholes as a function of
the line of sight. Here we simply assume that $n(D_L)$ is constant
($n(D_L) = n$):
\begin{eqnarray}
\tau &=& \pi n \int_0^{D_S}  \frac{\pi}{4} \left[ \frac{D_L (D_S - D_L)}{D_S} a^2 \right] ^{2/3}d D_L, \\
       &=& \sqrt[3]{\frac{\pi^5}{2^4}} n a^{4/3} D_S^{5/3} \int_0^1 \left[x ( 1- x)\right]^{2/3} dx, \\
       &\approx& 0.785 n a^{4/3} D_S^{5/3}.
\end{eqnarray}
The event rate expected for a source star $\Gamma$ is calculated
as
\begin{eqnarray}
\Gamma &=&  2 \int_0^{D_S} n(D_L) R_E v_T dD_L, \\
                &=& \sqrt[3]{2\pi} n v_T D_S^{4/3} a^{2/3} \int_0^1 \sqrt[3]{x (1 - x)} dx, \\
                &\approx& 0.978 n v_T a^{2/3} D_S^{4/3}. \label{eqn:gm3}
\end{eqnarray}

There is no reliable prediction of the number density of
wormholes. Several authors \cite{kra00, lob09} have speculated
that wormholes are very common in the universe, at least as
abundant as stars. Even if we accept such speculation, there are
still large uncertainties in the value of $n$ because the
distribution of wormholes is not specified. Here, we introduce two
possibilities. One is that wormholes are bound to the Galaxy and
the number density is approximately equal to the local stellar
density. The other possibility is that wormholes are not bound to
the Galaxy and are approximately uniformly distributed throughout
the universe. For the bound hypothesis, we use $n =
\rho_{Ls}/\langle M_{star} \rangle = 0.147$ pc$^{-3}$, where
$\rho_{Ls}$ is the local stellar density in the solar
neighborhood, $\rho_{Ls} = 0.044 \, M_\odot$ pc$^{-3}$, and
$\langle M_{star} \rangle$ is the average mass of stars. We use
$\langle M_{star} \rangle = 0.3 \, M_\odot$, a typical mass of an
M dwarf; {\it i.e.}, the dominant stellar component in the Galaxy.
For the unbound hypothesis, we assumed that the number density of
the wormholes is the same as the average stellar density of the
universe. The stellar density of the universe is estimated
assuming that the fraction of  baryonic matter accounted for by
star is the same as that of the solar neighborhood. Then we obtain
$n = \rho_c \Omega_b \rho_{Ls} / (\rho_{Lb} \langle M_{star}
\rangle) = 4.97 \times 10^{-9}$ pc$^{-3}$, where $\rho_c = 1.48
\times 10^{-7} \, M_\odot$ p$^{-3}$ is the critical density,
$\Omega_b = 0.042$ is the baryon density of the universe divided
by the critical density, and $\rho_{Ls} = 0.044 \, M_\odot$
pc$^{-3}$ and $\rho_{Lb} = 0.18 \, M_\odot$ pc$^{-3}$ are the
local star and local baryon densities, respectively.

Using these values, we calculated the optical depths and event
rates for bulge and LMC lensings. Table \ref{tbl-3} presents the
results for the bulge lensings. In an ordinary Schwarzschild
microlensing survey, observations are made of more than 10 million
stars. Thus, we can expect approximately $10^7 \Gamma$ events in a
year. However, the situation is different in a wormhole search. As
mentioned previously, the magnification of wormhole lensing is
less than that of Schwarzschild lensing, and a remarkable feature
of wormhole lensing is the decreasing brightness around the
Einstein radius crossing times. Past microlensing surveys have
mainly searched for stars that increase in brightness. The stars
monitored are those with magnitudes down to the limiting magnitude
or less. However, we need to find stars that decrease in
brightness in the wormhole search. To do so, we need to watch
brighter stars. Therefore, far fewer stars can be monitored than
in an ordinary microlensing survey. Furthermore, the detection
efficiency of the wormhole is thought to be less than that for
Schwarzschild lensing because of the low magnification. Here we
assume that the effective number of stars monitored to find a
wormhole is $10^6$. To expect more than one event in a survey of
several years, $\Gamma$ must be greater than $\sim 10^{-6}$. The
values in Table \ref{tbl-3} indicate that the detection of
wormholes with $a > 10^4$ km is expected in the microlensing
survey of the Galactic bulge in the case of the bound model. The
results of the optical depths and the event rates for LMC lensing
are presented in Table \ref{tbl-4}. On the basis of the same
discussion for bulge lensing, we expect $\Gamma > 10^{-6}$ to find
a wormhole. The event rates expected for LMC lensing are greater
than those for bulge lensing. We expect detection of a wormhole
event if $a > 10^2$ km for the bound model. If no candidate is
found, we can set upper limits of $\Gamma$ and/or $\tau$ as
functions of $t_E$. To convert these values to physical parameters
($n$ and $a$) requires the distribution of $v_T$. Right now, there
is no reliable model of the distribution except for using the
bound or unbound hypothesis. On the other hand, the event rates
for the unbound model are too small for the events to be detected.

In past microlensing surveys \cite{alc00, tis07, wyr09, sum03},
large amounts of data have already been collected for both the
bulge and LMC fields. Monitoring more than $10^6$ stars for about
10 years can be achieved by simply reanalyzing the past data.
Thus, discovery of wormholes can be expected if their population
density is as high as the local stellar density and $10^2 \leq a
\leq 10^7$ km. Such wormholes of astronomical size are large
enough for humans to pass through. Thus, they would be of interest
to people discussing the possibility of space--time travel. If no
candidate is found, the possibility of a rich population of
large-throat wormholes bound to the Galaxy can be ruled out. Such
a limit, however, may not affect existing wormhole theories
because there is no prediction of the abundance. However,
theoretical studies on wormholes are still in progress. The limit
imposed by observation is expected to affect future wormhole
theories. On the other hand, the discovery of unbound wormholes is
very difficult even if their population density is comparable to
that of ordinary stars. To discover such wormholes, the monitoring
of a much larger number of stars in distant galaxies would be
necessary. For example, $\Gamma =  1.7 \times 10^{-6}$ and $t_E
\approx 380$ days for the M101 microlensing survey ($D_S = 7.4$
Mpc) if the throat radius is $10^7$ km. To carry out such a
microlensing survey, observation from space is necessary because
the resolving of a large number of stars in a distant galaxy is
impossible through ground observations.

Only the Ellis wormhole has been discussed in this paper. There
are several other types of wormholes \cite{sh04, nan06, rah07} for
which deflection angles have been derived. These wormholes are
expected to have different light curves. To detect those
wormholes, calculations of their light curves are necessary. The
method used in this paper can be employed only when we know the
analytic solutions of the image positions. If no analytic solution
is found, the calculation must be made numerically.

\section{Summary}
The gravitational lensing of the Ellis wormhole is solved in the
weak-field limit. The image positions are calculated as real
solutions of simple cubic formulas. One image appears on the
source star side and outside the Einstein ring. The other image
appears on the other side and inside the Einstein ring. A simple
estimation shows that the weak-field hypothesis is a good
approximation for Galactic microlensing if the throat radius is
less than $10^{11}$ km. The light curve derived has characteristic
gutters immediately outside the Einstein ring crossing times.
Optical depths and event rates for bulge and LMC lensings are
calculated for simple bound and unbound hypotheses. The results
show that the bound wormholes can be detected by reanalyzing past
data if the throat radius is between $10^2$ and $10^7$ km and the
number density is approximately equal to the local stellar
density. If the wormholes are unbound and approximately uniformly
distributed in the universe with average stellar density,
detection of the wormholes is impossible using past microlensing
data. To detect unbound wormholes, a microlensing survey of
distant galaxies from space is necessary.

\bigskip
We would like to thank Professor Hideki Asada of Hirosaki
University, Professor Matt Visser of Victoria University, and
Professor Tomohiro Harada of Rikkyo University for discussions and
their suggestions concerning this study. We thank Professor Philip
Yock of Auckland University for polishing our manuscript. Finally,
we would like to thank an anonymous referee who pointed out the
existence of a second image.

\appendix

\begin{figure}[htb]
\begin{center}
%\epsscale{.90}
\epsfig{file=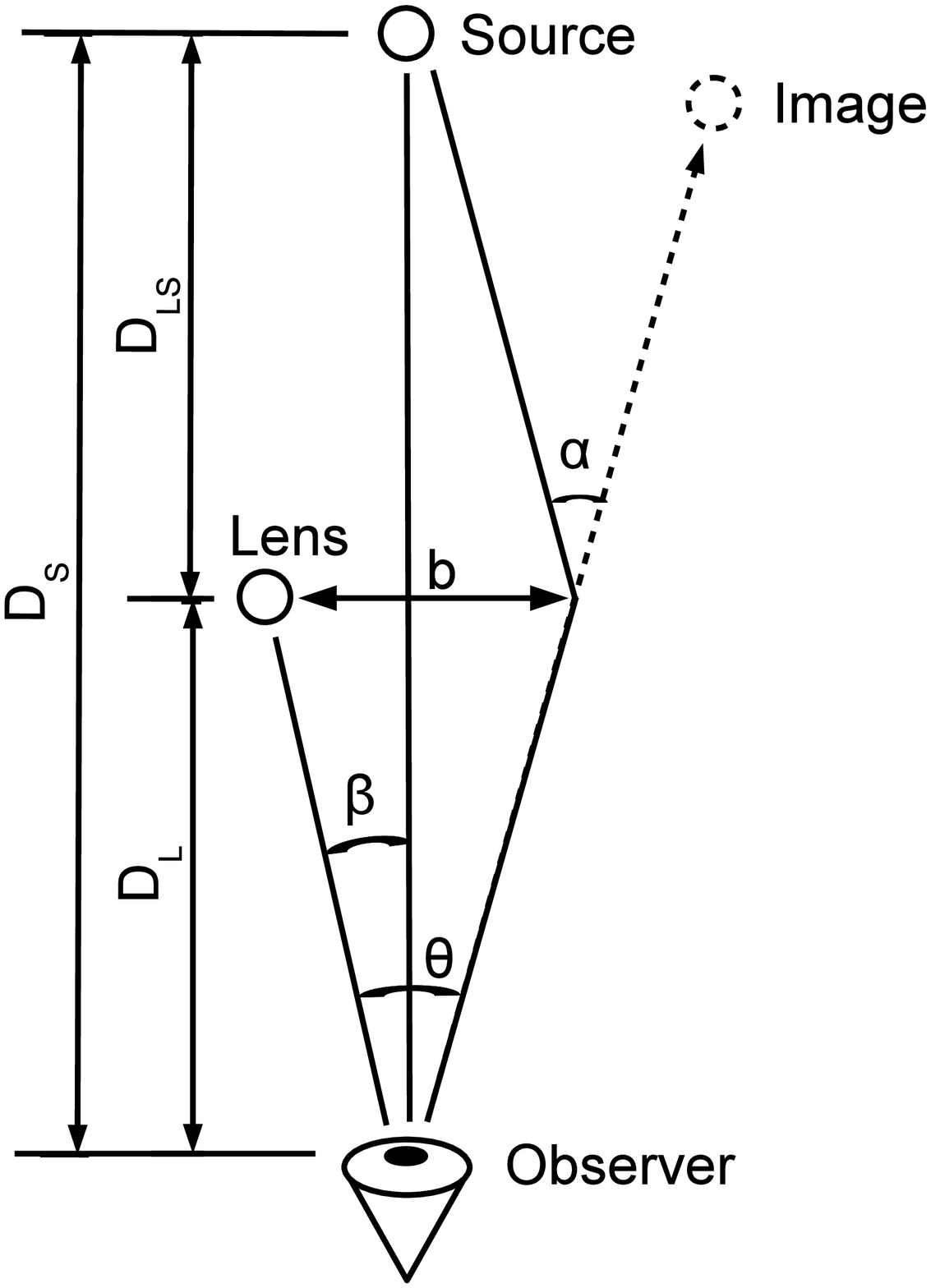,height=7.0in} \caption{Sketch of the
relation between the source star, lens (wormhole), and observer.
\label{fig1}}
\end{center}
\end{figure}

\begin{figure}
\includegraphics[angle=0,scale=1.0]{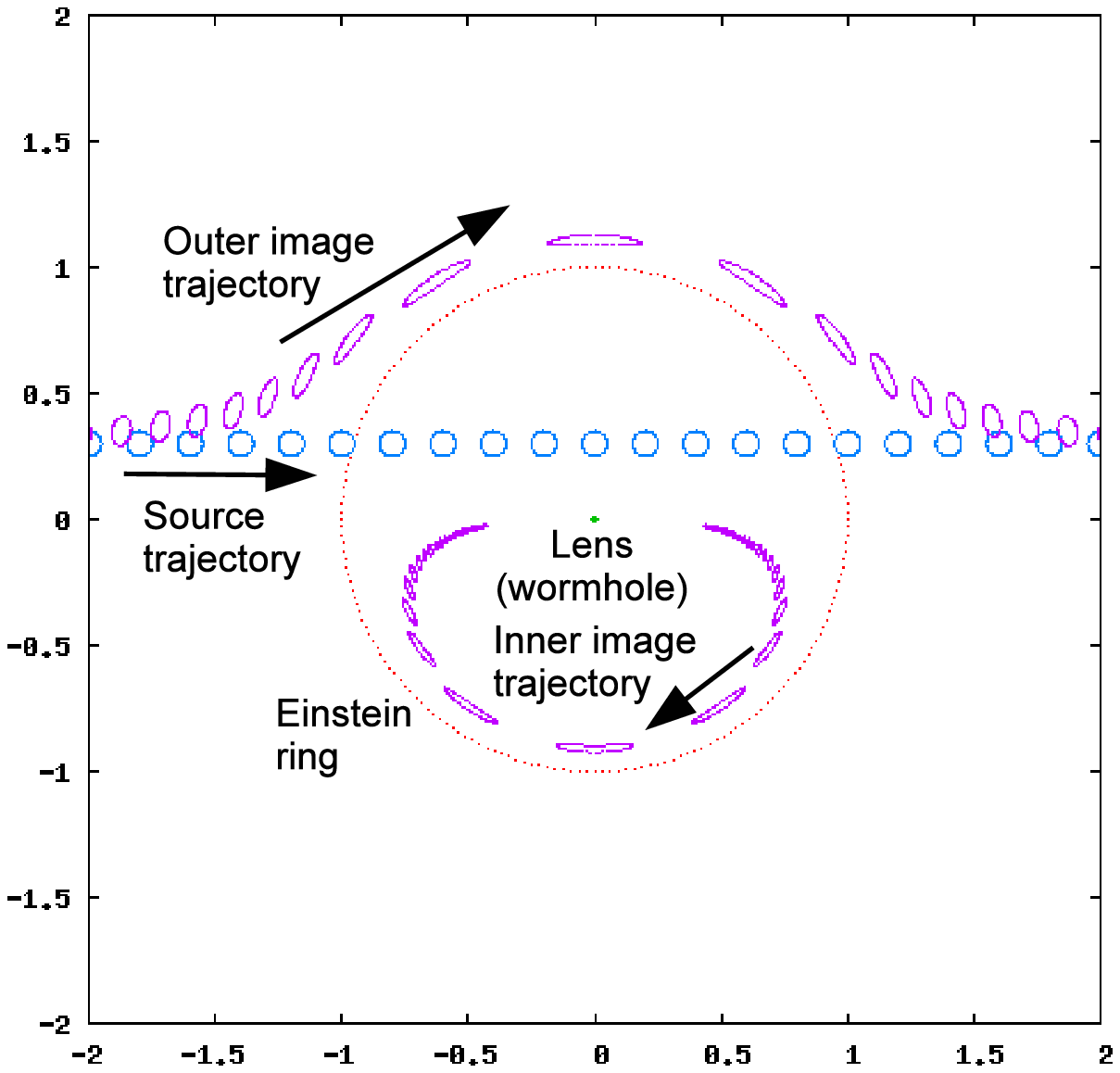}
\caption{Source and image trajectories in the sky from the
position of the observer.\label{fig2}}
\end{figure}

\begin{figure}
\includegraphics[angle=0,scale=0.8]{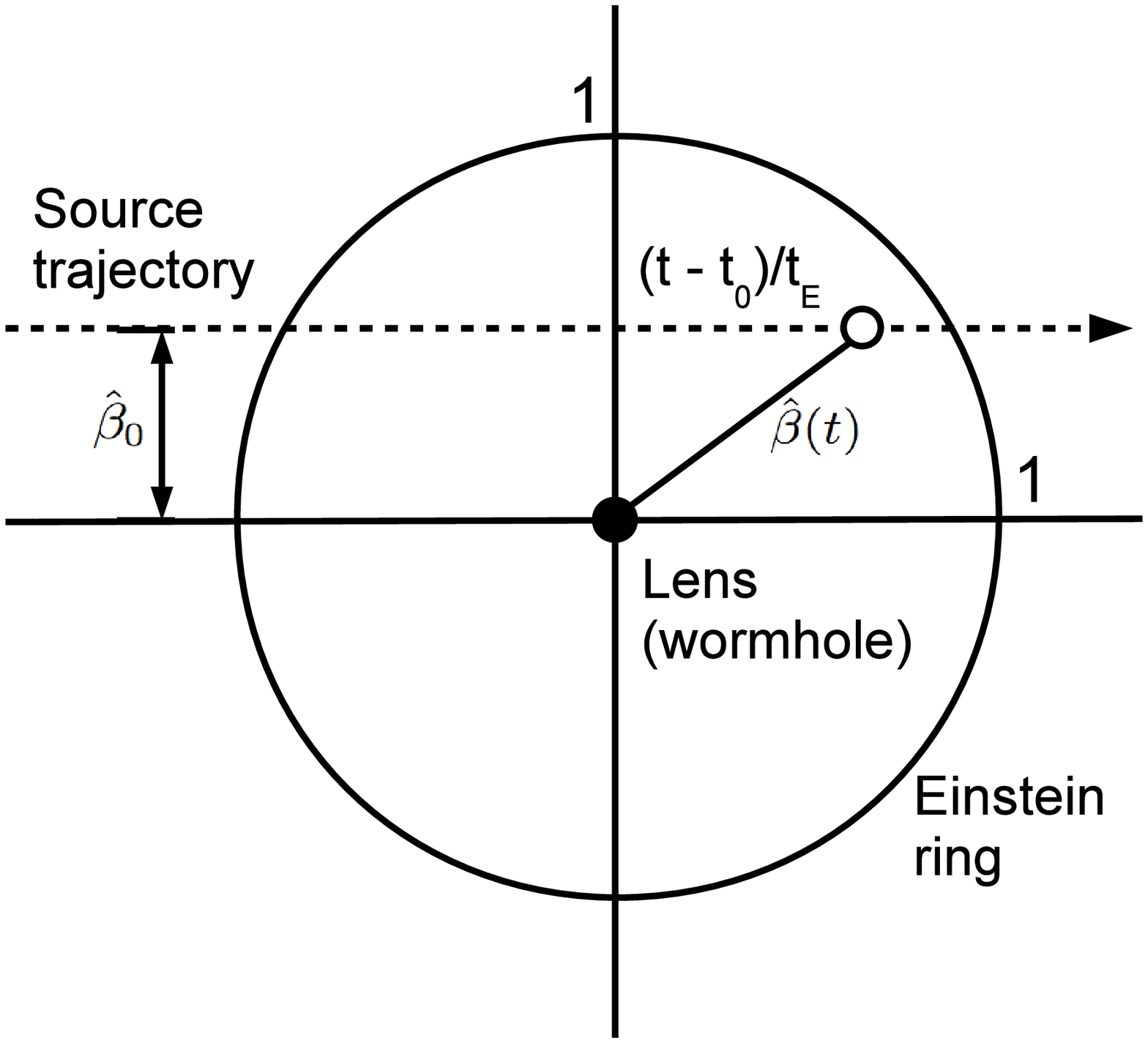}
\caption{Sketch of the relation between the source trajectory and
the lens (wormhole) in the sky. All quantities are normalized by
the angular Einstein radius $\theta_E$.\label{fig3}}
\end{figure}

\begin{figure}
\includegraphics[angle=0,scale=0.45]{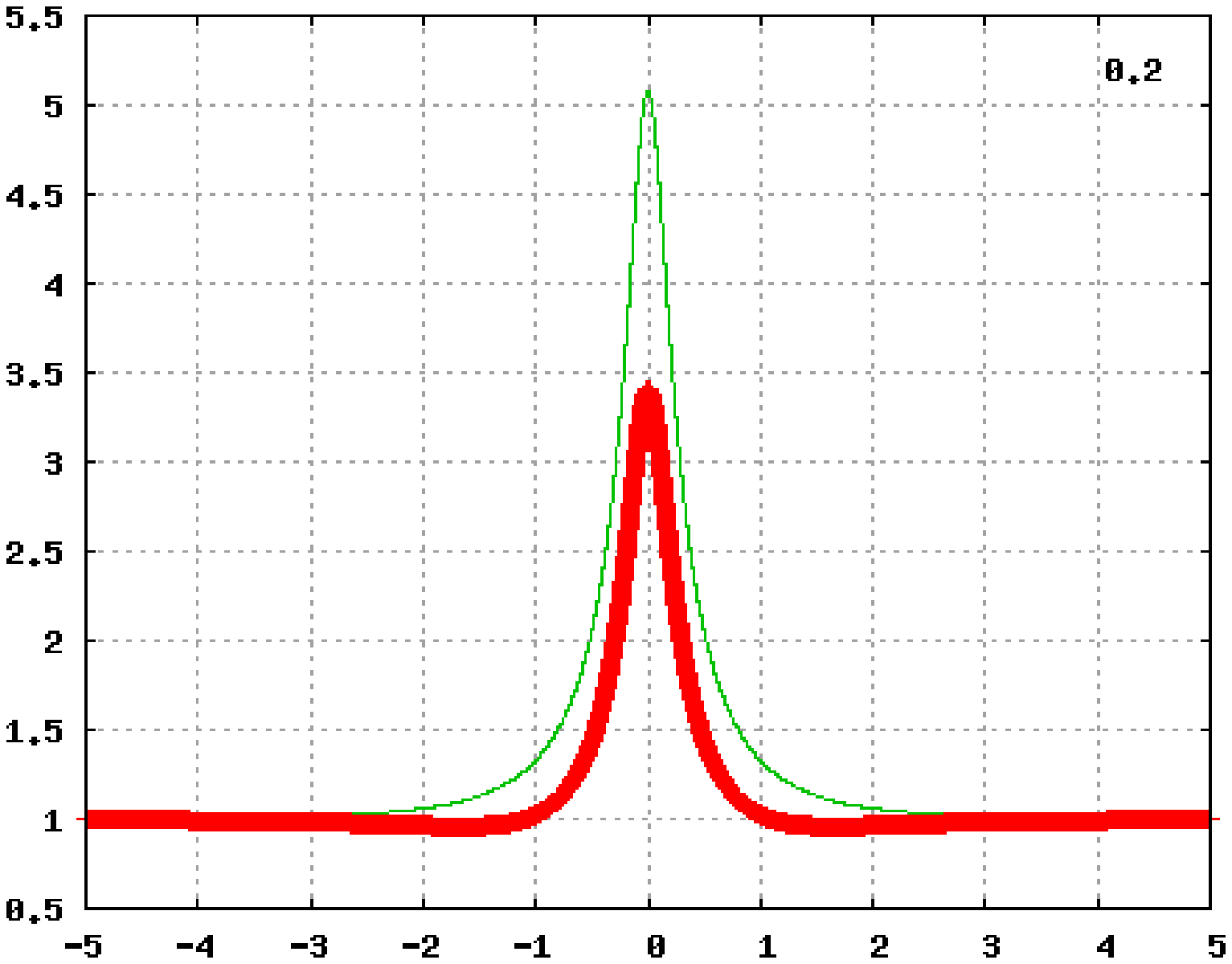}
\includegraphics[angle=0,scale=0.45]{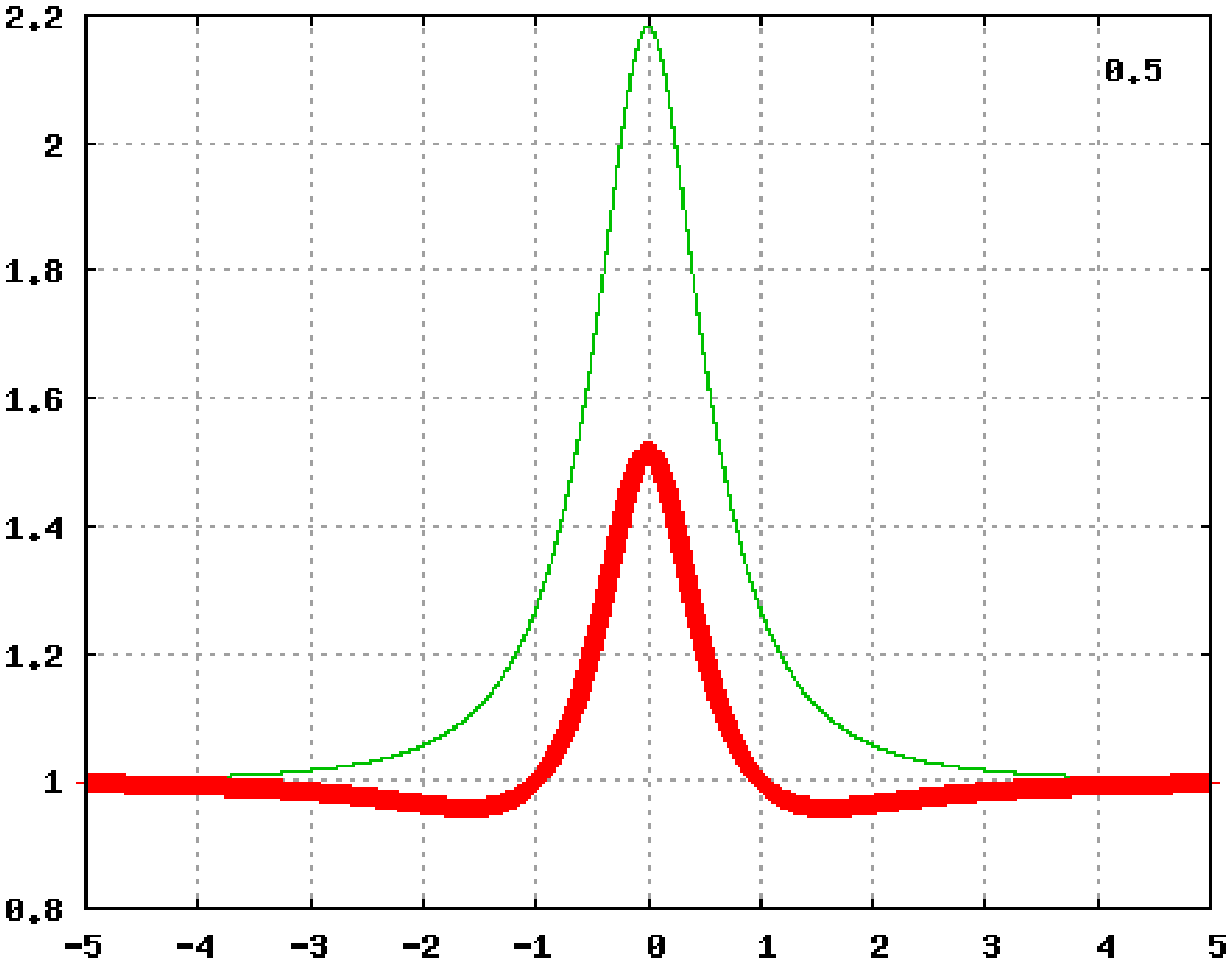}
\includegraphics[angle=0,scale=0.45]{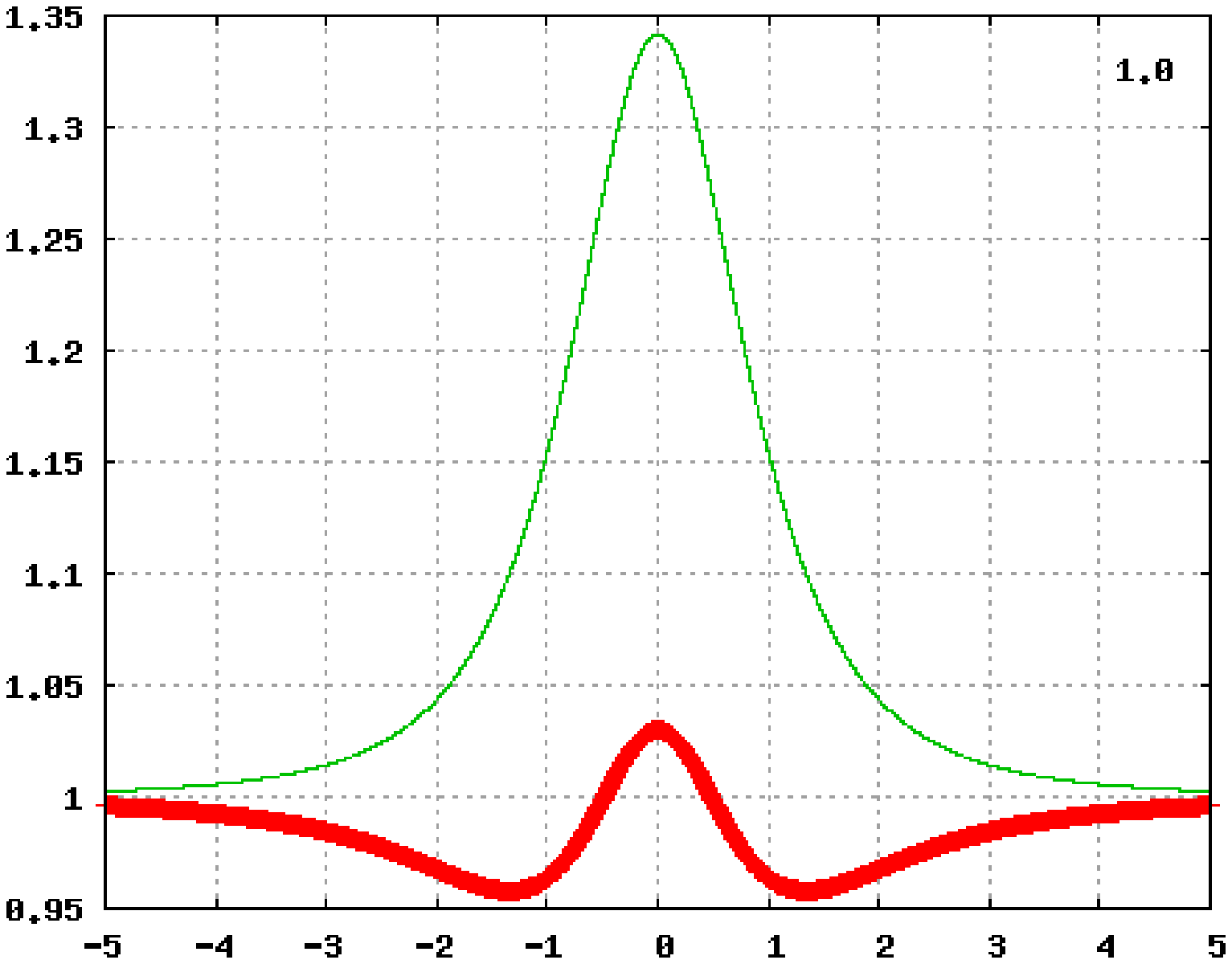}
\includegraphics[angle=0,scale=0.45]{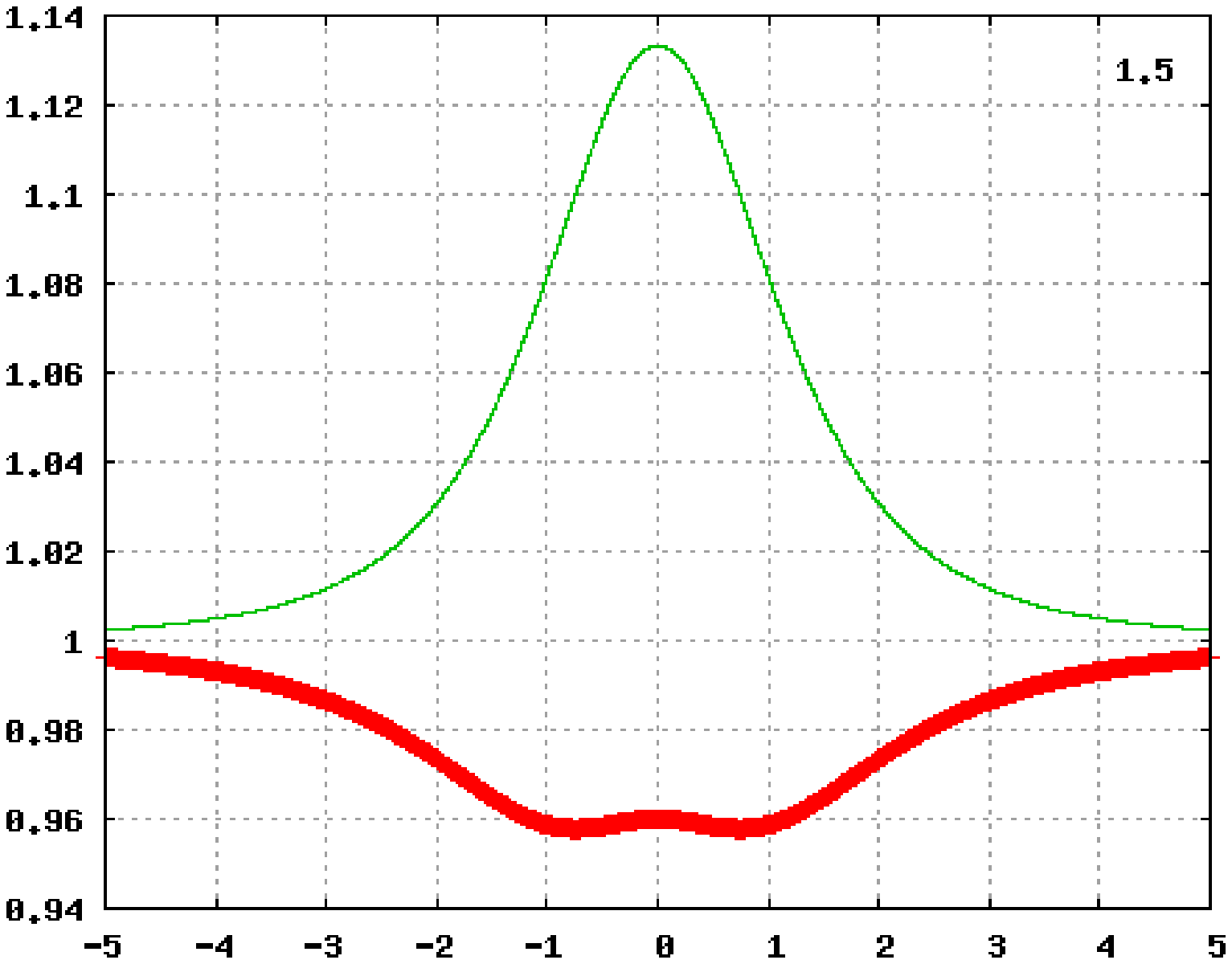}
\caption{Light curves for $\hat{\beta}_0 = 0.2$ (top left),
$\hat{\beta}_0 = 0.5 $ (top right), $\hat{\beta}_0 = 1.0$ (bottom
left), and $\hat{\beta}_0 = 1.5$ (bottom right). Thick red lines
are the light curves for wormholes. Thin green lines are
corresponding light curves for Schwarzschild lenses.\label{fig4}}
\end{figure}

\begin{center}
\begin{table}
\caption{Einstein radii for bulge and LMC lensings\label{tbl-1}}
\begin{tabular}{rrrrrrr}
\hline \hline
  & & \multicolumn{2}{c}{Bulge} & &  \multicolumn{2}{c}{LMC} \\ \cline{3-4} \cline{6-7}
  $a$ (km) & &  $R_E$ (km) & $\theta_E$ (mas) & & $R_E$ (km) & $\theta_E$ (mas) \\
 \hline
1            & & $3.64 \times 10^5$ & 0.001 & & $6.71 \times 10^5$  & $< 0.001$ \\
10          &  & $1.69 \times 10^6$ & 0.003 & & $3.12 \times 10^6$ & 0.001 \\
$10^2$ &  & $7.85 \times 10^6$ & 0.013 & & $1.45 \times 10^7$ & 0.004 \\
$10^3$ &  & $3.64 \times 10^7$ & 0.061 & & $6.71 \times 10^7$ & 0.018 \\
$10^4$ &  & $1.69 \times 10^8$ & 0.283 & & $3.12 \times 10^8$ & 0.083  \\
$10^5$ &  & $7.85 \times 10^8$ & 1.31    & & $1.45 \times 10^9$ & 0.387  \\
$10^6$ &  & $3.64 \times 10^9$ & 6.10    & &  $6.71 \times 10^9$ &  1.80 \\
$10^7$ &  & $1.69 \times 10^{10}$ & 28.3    & &  $3.12 \times 10^{10}$ &  8.35 \\
$10^8$ &  & $7.85 \times 10^{10}$ & 131    & &  $1.45 \times 10^{11}$ &  38.7 \\
$10^9$ &  & $3.64 \times 10^{11}$ & 610    & &  $6.71 \times 10^{11}$  &  180 \\
$10^{10}$ & & $1.69 \times 10^{12}$ & 2 832    & &  $3.12 \times 10^{12}$ &  835 \\
$10^{11}$ & & $7.85 \times 10^{12}$ & 13 143    & &  $1.45 \times 10^{13}$ &  3 874 \\
%% Text for table notes should follow after the \enddata but before
%% the \end{deluxetable}. Make sure there is at least one \tablenotemark
%% in the table for each \tablenotetext.
\hline \hline
\end{tabular}
%\tablecomments{$a$ is the throat radius of the wormhole, $R_E$ is the Einstein radius, and $\theta_E$ is the angular Einstein radius.}
%\tablenotetext{a}{$D_S = 8 kpc$ and $D_L = 4 kpc$ are assumed. }
%\tablenotetext{b}{$D_S = 50 kpc$ and $D_L = 25 kpc$ are assumed. }
\end{table}
\end{center}

\begin{center}
\begin{table}
\caption{Einstein radius crossing times for bulge and LMC
lensings\label{tbl-2}}
\begin{tabular}{rrrrrrrrr}
\hline \hline
  & & \multicolumn{2}{c}{Bulge} & & \multicolumn{2}{c}{LMC} \\
  $a$ (km) & & \multicolumn{2}{c}{$t_E$ (day)} & & \multicolumn{2}{c}{$t_E$ (day)}  \\ \cline{3-4} \cline{6-7}
 & & Bound & Unbound & & Bound & Unbound\\
 \hline
1            & & 0.019 & 0.001 & & 0.035 & 0.002   \\
10          & & 0.089 & 0.004 & & 0.164 & 0.007  \\
$10^2$ & & 0.413 & 0.018 & & 0.761 & 0.033  \\
$10^3$ & & 1.92   & 0.084 & & 3.53   & 0.155  \\
$10^4$ & & 8.90    & 0.392 & &16.4   & 0.721  \\
$10^5$ & & 41.3    & 1.82   & & 76.1   & 3.35  \\
$10^6$ & & 192    & 8.44    & & 353   & 15.5   \\
$10^7$ & & 890    & 39.2  & & 1 639   & 72.1 \\
$10^8$ & & 4 130    & 182    & & 7 608   & 335  \\
$10^9$ & & $> 10^4$    & 843    & & $> 10^4 $  & 1 553  \\
$10^{10}$ & & $> 10^4 $  & 3915    & & $> 10^4$  & 7 212 \\
%% Text for table notes should follow after the \enddata but before
%% the \end{deluxetable}. Make sure there is at least one \tablenotemark
%% in the table for each \tablenotetext.
\hline \hline
\end{tabular}
%\tablecomments{$a$ is the throat radius of the wormhole, $t_E$ is the Einstein radius crossing time.}
%\tablenotetext{a}{$D_S = 8 kpc$ and $D_L = 4 kpc$ are assumed. }
%\tablenotetext{b}{$D_S = 50 kpc$ and $D_L = 25 kpc$ are assumed. }
%\tablenotetext{c}{$v_T = 220 km/s$ is assumed.}
%\tablenotetext{d}{$v_T = 5000 km/s$ is assumed.}
\end{table}
\end{center}

\begin{center}
\begin{table}
\caption{Optical depths and event rates for bulge
lensing\label{tbl-3}}
\begin{tabular}{rrrrrrrrrrrrrrrrr}
\hline \hline
 & & \multicolumn{3}{c}{Bound} & & \multicolumn{3}{c}{Unbound} \\ \cline{3-5} \cline{7-9}
  $a$ (km) & & \multicolumn{1}{c}{$\tau $} & & $\Gamma$ (1/year) & &   \multicolumn{1}{c}{$\tau $} & & $\Gamma$ (1/year) \\
 \hline
 $10$  & & $8.24 \times 10^{-12}$ & & $2.45 \times 10^{-8}$ & & $2.78 \times 10^{-19} $ & & $1.88 \times 10^{-14}$ \\
 $10^2$  & & $1.77 \times 10^{-10}$ & & $1.14 \times 10^{-7}$ & & $6.00 \times 10^{-18} $ & & $8.73 \times 10^{-14}$ \\
 $10^3$  & & $3.82 \times 10^{-9}$ & & $5.27 \times 10^{-7}$ & & $1.29 \times 10^{-16} $ & & $4.05 \times 10^{-13}$ \\
 $10^4$  & & $8.24 \times 10^{-8}$ & & $2.45 \times 10^{-6}$ & & $2.78 \times 10^{-15} $ & & $1.88 \times 10^{-12}$ \\
 $10^5$  & & $1.77 \times 10^{-6}$ & & $1.14 \times 10^{-5}$ & & $6.00 \times 10^{-14} $ & & $8.73 \times 10^{-12}$ \\
 $10^6$  & & $3.82 \times 10^{-5}$ & & $5.27 \times 10^{-5}$ & & $1.29 \times 10^{-12} $ & & $4.05 \times 10^{-11}$ \\
 $10^7$  & & $8.24 \times 10^{-4}$ & & $2.45 \times 10^{-4}$ & & $2.78 \times 10^{-11} $ & & $1.88 \times 10^{-10}$ \\
 $10^8$  & & $1.77 \times 10^{-2}$ & & $1.14 \times 10^{-3}$ & & $6.00 \times 10^{-10} $ & & $8.73 \times 10^{-10}$ \\
 $10^9$  & & $3.82 \times 10^{-1}$ & & $5.27 \times 10^{-3}$ & & $1.29 \times 10^{-8} $ & & $4.05 \times 10^{-9}$ \\
 $10^{10}$  & & $8.24 $ & & $2.45 \times 10^{-2}$ & & $2.78 \times 10^{-7} $ & & $1.88 \times 10^{-8}$ \\

%% Text for table notes should follow after the \enddata but before
%% the \end{deluxetable}. Make sure there is at least one \tablenotemark
%% in the table for each \tablenotetext.
\hline \hline
\end{tabular}
%\tablecomments{$a$ is the throat radius of the wormhole, $\tau$ is the optical depth, $\Gamma$ is the event rate. $D_S = 8 kpc$ is assumed. }
%\tablenotetext{a}{$v_T = 220 km/s$ and $n = 0.147 pc^{-3}$ are assumed.}
%\tablenotetext{b}{$v_T = 5000 km/s$ and $n = 4.97 \times 10^{-9} pc^{-3}$ are assumed.}
\end{table}
\end{center}

\begin{center}
\begin{table}
\caption{Optical depths and event rates for LMC
lensing\label{tbl-4}}
\begin{tabular}{rrrrrrrrrrrrrrrrr}
\hline \hline
 & & \multicolumn{3}{c}{Bound} & & \multicolumn{3}{c}{Unbound} \\ \cline{3-5} \cline{7-9}
  $a$ (km) & & \multicolumn{1}{c}{$\tau $} & & $\Gamma$ (1/year) & &   \multicolumn{1}{c}{$\tau $} & & $\Gamma$ (1/year) \\
 \hline
 $10$  & & $1.75 \times 10^{-10}$ & & $2.82 \times 10^{-7}$ & & $5.90 \times 10^{-18} $ & & $2.17 \times 10^{-13}$ \\
 $10^2$  & & $3.76 \times 10^{-9}$ & & $1.31 \times 10^{-6}$ & & $1.27 \times 10^{-16} $ & & $1.01 \times 10^{-12}$ \\
 $10^3$  & & $8.11 \times 10^{-8}$ & & $6.07 \times 10^{-6}$ & & $2.74 \times 10^{-15} $ & & $4.67 \times 10^{-12}$ \\
 $10^4$  & & $1.75 \times 10^{-6}$ & & $2.82 \times 10^{-5}$ & & $5.90 \times 10^{-14} $ & & $2.17 \times 10^{-11}$ \\
 $10^5$  & & $3.76 \times 10^{-5}$ & & $1.31 \times 10^{-4}$ & & $1.27 \times 10^{-12} $ & & $1.01 \times 10^{-10}$ \\
 $10^6$  & & $8.11 \times 10^{-4}$ & & $6.07 \times 10^{-4}$ & & $2.74 \times 10^{-11} $ & & $4.67 \times 10^{-10}$ \\
 $10^7$  & & $1.75 \times 10^{-2}$ & & $2.82 \times 10^{-3}$ & & $5.90 \times 10^{-10} $ & & $2.17 \times 10^{-9}$ \\
 $10^8$  & & $3.76 \times 10^{-1}$ & & $1.31 \times 10^{-2}$ & & $1.27 \times 10^{-8} $ & & $1.01 \times 10^{-8}$ \\
 $10^9$  & & $8.11 $ & & $6.07 \times 10^{-2}$ & & $2.74 \times 10^{-7} $ & & $4.67 \times 10^{-8}$ \\
 $10^{10}$  & & $175 $ & & $2.82 \times 10^{-1}$ & & $5.90 \times 10^{-6} $ & & $2.17 \times 10^{-7}$ \\

%% Text for table notes should follow after the \enddata but before
%% the \end{deluxetable}. Make sure there is at least one \tablenotemark
%% in the table for each \tablenotetext.
\hline \hline
\end{tabular}
%\tablecomments{$a$ is the throat radius of the wormhole, $\tau$ is the optical depth, $\Gamma$ is the event rate. $D_S = 8 kpc$ is assumed. }
%\tablenotetext{a}{$v_T = 220 km/s$ and $n = 0.147 pc^{-3}$ are assumed.}
%\tablenotetext{b}{$v_T = 5000 km/s$ and $n = 4.97 \times 10^{-9} pc^{-3}$ are assumed.}
\end{table}
\end{center}

\Chapter{The deflection of light ray in strong field: \textit{a
material medium approach}}
            {The deflection of light ray in strong field}{Asoke Kumar Sen}
\bigskip\bigskip

\addcontentsline{toc}{chapter}{{\it Asoke Kumar Sen}}
\label{polingStart}

\begin{raggedright}
{\it Asoke Kumar Sen\index{author}{Sen, Asoke Kumar}\\
Department of Physics\\
Assam University\\
India}
\bigskip\bigskip
\end{raggedright}

\section{Introduction}
The first correct expression for gravitational deflection of light
due to a gravitational mass (M) was derived by Einstein in 1915,
with an expression for bending  given by $ 4GM/(c^2
r_{\bigodot})$, where $r_{\bigodot}$ is the closet distance of
approach, which is approximately the solar radius. The exact
amount of deflection for a ray of light can be worked out from the
null geodesic, which a ray of light follows [1,2,3,4]. The
deflection of a light ray passing close to a gravitational mass
can be alternately calculated by following an approach, where the
effect of gravitation on the light ray is estimated by considering
the light ray to be passing through a \textit{material
medium}[5-9]. The value of the refractive index of that medium in
this case  is decided by the strength of gravitational field [10].

Fischbach and Freeman [7], derived the effective refractive index
of the material medium and calculated the second order
contribution to the gravitational deflection. Sereno [8] had also
used the material medium approach for gravitational lensing
calculations  by drawing trajectory of light ray by Fermat's
principle.  More recently Ye and Lin [9], using similar approach
calculated the gravitational time delay and the effect of lensing.
On the other hand, the calculation of higher order deflection
terms, due to Schwarzschild Black hole, from the null geodesic,
have been performed by various authors [11-14].

With the above background, in the present work,  the {\it material
medium} approach is followed, to calculate a more exact expression
for the deflection term due to a non-rotating sphere
(Schwarzschild geometry), without any weak field approximation.

\section{The trajectory of a light ray described by material medium}

The Schwarzschild  equation  expressed in  isotropic form is given
by[10]

\begin{equation}
ds^2= (\frac{1-r_g/(4\rho)}{1+r_g/(4\rho)})^2 c^2 dt^2 - (1+
\frac{r_g}{4 \rho})^4(d\rho ^2 + \rho ^2 ( sin ^2 \theta  d \phi ^
2 + d \theta ^ 2))
\end{equation}

By setting $ds=0$,  the  velocity of light from above can be
identified  as :

\begin{equation}
v( \rho ) = \frac {(1-\frac{r_g}{4\rho })  c }{(1+ \frac {r_g}{4
\rho})^3}
\end{equation}

and this results in an effective value of refractive index
$n(r)=\frac{c}{v(r)}=\frac{r}{r-r_g} $, where $r_g= 2GM/c^2$ is
Schwarzschild radius.  Here the trajectory of the light ray and
gravitational mass together define a plane and the equation of
light ray in polar co-ordinate system can be written as[15]:

\begin{equation}
\theta = A. \int^{\infty}_{r_{\bigodot}}{\frac {dr}{r
\sqrt{n^2r^2-A^2}}}
\end{equation}

The trajectory is such that $n(r).d$ always remains a $constant$,
where $d$ is the perpendicular distance between the trajectory of
the light ray from the origin and the \textit{constant} is taken
here as $A$ [15]. From geometry we identify $A=
n(r_{\bigodot})r_{\bigodot}$ and the value of deflection
($\triangle \phi$), can be accordingly written as :

\begin{equation}
\triangle \phi = 2 \int^{\infty}_{r_{\bigodot}} { \frac {dr}{r
\sqrt{(\frac{n(r).r}{n(r_{\bigodot}).r_{\bigodot}})^2-1}}} - \pi
\end{equation}

After certain mathematical steps we can write:

\begin{equation}
\triangle \phi = 2 D \int^{a}_{0} { \frac {x dx}{\sqrt{
1-D^2x^2(1-x)^2}}}
\end{equation}

where $D =\frac{r_{\bigodot} ^2}{r_g( r_{\bigodot} - r_g)}$ and
$a=r_g/r_{\bigodot}$.

Using Mathematica to perform integration,   we can write the final
expression for gravitational deflection  as :

\begin{equation}
\triangle \phi
=4\left\{\frac{(\sqrt{D}+\sqrt{D-4})E-(2\sqrt{D-4})F}{(\sqrt{D+4}-\sqrt{D-4})}\right\}^{x=a}_{x=0}
\end{equation}

where $E\equiv E(p,q^2)$ is the Elliptic Integral of first kind
and $F\equiv F(-q,p,q^2)$ is Incomplete Elliptic Integral of Third
kind. The arguments p,$q^2$,-q,p,$q^2$ are expressed by the
following mathematical relations:

\begin{eqnarray}
p=&& \arcsin {\sqrt{\frac
{(\sqrt{D-4}-\sqrt{D+4})(\sqrt{D-4}+(2x-1)\sqrt{D})}
{(\sqrt{D-4}+\sqrt{D+4})(\sqrt{D-4}-(2x-1)\sqrt{D} )}}} \nonumber\\
\end{eqnarray}

\begin{eqnarray}
q=&&\frac{(\sqrt{D-4}+\sqrt{D+4})}{(\sqrt{D-4}-\sqrt{D+4})} \nonumber\\
\end{eqnarray}

This mathematical expression for deflection, in strong filed is
claimed to be more exact than all other expressions derived so far
using \textit{material medium} approach. It has been tested under
various boundary conditions.

\Chapter{Rapidly Rotating Lenses: Repeating orbital motion
features in close binary microlenses}
            {Rapidly Rotating Lenses}{Penny et al.}
\bigskip\bigskip

\addcontentsline{toc}{chapter}{{\it Penny et al.}}
\label{polingStart}

\begin{raggedright}

{\it Matthew T. Penny\index{author}{Penny, Matthew T.}, Eamonn Kerins\index{author}{Kerins, Eamonn}\\
Jodrell Bank Centre for Astrophysics, University of Manchester,
UK\\}
\bigskip
{\it Shude Mao\index{author}{Mao, Shude}\\
Jodrell Bank Centre for Astrophysics, University of Manchester,
UK\\ National Astronomical Observatories, Chinese Academy of
Sciences \\
China}

\bigskip
\end{raggedright}

Most binary lenses detected in microlensing events complete only a
small fraction of their orbits during the event, as binaries with
shorter period orbits tend not to show detectable binary
signatures. However, we show that there exist some lenses (rapidly
rotating lenses or RRLs) with detectable binary signatures that
can complete several orbits while their binary lens features are
detectable, and hence the same features can be seen to repeat in a
lightcurve. Using simple analytical constraints we show there
exist regions of the total lens mass-semimajor axis plane where
repeating features are detectable, and confirm this with numerical
calculations and example lightcurves. With a photometric precision
of $1$~percent, the region of detectability covers total masses
$\sim 0.1 M_{\odot}$ and above, and semimajor axes $\sim
0.2$--$2$~AU for typical stellar binary lenses. The detectability
increases with decreasing lens-source velocities and increasing
lens distances. We also present a method of modelling microlensing
events with repeating features, by timing the occurrence of
lightcurve features, which reduces the need for computationally
expensive lightcurve fitting. If fitted successfully, the
combination of finite source effects, and the orbital motion of
the lens can lead to a mass measurement \cite{Dominik:1998mrb},
and possibly a determination of orbital parameters such as
inclination and eccentricity.

%%
%%  Figures
%%

\begin{figure}
\epsfig{file=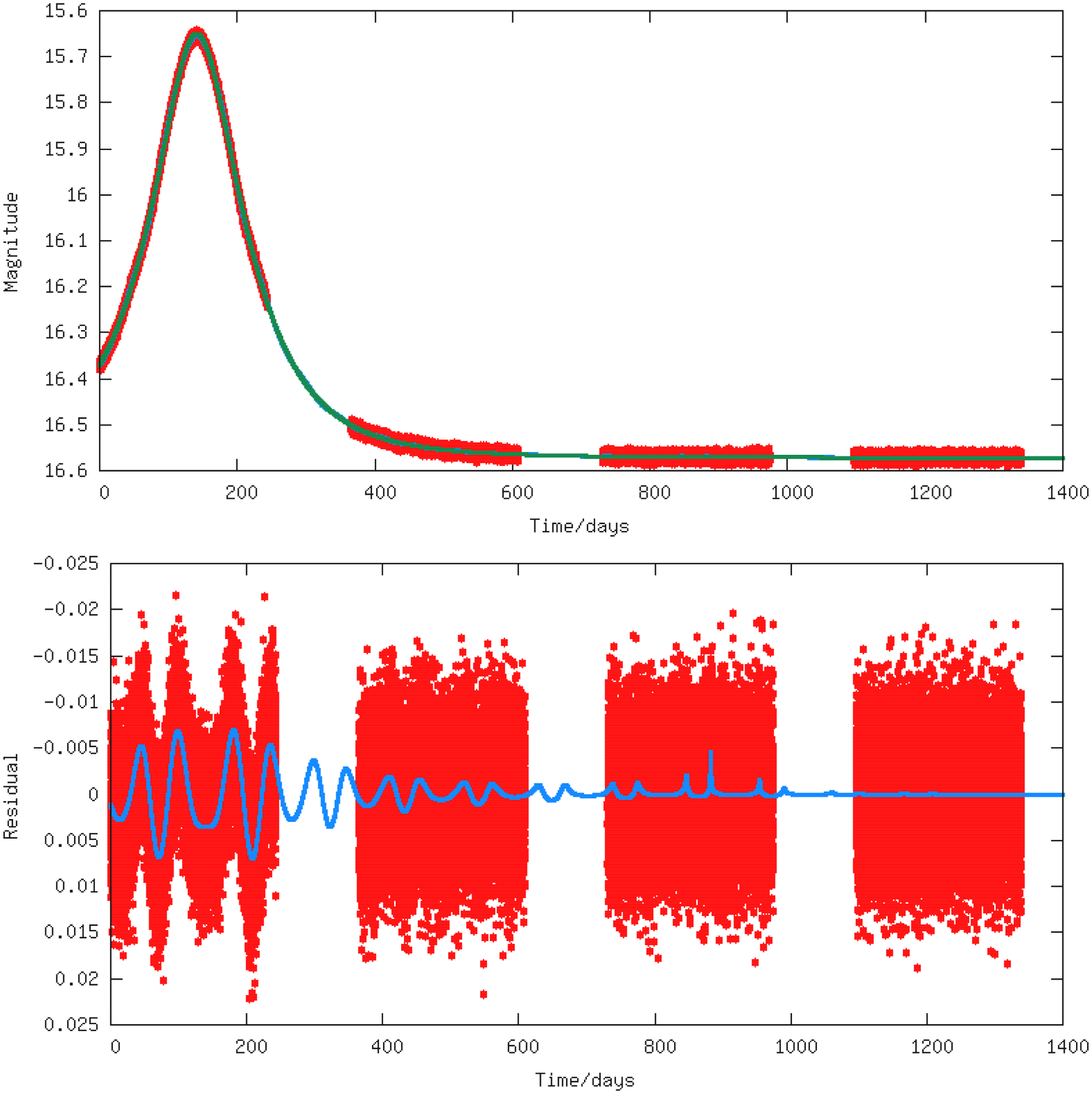,width=\textwidth} \caption{An example
of an RRL. The lightcurve is plotted above, and
  its residual to a single lens lightcurve below. Red points show
  simulated data that would be typical for a space based microlensing
  survey, while the blue line shows the model RRL lightcurve and the
  green line shows the single lens model.}
\label{exampleLC}
\end{figure}

\begin{figure}
\epsfig{file=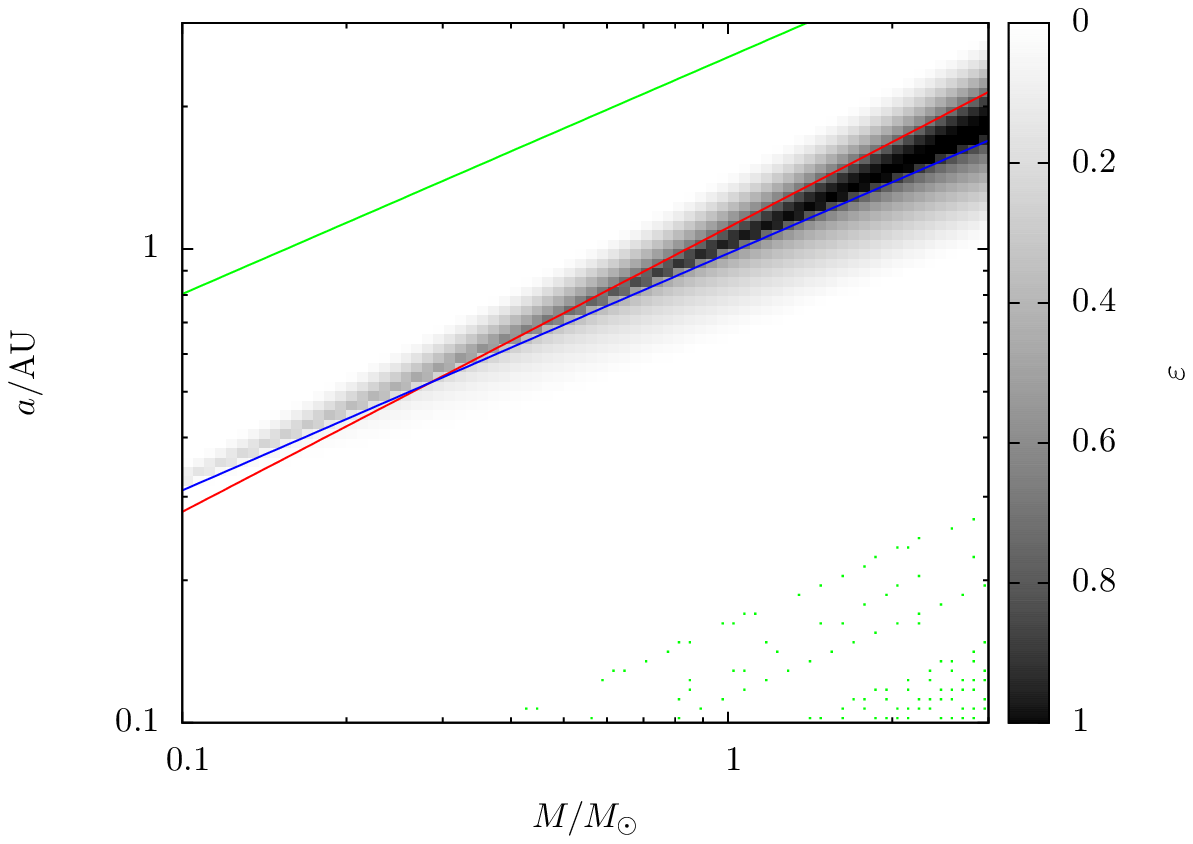,width=\textwidth} \caption{The
region of the total mass -- semimajor axis plane where
  RRLs are detectable, for a typical set of lensing
  parameters. The red and blue lines show the analytically
  derived region of detectability, which agrees well with the region
  of numerical determined region of guaranteed detectability (where $\varepsilon=1$).}
\label{detectability}
\end{figure}

%%%%%%%%%%%%  Section 6   %%%%%%%%%%%%%%%%%%%%%%%%%%%%
\emptyheads

\begin{center}
\begin{large}
{\Huge \sffamily Towards the future: new
facilities/instrumentation/procedures}
\end{large}
\end{center}
\bigskip

\begin{tabbing}
 Brief Reports from the \=B factories xxxxx \= xxxxxxxxx\= \kill
\\
\\
Microlensing with the SONG global network \\
\hspace{0.1cm}{\it Uffe G. J{\o}rgensen, Kennet B. W. Harps{\o}e,
Per K. Rasmussen, Michael I.\,Andersen,} \\
{\it Anton N. S{\o}rensen, J{\o}rgen Christensen-Dalsgaard,
S{\o}ren Frandsen, Frank Grundahl,}\\
{\it Hans Kjeldsen}\\
\\
Next-generation microlensing pilot planet search and the frequency
of planetary systems \\
\>\> {\it Dan Maoz \& Yossi Shvartzvald}\\
\\
Kohyama Astronomical Observatory: current status \\
\hspace{0.5cm} {\it Atsunori Yonehara, Mizuki Isogai, Akira Arai, Hiroki Tohyama}\\
\\
Optimal imaging for gravitational microlensing \\
\hspace{0.1cm}{\it Kennet B. W. Harps{\o}e, Uffe G. J{\o}rgensen,
Per K. Rasmussen, Michael I. Andersen,} \\
{\it Anton N. S{\o}rensen, J{\o}rgen Christensen-Dalsgaard,
S{\o}ren Frandsen, Frank Grundahl,} \\
{\it Hans Kjeldsen} \\
\\
IPAC's role as the science center for NASA's WFIRST mission \\
\>\> {\it Kaspar von Braun}\\
\\
EUCLID microlensing planet hunt \\
\>\> {\it Jean-Philippe Beaulieu \& Matthew Penny}\\
\\
Space-based microlensing exoplanet survey: WFIRST and/or Euclid \\
\>\> {\it David Bennett}\\
\\
Microlensing with Gaia satellite \\
\>\> {\it {\L}ukasz Wyrzykowski}\\

\\
\end{tabbing}

\fancyheads

\Chapter{Microlensing with the SONG global network}
            {Microlensing with the SONG global network}{J{\o}rgensen et al.}
\bigskip\bigskip

\addcontentsline{toc}{chapter}{{\it J{\o}rgensen et al. }}
\label{polingStart}

\begin{raggedright}

{\it Uffe Gr{\aa}e J{\o}rgensen\index{author}{J{\o}rgensen, Uffe
Gr{\aa}e}, Kennet B. W. Harps{\o}e\index{author}{Harps{\o}e,
Kennet B. W.}, Per Kj{\ae}rgaard\index{author}{Kj{\ae}rgaard,
Per}, Michael I.\,Andersen\index{author}{Andersen, Michael I.},
Anton Norup S{\o}rensen\index{author}{S{\o}rensen, Anton Norup}\\
Niels Bohr Institute and Centre for Star and Planet Formation \\
University of Copenhagen \\
Denmark}
\bigskip

{\it J{\o}rgen Christensen-Dalsgaard\index{author}{JCD}, S{\o}ren Frandsen\index{author}{SF},
Frank Grundahl\index{author}{FG}, Hans Kjeldsen\index{author}{HK}\\
Dep. Phys. Astronomy, Aarhus University, Aarhus\\
Denmark\\}

\bigskip

\end{raggedright}

\section{Introduction}

SONG (Stellar Observations Network Group) is a network of 1m
robotic telescopes, designed for microlensing exoplanet follow-up
observations and long time-series of high-precision radial
velocity observations of bright stars for asteroseismology and
exoplant research. The prototype of the telescope is under
installation at Teide observatory and will have first light
mid-2011. Two more nodes are under construction in China and
Argentina, and several national applications are in the process
for further nodes around the globe. The telescope design has been
described in detail recently by Grundahl et al.\,2009~\cite{FG09}.
Here we will describe the design and organization of the network
with emphasis on the microlensing aspects.

\section{The telescope, dome and network construction}
All nodes in the network will have a 1m telescope with two lucky
imaging cameras at the Nasmyth and a spectrograph at the Coud{\'e}
focus. The telescope is produced by Astelco in Germany, the
cameras by Andor in UK, and the spectrograph for the prototype is
build in our workshop at Aarhus University in Denmark. A movable
mirror will direct the light either to the cameras at the Nasmyth
platform or pass it through a vacuum tube into the Coud{\'e} room
in a temperature stabilized adherent shipping container. In the
end of the container facing away from the telescope, and separated
from the Coud{\'e} room, is a small control room with the
connecting computers. The data from each telescope will be
uploaded on-line to a central data centre which is under
construction. All data will be stored at the data centre at least
for the lifetime of the project, and all participants in the
project can access and download all data. A group or an individual
becomes a member of the project by supplying a substantial
contribution to the network, typically a node or the operation of
a node, and is then automatically member of the steering
committee, whose role is to coordinate the scientific and
operational decisions.

\section{The microlensing observations}
Follow-up microlensing exoplanet observations have until today
been performed by networks of existing all-purpose and amateur
telescopes. This situation will change in the near future with 3
modern network coming on-line, the Las Cumbres Observatory Global
Telescope Network, LCOGT (http://lcogt.net/), the Korean
Microlensing Telescope Network, KMT-Net, and SONG
(http://song.phys.au.dk/). While LCOGT has a broad scientific and
educational goal, KMT-Net is focused on microlensing, and has with
its 20Kx20K 0.36$^{\prime\prime}$ pixel size a large survey
capacity together with a high cadence. SONG differs from the two
other networks by being designed for high-spatial-resolution
dedicated high-cadence follow-up microlensing observations. For
this purpose, each of the SONG telescopes have two lucky imaging
cameras at the Nasmyth focus, with a pixel size of only
0.09$^{\prime\prime}$ in order to sample diffraction limited
images with two pixels. Experiments with high frame rate
techniques (also called frame selection, and including the lucky
imaging technique) have demonstrated the ability to obtain
near-diffraction limited images from the ground (e.g.\ Baldwin et
al.\,2001~\cite{B01}\index{B01}, 2008~\cite{B08}\index{B08}, Smith
et al.\,2009~\cite{S09}\index{S09}, Tokovinin et
al.\,2010~\cite{T10}\index{T10}). The advantage of the obtainable
high spatial resolution is in being able to resolve dimmer source
stars in crowded fields. This is the same advantage which has been
pointed out in connection with possible space born microlensing
missions, but here at a few per cent of the cost of a space
mission.

Our present network, MiNDSTEp (http://www.mindstep-science.org/),
was during 2008-2010 operated from the Danish 1.54m telescope
(build in 1975) at ESO's La Sila Observatory, and will from 2011
include also the MONET 1.2m telescope in South Africa
(http://monet.uni-goettingen.de/foswiki) with further support from
the MONET McDonald telescope, the SONG prototype at Tenerife, a
Chinese 1.5m telescope and two survey telescopes (VYSOS,
http://vysos.ifa.hawaii.edu/) in Chile and at Hawaii. The 1.54m
telescope CCD has a pixel size of 0.4$^{\prime\prime}$, and
experience with microlensing research near the Galactic centre has
shown us that the corresponding resolution of 0.8$^{\prime\prime}$
determines a de facto limiting source star magnitude of I$\sim$18.
This corresponds to a limiting stellar radius of a typical K-giant
source star of 10 $R_{\odot}$ at the distance of the Galactic
centre. Experiments with a lucky imaging camera mounted at the
1.54m telescope has resulted in 0.35$^{\prime\prime}$ resolution.
This limit is set by the optics of the telescope, and we expect to
routinely reach 0.25$^{\prime\prime}$ with the SONG telescopes,
which are designed for the higher spatial resolution. This
corresponds to collecting 10 times more photons per area of sky
per time unit with SONG than with the 1.54m telescope with a
conventional CCD, since the higher throughput will compensate for
the smaller mirror size. On top of this, the 4 times broader
filters, the faster slew time, and the more efficient robotic
observing schedule are expected to give us a gain of additional a
factor 10 in number of collected photons per time unit, so
efficiently 100 times more photons per sky area per time unit.
This gain will be divided into observing more targets at lower
exposure times. Allowing to go 2.5 magnitudes deeper in the field
(a factor 10 in number of collected photons) corresponds to a
factor 10 smaller $(\Theta_s)^2$ (with $\Theta_s$ the angular
source radius), and hence a factor 10 smaller reachable
$(\Theta_E)^2$ (with $\Theta_E$ the exoplanetary Einstein radius)
from identical photometric accuracy, corresponding to a factor 10
smaller reachable exoplanetary mass. Present surveys with
conventional CCDs, like the one at the Danish 1.54m telescope,
have allowed detections of exoplanets typically in the mass range
100 to 5 Earth masses, which makes us expect a similar detection
efficiency with SONG for 10 Earth masses to planets smaller than
Earth. Theoretical models predict 50 times higher abundance for
planets in this latter mass range (in the terrestrial zone) than
in the mass range which has been accessible with the present
microlensing surveys, making the coming years a very promising
time for microlensing exoplanetary research.

The present microlensing campaigns have high sensitivity to
Jupiter-Saturn like exoplanets in Jupiter-Saturn like orbits. With
the high sensitivity to low-mass exo\-planets in terrestrial like
orbits expected from coming surveys, and an increasing sensitivity
to giant planets in larger orbits, microlensing campaigns in the
coming decade will be able to answer the question of how abundant
solar system analogues are among the planetary systems in our
galaxy. Also the question about the abundance of exoplanets in
orbits where life as we know it could exist at the planetary
surface is expected to be answered by microlensing in the coming
decade(s), and here we expect SONG to play a central role, due to
its high sensitivity to low-mass planets in the terrestrial zone
(see also J{\o}rgensen 2008~\cite{J08}\index{J08}). In this
connection it is important to stress that the habitable zone most
often is described as the orbital distances where liquid water can
exist at the planetary surface, even when people actually think of
the distances where liquid water could exist at the surface if the
planet was the size of the Earth, had an Earth-like atmosphere,
and rotated as the Earth. In particular the latter point is
important, because it implies that a truly habitable zone might
only exist around solar type stars. Planets orbiting stars with
less than half a solar mass will be tidally locked if they are in
the habitable zone distance, and stars above 2 solar masses have a
main sequence lifetime short compared to biological timescales.
Radial velocity and transit studies are, for technical reasons,
biased toward finding small-mass planets around M-dwarfs.
Microlensing, on the other hand is not biased toward any
particular stellar lens-mass range, and statistically one fourth
of the lensing stars will be solar like (i.e. between half and two
solar masses). SONG will therefore be highly sensitive toward
detecting Earth-mass exoplanets in the (truly) habitable zone
around such lensing stars.

\Chapter{Next Generation Microlensing and the Frequency of
Planetary Systems: Preliminary Results from a Pilot Experiment}
            {Next Generation Microlensing }{Maoz \& Shvartzvald}
\bigskip\bigskip

\addcontentsline{toc}{chapter}{{\it Maoz \& Shvartzvald}}
\label{polingStart}

\begin{raggedright}

{\it Dan Maoz\index{author}{Maoz, Dan} \& Yossi
Shvartzvald\index{author}{Shvartzvald, Yossi}\\
Tel-Aviv University\\
Israel}
\bigskip\bigskip

\end{raggedright}

\section{Introduction}

Current microlensing planet surveys have been based on wide-field,
low-cadence, searches for microlensing events toward the bulge of
the Galaxy by two main surveys: OGLE (Udalski et al. 2008) and MOA
(Sako et al. 2008). These surveys send out public ``alerts'' when
they detect lensing events in which fits to the initial light
curve  suggest high peak magnifications, of order 100 or more. The
high-magnification events constitute about 1 per cent of all
lensing events. Events that appear promising to be of high
magnification trigger intensive photometric followup efforts by
networks of small telescopes (including dedicated amateurs) to
obtain coverage, over the peak of the event, that is as continuous
as possible. The motivation for this observing strategy has been
that, near the peak of a high magnification event, the source star
is distorted into a nearly complete Einstein ring that covers a
large area in the lens plane. The snow lines of typical lens stars
happen to roughly coincide with their Einstein radii. Thus,
planets, if they exist around a lens, and if projected in the
vicinity of the snow line, will always produce significant
perturbations to the light curves of high-magnification lensing
events. There are about 700 lensing events per yearly bulge
season, and of order 10 are of high magnification. Fair coverage
of these events over  their peaks by the current generation of
followup observations has yielded 2-3 planets per year, suggesting
that a fraction of roughly 1/10 to 1/3 of lens stars have planets
near their snow lines (Gould et al. 2010; Sumi et al. 2010). But,
this low detection rate also explains why the current strategy
has, over the past 5 years, produced only about a dozen
microlensing planet discoveries, a tiny fraction of the 500 known
extrasolar planets.

As opposed to the current microlensing surveys, which have focused
on the rare high-magnification lensing events that probe the
``central'' caustics of host+planet lenses, ``Generation-2''
surveys will go after the much more common, low-magnification,
events and the perturbations caused by their ``planetary"
caustics. To harvest this potential new crop of
microlensing-discovered planets requires the combination of a wide
survey area with dense temporal sampling. This has now become
possible with the availability of degree-scale CCD imaging mosaics
on several 1m- to 2m-class telescopes. In Generation-2
microlensing, such telescope-camera combinations, situated around
the globe, continuously monitor about 10~deg$^2$ of the Galactic
bulge with cadences of less than one hour. A large fraction of the
700 or so lensing events that occur per bulge season are
monitored. Among those lenses that host planets, a perturbation,
which can last as little as several hours,
 is detected in the light curve
in those  events where a lensed image happens
 to pass close
to the planet. Simulations by us (see below) and by others (e.g.
Penny et al. 2010) show that this observing approach should
produce dozens of planet discoveries per year. Over several years,
the sample of microlensing-discovered planets could then start to
be comparable to those from other planet-detection methods, but
would probe very different regions of planetary parameter space
than those probed by the other methods.

\section{A Generation-2 Pilot Survey}
In the summer of 2010, we launched the pilot stage of a
Generation-2 microlensing planet survey. During two 3-week
periods, we observed the bulge fields having the highest past
lensing event rates using a network of four telescopes Up to the
unavoidable but moderate weather gaps, we obtained continuous
coverage for 81 microlensing events, with full coverage of the
event for about 15. We plan to continue and expand this survey in
the coming years and to address the following questions: What is
the frequency of occurrence of planets and planetary systems
around disk and bulge stars in the Milky Way? Are most of the
systems Solar analogs, or are they clearly distinct, and in what
ways? What is the emerging picture regarding planetary formation
and evolution, when combining these results with those from the
complementary, ongoing, radial-velocity
 and transit
surveys (e.g. Howard et al. 2010)?

The pilot phase took place in June-July 2010 during two 3-week
periods (separated by a one-week break around the full moon), when
we monitored the $8$~deg${^2}$ of the Galactic bulge that have the
highest microlensing event rates. We used the 1m-telescope at Wise
Observatory in Israel with the LAIWO camera (1 deg$^2$ field of
view). Our observations were coordinated with those of the  1.3m
Warsaw University telescope in Chile with the OGLE-IV camera (1.4
deg$^2$ field of view); the  1.8m MOA-II telescope in New Zealand
with the MOA-cam3 camera (2.2 deg$^2$ field of view); and the
Palomar Observatory 1.2m Oschin telescope in California with the
PTF camera (7.8 deg$^2$; Law et al. 2009; this telescope
participated only in the second 3-week run). Observing cadence at
each telescope was between 20 and 40 minutes.
 During each of the 3 weeks, coverage
was about 90\% of fully continuous. During our pilot, 81 lensing
events (identified as such by MOA) were covered or partly covered
by our observations. We are performing difference-image analysis
(DIA) on the data for these events, to obtain the most accurate
photometry for each event. We will show some of the emerging light
curves, based on preliminary analysis of the data from Wise, MOA,
and PTF (data obtained by OGLE are still being reduced).

\section{Numerical Simulations and Future Prospects}

In parallel, we have carried out numerical ray-tracing
simulations, both for planning the experiment and with a view
toward analysing the results, whether statistically or in terms of
modeling individual events. We begin by producing large numbers of
simulated light curves. The physical parameters -- lens mass,
distance, and the effective lens-source velocities -- are drawn
from the distributions of those parameters that, when combined
with Galactic structure models, reproduce the lens-crossing
timescale distributions observed in real lensing events of the
past years (Dominik 2006). Impact parameters
 and the magnitudes of the source stars (and hence the photometric
 errors, see below) are also chosen to reproduce
the observed distributions. The interesting but unknown physical
parameters -- number, orbital separation, and masses of the
planets, are input according to various possible prescriptions.
For example, one possibility is that some fraction $f$ of all
stars hosts planetary systems that mimic the Solar System, but the
planets are scaled in mass and orbital radius with the host mass
$M$, according to the snow-line radius, with some power-law of
index $s$, $R_{\rm snow}\propto M^s$. Thus, for a choice of $s$
and $f$, we can produce many systems with random orbital phases of
their planets and random line-of-sight inclinations of their
orbital planes. We then ray trace through them to obtain simulated
light curves. The light curves from a particular choice of model
parameters are then ``observed'', in terms of including all the
observational effects:
 we temporally sample the light curves
 with observing sequences drawn from the real existing
observations, and scatter the photometry based on a realistic
measurement error distribution, as observed in the real data (and
thus including all error types -- Poisson, backgrounds from bright
stars, outliers due to improperly subtracted DIA residuals). We
then search these simulated light curves for planetary deviations.
For a given combination of physical parameters, the simulations
can give us the probability of the observed lensing statistics,
such as
 number and type of perturbations,
and will thus permit to constrain the interesting physical
parameter space.

We will show
 predictions from our Monte Carlo simulations, in
  which, for various combinations of planetary population parameters
and experiment parameters, a
  Generation-2 survey is
  repeated many times.
This gives
  the Monte-Carlo distribution of the number of detected
  planetary events per season, for a combination of seasonal experiment
  duration, observing cadence,
fraction $f$ of stars hosting Solar-like planetary
  systems, and power-law index $s$ relating system size to host-star
  mass.
 The results of the real experiment, when
  compared with the
  distributions from such mock experiments, will permit
  constraining the physical parameter space.
Our current calculations indicate that we can realistically expect
to discover up to a few dozen planets per bulge season, and up to
the order of 100 planets in the course of several years.

\def\mnras{Monthly Notices of the Royal Astronomical Society}
\def\apj{Astrophysical Journal}
\def\apjs{Astrophysical Journal Supplement}
\def\apjl{Astrophysical Journal Letters}
\def\aj{Astronomical Journal}
\def\aap{Astronomy \& Astrophysics}
\def\pasp{Publications of the Astronomical Society of the Pacific}
\def\actaa{Acta Astronomica}

\Chapter{Kohyama Astronomical Observatory: Current Status}
            {Kohyama Astronomical Observatory: Current Status}{Yonehara et al.}
\bigskip\bigskip

\addcontentsline{toc}{chapter}{{\it Yonehara et al.}}
\label{polingStart}

\begin{raggedright}

{\it Atsunori Yonehara\index{author}{Yonehara, Atsunori}, Mizuki
Isogai\index{author}{Isogai, Mizuki},
Akira Arai\index{author}{Arai, Akira}, Hiroki Tohyama\index{author}{Tohyama, Hiroki}\\
Department of Physics, Faculty of Science, Kyoto Sangyo
University, JAPAN\\}

\bigskip\bigskip

\end{raggedright}

\section{Introduction}

Kohyama Astronomical Observatory locates at north rim of Kyoto
city. The location is not so ideal for astronomical observation,
especially for observing southern sky, but quite convenient for
any urgent observation program such as follow-up observations of
galactic microlensing. Construction of our observatory has been
finished on 22nd December 2009, and we have 1.3-meter, optical and
near-infrared, telescope (we call it Araki Telescope) in our
observatory.

\section{Dual Band Imager, ADLER}

After the construction has been finished, we equipped several
observational instruments to our telescope. The most important
instrument for microlensing observation is a dual band imager,
ADLER (Araki telescope DuaL band imagER). ADLER is equipped at
Cassegrain focus of our telescope, and the wavelength covergae is
from 380nm to 900nm. Photons are divided in two wavelength regime,
from 380nm to 670nm and from 670nm to 900nm by dichroic mirror.
Gunn-g' filter is available at shorter wavelength regime, and
Gunn-i' and -z' filters are available at longer wavelength regime
currently. The field of view is $12^\prime \times 12^\prime$ which
is covered by 2k $\times$ 2k CCD camrea, Spectral Instruments 850
series. Recently, regular observations by using ADLER starts from
this October, and now we have several scientific targets for
observations with ADLER.

For microlensing comunity, it should be mentioned that galactic
and quasar microlensing observations are one of the most important
targets in our observatory, since other targets are cataclysmic
variables, comets, and transits. Thanks to the dual band imaging
and wide field of view, we are able to perform DIA photometry for
galactic microlening with two waveband at the same time. Dual band
imaging is not so essential for planet microlensing huntings, but
chromaticity during quasar microlensings provides us important
information for quasar central engine.

\section{Current Status}

We have already made several photometries for cataclysmic
variables and transits, and we are able to achive less than 1
percent accuracy for relative photometry with ADLER in both
wavelength ragime. Sky brigntness strongly depends on wavelength
regime and sky direction. Shorter wabelength regime and southern
sky is worse due to Kyoto city lights, and some quantitative
studies for such dependence will be presented. Now we are
accumulating various targets for scientific purpose and for
checking capabilities of our telescope/instruments.

Further, some observational data (transits, quasars), weather
conditions (typically 30 percent fine weather) and seeing
(typically a few arcsec) in a half year or so will be shown in
presentation.

\bigskip
I am grateful to other members in Kohyama Astronomical
Observatory.

%\begin{thebibliography}{99}
%
%%
%%  bibliographic items can be constructed using the LaTeX format in SPIRES:
%%    see    http://www.slac.stanford.edu/spires/hep/latex.html
%%
%
%\end{thebibliography}

\Chapter{Optimal Imaging for Gravitational Microlensing}
            {Optimal Imaging for Gravitational Microlensing}{Harps{\o}e et al.}
\bigskip\bigskip

\addcontentsline{toc}{chapter}{{\it Harps{\o}e et al.}}
\label{polingStart}

\begin{raggedright}

{\it Kennet B.W. Harps{\o}e\index{author}{Harpsoe, K.},
Uffe G. J\o rgensen\index{author}{Jorgensen, U.},
Per K. Rasmussen\index{author}{Rasmussen, P.},
Michael I. Andersen\index{author}{Andersen, M.},
Anton N. S{\o}rensen \index{author}{Sorensen, A.}\\
The Niels Bohr Institute, University of Copenhagen \\
Denmark \\}
\bigskip

{\it J{\o}gen
Christensen-Dalsgaard\index{author}{Christensen-Dalsgaard, J.},
S{\o}ren Frandsen\index{author}{Frandsen, S.},
Frank Grundahl\index{author}{Grundahl, F.}, Hans Kjeldsen\index{author}{Kjeldsen, H}\\
Dep.Phys.Astronomy, Aarhus University\\
Denmark}

\bigskip\bigskip
\end{raggedright}

\section{Introduction}
High framerate imaging of astronomical sources offers new
possibilities for improved resolution and short timescale
photometry. Image motion due to turbulence is the most important
contribution to atmospheric seeing. By taking images at a
framerate comparable to the dynamic timescale of the atmosphere
(10-100Hz), it is possible to ``freeze'' the motion and correct
for it, thereby improving the image quality significantly. One
well known variations of this is the lucky imaging method, where
only nearly diffraction limited frames are kept and co added. The
main obstacle for pursuing this path in astronomy has been read
out noise in CCD's. The large number of fames needed to collect
appreciable amounts of photons from faint astronomical sources at
the required framerates implies that the signal is completely
overwhelmed by the read out noise typical of CCD's.

\section{The EMCCD}

The invention of the EMCCD (Electron Multiplying CCD) has made
read out noise on the sub electron level possible. The read out
register of an EMCCD has been extended into a solid state version
of a photomultiplier. This makes it possible to amplify a single
photoelectron into hundreds of electrons, thereby overcoming the
read out noise. But the electron cascades resulting from the
amplification are stochastic, and new methods must be employed to
reduce the images for extracting photometric information. It is
reasonable to assume that the distribution of the final size of
the cascades is exponential, with some decay constant called the
EM gain. For instance, spurious charge is present but undetectable
in all conventional CCDs. It is visible in an EMCCD, making it
impossible to estimate the bias from a simple average of bias
frames.

%%%%%%%%%%%%%%%%%%%%%%%%%%%%%%%%%%%%%%%%%%%%%%%%%%%%%%%%%%%%%%%%%%%%%%%%%
%%
%%   use this format to include an .eps figure into your paper
%%
%\begin{figure}[htb]
%\begin{center}
%\epsfig{file=rgb.eps,height=1.5in}
%\caption{Plan of the magnet used in the Mesmeric studies.}
%\label{fig:magnet}
%\end{center}
%\end{figure}
%%%%%%%%%%%%%%%%%%%%%%%%%%%%%%%%%%%%%%%%%%%%%%%%%%%%%%%%%%%%%%%%%%%%%%%%%%%

EMCCD's can be operated in two distinct ways: a photon counting
mode with only shot noise, but with coincidence losses at high
flux levels; or a linear mode with shot noise inflated by a factor
of two. At present, we are employing the linear mode. But real
photon counting optimal imaging is theoretically possible and a
very interesting concept.

%%%%%%%%%%%%%%%%%%%%%%%%%%%%%%%%%%%%%%%%%%%%%%%%%%%%%%%%%%%%%%%%%%%%%%%%%
%%
%%   use this format to include a LaTeX table  into your paper
%%
%\begin{table}[b]
%\begin{center}
%\begin{tabular}{l|ccc}
%Patient &  Initial level($\mu$g/cc) &  w. Magnet &
%w. Magnet and Sound \\ \hline
% Guglielmo B.  &   0.12     &     0.10      &     0.001  \\
% Ferrando di N. &  0.15     &     0.11      &  $< 0.0005$ \\ \hline
%\end{tabular}
%\caption{Blood cyanide levels for the two patients.}
%\label{tab:blood}
%\end{center}
%\end{table}
%%%%%%%%%%%%%%%%%%%%%%%%%%%%%%%%%%%%%%%%%%%%%%%%%%%%%%%%%%%%%%%%%%%%%%%%%%%

\section{Optimal Imaging Methods}

If a standard Kolmogorov model for atmospheric turbulence is
assumed, the six most dominant Zernike terms aberrating an image
will be piston, tip and tilt, focus and the two astigmatism terms,
listed in descending order of power \cite{Tatarski}. Piston is
unimportant for imaging, and tip and tilt result in an overall
solid-body translation of the image that can be corrected simply
by shifting the images and co-add. This methods colloquially know
as shift and add.

Focus, astigmatism, and higher order terms will blur the images.
The amount of blurring in the individual frames will be random;
sometimes one will be lucky and obtain an image near the
diffraction limit. By collecting only these images, then shifting
and co-adding them, nearly diffraction limited images with useful
signal to noise ratios can be obtained, hence this method is known
as lucky imaging \cite{Tubbs}.

All of the images obtained in a high framerate sequence will
contain information, though, and for every combination of target
and desired measurement there must exist an optimal weighting
scheme for combining the images.

\section{Optimal Imaging in Microlensing}

The optimal imaging techniques are highly interesting for
gravitational microlensing as it can resolve much of the flux
blending in the dense fields where microlenses are most often
observed. Furthermore the technique works best on the small size
telescopes often used for microlensing followups, and it is
relatively inexpensive and straightforward to implement on
telescopes.

From experience we know that with the DFOSC conventional CCD
camera, at the Danish 1.54m telescope at the ESO La Silla
observatory, we can get useful photometry from stars down to I=18,
but not fainter due to crowding. With the improved resolution from
the Lucky Imaging we expect to be able to do photometry on stars
down to about I=21 \cite{Holtzman}. Assuming a constant distance
and surface temperature of the source stars, this allows us to
track microlensing event in stars with ten times smaller
${\Theta_s}^2$, where $\Theta_s$ is the angular source size. Which
means that planets with squared Einstein radii ten times smaller
will be detectable without being washed out by finite source
effects; this corresponds to being able to detect planets with one
tenth of the mass. \cite{Jorgensen}.

\Chapter{IPAC's Role as the Science Center for NASA's WFIRST
Mission}
            {IPAC's Role as the Science Center for NASA's WFIRST Mission}{Kaspar von Braun}
\bigskip\bigskip

\addcontentsline{toc}{chapter}{{\it Kaspar von Braun}}
\label{polingStart}

\begin{raggedright}

{\it Kaspar von Braun\index{author}{von Braun, Kaspar}\\
NASA Exoplanet Science Institute\\
California Institute of Technology \\
USA\\}

\bigskip\bigskip
\end{raggedright}

I will discuss the roles NASA's Infrared Processing and Analysis
Center (IPAC) and the NASA ExoPlanet Science Institute (NExScI)
will play in the WFIRST mission. As the Science Center for WFIRST,
IPAC will work with the project, the Science Definition Team, and
scientific community to develop all three of WFIRST's scientific
themes in preparation for a new star later in the decade: Dark
Energy, a near-IR survey, and a microlensing survey to conduct a
census of planets covering a wide range of masses and orbital
periods. In the near-term, NExScI will continue to develop its
archive of microlensing events and lightcurves analogous to its
archives of transit datasets. In addition, the topic of NExScI's
next Sagan Exoplanet Summer Workshop will be planet searches using
microlensing (http://nexsci.caltech.edu/workshop/2011/). IPAC will
be working with the microlensing community as their science
advocate within the WFIRST project, developing relevant archives
and tools in the decade leading up to the launch of WFIRST, and
ultimately in developing mission products that will take full
advantage of WFIRT's remarkable capabilities.

\Chapter{EUCLID: From frozen Mars to habitable Earth via
microlensing}
            {EUCLID: From frozen Mars to habitable Earth via microlensing }{Beaulieu \& Penny}
\bigskip\bigskip

\addcontentsline{toc}{chapter}{{\it Beaulieu \& Penny}}
\label{polingStart}

\begin{raggedright}

{\it Jean-Philippe Beaulieu \index{author}{Beaulieu, Jean-Philippe}\\
Institut d'Astrophysique de Paris\\
France\\}
\bigskip
{\it Matthew Penny \index{author}{Penny, Matthew}\\
Manchester University\\
UK\\}

\bigskip\bigskip
\end{raggedright}

In the last fifteen years, astronomers have found over 500
exoplanets including some in systems that resemble our very own
solar system. These discoveries have already challenged and
revolutionized our theories of planet formation and dynamical
evolution.  Several different methods have been used to discover
exoplanets, including radial velocity, stellar transits, direct
imaging, pulsar timing, astrometry, and gravitational microlensing
which is based on Einstein's theory of general relativity. So far
10 exoplanets have been published with this method. While this
number is relatively modest compared with that discovered by the
radial velocity method, microlensing probes a part of the
parameter space (host separation vs. planet mass) not accessible
in the medium term to other methods. The mass distribution of
microlensing exoplanets has already revealed that  cold
super-Earths (at or beyond the ``snow line''  and with a mass of
around 5 to 15 Earth mass appear to be common (Beaulieu et al.,
2006, Gould et al., 2006, Sumi et al. 2010) . We detected a scale
1/2 model of our solar system (Gaudi et al., 2008), several hot
Neptunes/Super Earth, shown that our detection efficiencies
extends to 1 Earth mass planets (Batista et al., 2009). We have
made the first measurement of the frequency of ice and gas giants
beyond the snow line, and have shown that this is about 7 times
higher than closer-in systems probed by the Doppler method.  This
comparison provides strong evidence that most giant planets do not
migrate very far (Gould et al. 2010).

Microlensing is currently capable of detecting cool planets of
super-Earth mass from the ground (and on favourable circumstances
down to 1 Earth), with a network of wide-field telescopes
strategically located around the world, could routinely detect
planets with mass as low as the Earth. Old, free-floating planets
can also be detected; a significant population of such planets are
expected to be ejected during the formation of planetary systems.
Microlensing is roughly uniformly sensitive to planets orbiting
all types of stars, as well as white dwarfs, neutron stars, and
black holes, while other method are most sensitive to FGK dwarfs
and are now extending to M dwarfs. It is therefore an independent
and complementary detection method for aiding a comprehensive
understanding of the planet formation process. Ground-based
microlensing mostly probes exoplanets outside the snow line, where
the favoured core-accretion theory of planet formation predicts a
larger number of low-mass exoplanets (Ida \& Lin 2005). The
statistics provided by microlensing will enable a critical test of
the core accretion model.  Exoplanets probed by microlensing are
much further away than those probed with other methods. They
provide an interesting comparison sample with  nearby exoplanets,
and allow us to study the extrasolar population throughout the
Galaxy.

Ultimately, a comprehensive census of cold planets below Earth
masses, and habitable Earth mass planets requires a space-based
microlensing survey. The remarkable synergy between Dark Energy
probes by Cosmic shear has been realized and implemented in Europe
since 2006 with the DUNE proposal, and has been developed in
EUCLID. A 3 month microlensing program is part of the additional
science of EUCLID aiming at low mass telluric planets down to the
mass of Mars at the snow line and first hint on habitable super
Earth (described in the yellow book and in Beaulieu et al. 2010).
There is a proposition for an extension of the initial 3 month
``additional science'' program by a 9 months ``legacy program''
with the objective of measuring the abundance of habitable Earth
around solar like stars. This legacy program could take place once
the Dark Energy objective will have been reached, at the end of
the mission. We suggest to have a first survey of 2 months shortly
after the launch to guaranty high profiles results (detection of
planets down to the mass of Mars) early in the mission. Results
with high visibility would be valuable for EUCLID within the first
year of operation. It is important to have observations early in
the mission life time and after few years to maximize the baseline
in order to be able to have better constraints on host masses.

With the 9 month legacy program in addition to the 3 month that
are currently in the additional science of EUCLID we will provide:
a) a complete census of planets down to Earth mass with
separations exceeding 1 AU b) complementary coverage to Kepler of
the planet discovery space c) sensitivity to planets down to 0.1
$M_{\odot}$, including all Solar System analogues except for
Mercury d) complete lens solutions for most planet events,
allowing direct measurements of the planet and host masses, and
distance from the observer.

\Chapter{Space-Based Microlensing Exoplanet Survey: WFIRST and/or
Euclid}
            {Space-Based Microlensing Exoplanet Survey: WFIRST and/or Euclid}{David P. Bennett}
\bigskip\bigskip

\addcontentsline{toc}{chapter}{{\it David P. Bennett}}
\label{polingStart}

\begin{raggedright}

{\it David P. Bennett\index{author}{Bennett, David P.}\\
University of Notre Dame, Notre Dame, IN\\
USA\\}

\bigskip\bigskip
\end{raggedright}

The US Astro2010 Decadal Survey report, ``New Worlds, New Horizons
in Astronomy and Astrophysics'' recommended a new mission called
WFIRST as its top ranked large space mission for the next decade.
The WFIRST mission is to have two major science programs that will
drive the design requirements: a dark energy program and a
microlensing planet search program, based on the Microlensing
Planet Finder (MPF) concept. ESA is considering a similar mission,
known as Euclid, which addresses dark energy, but also has a
modest microlensing program that could be expanded. Both NASA and
ESA have expressed interest in a possible joint program to address
both the WFIRST and Euclid science goals, assuming that Euclid is
selected. However, there are a number of political and technical
issues that would complicate such a joint program.

I discuss the science justification for the WFIRST microlensing
program, and consider a variety of mission designs that might be
considered by the WFIRST Science Definition Team.

\Chapter{Microlensing with Gaia satellite}
            {Microlensing with Gaia satellite}{Wyrzykowski \& Belokurov}
\bigskip\bigskip

\addcontentsline{toc}{chapter}{{\it Wyrzykowski \& Belokurov}}
\label{polingStart}

\begin{raggedright}

{\it {L}ukasz Wyrzykowski\index{author}{Wyrzykowski, {L}ukasz}, Vasily Belokurov\index{author}{Belokurov, V.}\\
Institute of Astronomy, University of Cambridge \\
UK}
\bigskip\bigskip

\end{raggedright}

% Gaia - what is it
Gaia is an ambitious astrometric survey satellite that will be
launched by ESA in 2012/2013. Its unique combination of precise
astrometry, photometry and spectroscopy will have a tremendous
impact on many fields of astrophysics, from Galaxy structure,
stellar models, variable stars to asteroids and extra-solar
planets \cite{Perryman2001}.

Gaia offers a unique contribution to the field of microlensing
with its uniform continuous all-sky coverage over 5 years and a
superb astrometry as its primary goal. When employed in the
microlensing studies it would provide an one-off opportunity for
studying microlensing events. Sole ground-based photometric
observations only in very special cases allow for solving and
working out the combination of parameters of the microlensing
event participating objects.

The astrometric accuracy of a single 1 dimensional measurement in
Gaia is about 50 $\mu$as at 14 mag and 400 $\mu$as at 18 mag. It
is enough to detect typical displacement of the centroid of the
images of 1 mas. About 15,000 astrometric events are expected  to
occur over the whole sky during the 5 year of the mission, however
only in few percent of them the mass of the lens can be recovered
with a reasonable accuracy \cite{Belokurov2002}.

\begin{figure}[htb]
\begin{center}
\epsfig{file=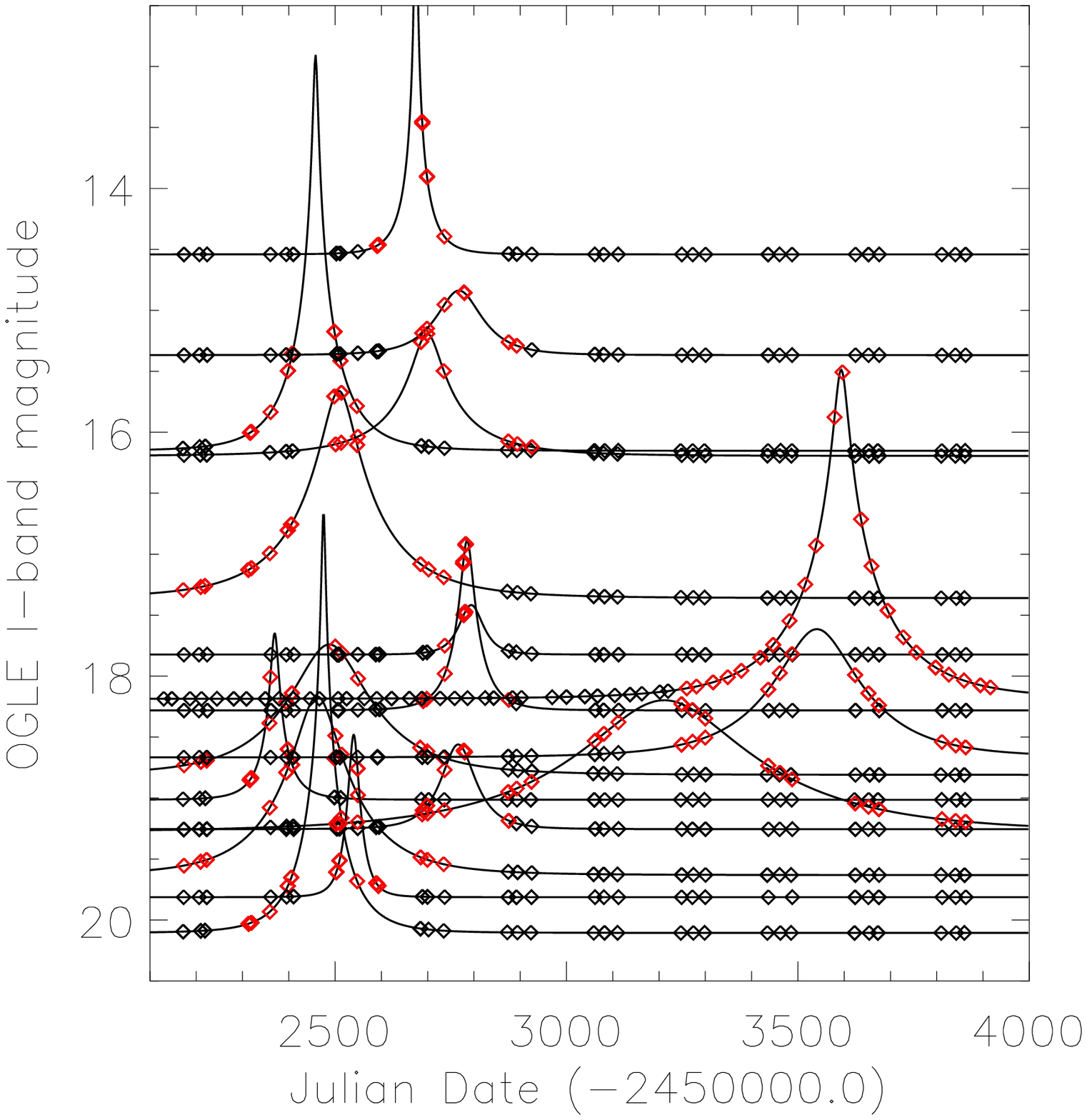,height=3.5in} \caption{Selected OGLE
microlensing events as would have been observed by Gaia. Red
indicate data points taken during the actual event. Only shown are
events with the number of data points during the event more than
20.} \label{fig:ogleevents}
\end{center}
\end{figure}

% Science Alerts - design, operation
Gaia sparse sampling prevents from obtaining detailed light curves
of the events, see example in Fig. \ref{fig:ogleevents}, hence the
characterisation of the events is difficult. Photometric and
astrometric microlensing deviations will be detected by the Gaia
Science Alerts system, currently being developed and installed in
Cambridge. The system will operate in near-real-time, analysing
all data incoming from the satellite as soon as they are arrive
and are preliminarily processed, usually between couple and 24
hours after observation. It will utilise all Gaia data available
at these time-scales, including low-resolution spectroscopy.
Thanks to that the classification of the observed deviations will
be less affected by false-positives.

Gaia Science Alerts will therefore release alerts to the community
about plausible microlensing photometric and astrometric
deviations. However, without well-organized ground-based all-sky
follow-up of the events detected by Gaia it will not be possible
to fully exploit the scientific potential of these microlensing
events. When Gaia data are supplemented by ground-based
observations the number of measured masses increases by several
factors.

Due to characteristic sampling of Gaia vast majority of alerted
events will usually have long time-scales. The detection
efficiency reaches 100\% for events with time-scales of around 100
days. These will be mainly caused by relatively nearby disk
lenses, exhibiting a parallax effect and allowing for probing the
local mass function. Moreover, Gaia will create an opportunity to
detect massive stellar remnants, like neutron stars or
black-holes, acting as lenses in the detected events.

Gaia Science Alerts Working Group wiki is available at this address:\\
{\it http://www.ast.cam.ac.uk/research/gsawg}.

This is a platform containing important information about the
design and operation of the alerting system. It is also a site
where we gather all relevant information about possible triggers
of alerts. One can also find there archived talks and presentation
from the first Science Alerts workshop held in June 2010 in
Cambridge.

The website also provides information about the forthcoming
workshop in June 2011, dedicated to the follow-up and verification
of the Gaia Science Alerts.

%%%%%%%%%%%%  Section 7   %%%%%%%%%%%%%%%%%%%%%%%%%%%%
\emptyheads

\begin{center}
\begin{large}
{\Huge \sffamily Poster session}
\end{large}
\end{center}
\bigskip

\begin{tabbing}
 Brief Reports from the \=B factories xxxxx \= xxxxxxxxx\= \kill
\\
\\
Critical curve topology in special triple lens configurations \\
\>\> {\it Kamil Danek}\\
\\
PAndromeda - A Dedicated Deep Survey of M31 with Pan-STARRS 1 \\
\hspace{0.5cm}{\it Chien-Hsiu Lee, Arno Riffeser, Stella Seitz, Ralf Bender, Johannes Koppenhoefer}\\

\\
\end{tabbing}

\fancyheads

\Chapter{Critical curve topology in special triple lens
configurations}
            {Critical curve topology in special triple lens configurations}{Dan\v{e}k \& Heyrovsky}
\bigskip\bigskip

\addcontentsline{toc}{chapter}{{\it Dan\v{e}k \& Heyrovsky}}
\label{polingStart}

\begin{raggedright}

{\it Kamil Dan\v{e}k\index{author}{Dan\v{e}k, Kamil}, \& David Heyrovsky\index{author}{Heyrovsky, David}\\
Institute of Theoretical Physics, Charles University in Prague \\
Czech Republic}

\bigskip\bigskip
\end{raggedright}

Inspired by the Erdl \& Schneider \cite{Singularities 1-2-point
lens} analysis of the parameter dependence of binary lensing
topologies,  we extend their approach to special cases of the
triple lens. While the binary lens is characterised by two
parameters, three more parameters are needed to describe the
triple lens. We analysed several two-dimensional cuts through the
five-dimensional parameter space,  identifying the boundaries of
regions with different critical curve topology. For each region we
present corresponding critical curves and caustics.

\Chapter{PAndromeda - A Dedicated Deep Survey of M31 with
Pan-STARRS 1}
            {PAndromeda - A Dedicated Deep Survey of M31 with Pan-STARRS 1}{Lee et al.}
\bigskip\bigskip

\addcontentsline{toc}{chapter}{{\it Lee et al.}}
\label{polingStart}

\begin{raggedright}

{\it Chien-Hsiu Lee\index{author}{Lee, Chien-Hsiu}, Arno
Riffeser\index{author}{Riffeser, Arno}, Stella
Seitz\index{author}{Seitz, Stella}, Ralf
Bender\index{author}{Bender, Ralf},
Johannes Koppenhoefer\index{author}{Koppenhoefer, Johannes }\\
Max Planck Institute for Extraterrestrial Physics, Garching \\
Germany\\}

\bigskip

\end{raggedright}

{\it Pan-STARRS 1 Science Consortium\index{PS1SC}\\
  University of Hawaii, Pan-STARRS Project Office,
  Max Planck Institute for Astronomy, Max Planck
  Institute for Extraterrestrial Physics, Johns Hopkins
  University, University of Durham, University of Edinburgh,
  Queens University of Belfast, Harvard-Smithsonian Center
  for Astrophysics, Los Cumbres Observatory Global Telescope
  Network, National Central University of Taiwan\\}

\bigskip
The Pan-STARRS 1 (PS1) survey of M31 (PAndromeda) is designed to
identify gravitational microlensing events, caused by bulge and
disk stars (self-lensing) and by compact matter in the halos of
M31 and the MW (halo lensing, or lensing by MACHOs).  PAndromeda
will improve our understanding of the M31 distance, the mix of
stellar ages and metalicities of the M31 disk, bulge, halo,
stellar streams \& dwarfs, it will constrain the extinction within
M31 and will provide insight into stellar population properties
based on the color profiles, SFB-fluctuations, resolved stars and
variables. These results are required for an accurate
interpretation of the microlensing events.  Here we present
photometric observations of gravitational microlensing candidates
discovered during the first season in 2010.

%%%%%%%%%%%%%%%%%%%%%%%%%%%%%
%       BACKMATTER
%%%%%%%%%%%%%%%%%%%%%%%%%%%%%
\emptyheads

\end{document}